\pdfoutput=1

\documentclass[sensors,article,submit,moreauthors,pdftex]{mdpi}

\firstpage{1} 
\pubvolume{xx}
\issuenum{1}
\articlenumber{5}
\pubyear{2020}
\copyrightyear{2020}
\history{Received: date; Accepted: date; Published: date}

\usepackage{amsfonts}
\usepackage{amssymb}
\usepackage{amstext}
\usepackage[separate-uncertainty,multi-part-units = single, allow-number-unit-breaks]{siunitx}
\usepackage{subfig}
\usepackage{wasysym}

\Title{Dosimetry and calorimetry performance of a scientific CMOS camera for environmental monitoring}

\Author{Alexis Aguilar-Arevalo\orcidE{} $^{4}$, Xavier Bertou\orcidH{} $^{1}$, Carles Canet\orcidI{} $^{2}$, Miguel Angel Cruz-P\'erez $^{6}$, Alexander Deisting\orcidA{} $^{9}$*, Adriana Dias\orcidM{} $^{9}$, Juan Carlos D'Olivo $^{4}$, Francisco Favela-P\'erez $^{1,4}$, Estela A. Garc\'es\orcidD{} $^{5}$, Adiv Gonz\'alez Mu\~noz $^{5}$, Jaime Octavio Guerra-Pulido\orcidJ{} $^{4}$, Javier Mancera-Alejandrez $^{3}$, Daniel Jos\'e Mar\'{i}n-L\'ambarri\orcidG{} $^{5}$, Mauricio Martinez Montero $^{4}$, Jocelyn Monroe\orcidK{} $^{9}$, Sean Paling $^{7}$, Simon J. M. Peeters\orcidB{} $^{8}$, Paul Scovell $^{7}$, Cenk T\"urko\u{g}lu\orcidC{} $^{8,\dagger}$, Eric V\'azquez-J\'auregui\orcidF{} $^{5}$, Joseph Walding\orcidL{} $^{9}$}

\AuthorNames{Alexis Aguilar-Arevalo and Xavier Bertou and Carles Canet and Miguel Angel Cruz-P\'erez and Alexander Deisting and Adriana Dias and Juan Carlos D'Olivo and Francisco Favela-P\'erez and Estela A. Garc\'es and Adiv Gonz\'alez Mu\~noz and Jaime Octavio Guerra-Pulido and Javier Mancera-Alejandrez and Daniel Jos\'e Mar\'{i}n-L\'ambarri and Mauricio Martinez Montero and Jocelyn Monroe and Sean Paling and Simon J. M. Peeters and Paul Scovell and Cenk T\"urko\u{g}lu and Eric V\'azquez-J\'auregui and Joseph Walding}

\address{
\noindent Argentina:\\ 
$^{1}$ \quad Centro At\'omico Bariloche, CNEA/CONICET/IB, Bariloche;\\
Mexico:\\
$^{2}$ \quad Centro de Ciencias de la Atm\'osfera. Universidad Nacional Aut\'onoma de M\'exico, CDMX, 04110 M\'exico\\
$^{3}$ \quad Facultad de Ingenier\'ia, Universidad Nacional Aut\'onoma de M\'exico\\
$^{4}$ \quad Instituto de Ciencias Nucleares, Universidad Nacional Aut\'onoma de M\'exico, A. P. 70-543, M\'exico, D.F. 04510\\
$^{5}$ \quad Instituto de F\'isica, Universidad Nacional Aut\'onoma de M\'exico, A. P. 20-364, M\'exico D. F. 01000\\
$^{6}$ \quad Programa de Posgrado en Ciencias de la Tierra. Universidad Nacional Aut\'onoma de M\'exico, Ciudad Universitaria, 04510 Coyoac\'an\\
United Kingdom:\\
$^{7}$ \quad Boulby Underground Laboratory, Boulby Mine, Saltburn-by-the-Sea\\
$^{8}$ \quad Department of Physics and Astronomy, University of Sussex, Brighton\\
$^{9}$ \quad Royal Holloway, University of London, Egham Hill}

\corres{Correspondence: alexander.deisting@cern.ch}

\firstnote{Current address: Particle Astrophysics Science And Technology Centre, Warsaw, Poland} 

\abstract{This paper explores the prospect of CMOS devices to assay lead in drinking water, using calorimetry. Lead occurs together with traces of radioisotopes, \textit{e.g.} \pb{210}, producing $\gamma$-emissions with energies ranging from \SI{10}{\kilo\electronvolt} to several \SI{100}{\kilo\electronvolt} when they decay; this range is detectable in silicon sensors. In this paper we test a CMOS camera (\textsc{Oxford Instruments} Neo 5.5) for its general performance as a detector of x-rays and low energy $\gamma$-rays and assess its sensitivity relative to the World Health Organization upper limit on lead in drinking water. Energies from \SI{6}{\kilo\electronvolt} to \SI{60}{\kilo\electronvolt} are examined. The CMOS camera has a linear energy response over this range and its energy resolution is for the most part slightly better than \SI{2}{\%}. The Neo sCMOS is not sensitive to x-rays with energies below $\sim\!\!\SI{10}{\kilo\electronvolt}$. The smallest detectable rate is \SI{40(3)}{\milli\hertz}, corresponding to an incident activity on the chip of \SI{7(4)}{\becquerel}. Taking calorimetric information into account we measure a minimal detectable rate of \SI{4(1)}{\milli\hertz} (\SI{1.5(1)}{\milli\hertz}) for \SI{26.3}{\kilo\electronvolt} (\SI{59.5}{\kilo\electronvolt}) $\gamma$-rays, which corresponds to an incident activity of \SI{1.0(6)}{\becquerel} (\SI{57(33)}{\becquerel}). The measured efficiency at this energy is \SI{0.08(2)}{\%} (\SI{0.0011(2)}{\%}). Toy Monte Carlo and Geant4 simulations agree with these results. These results show this CMOS sensor is well-suited as a $\gamma$- and x-ray detector with sensitivity at the few to \SI{100}{ppb} level for \pb{210} in a sample.}


\newcommand{\am}[1]{$^{#1}\textrm{Am}$}
\newcommand{\pb}[1]{$^{#1}\text{Pb}$}
\newcommand{\fe}[1]{$^{#1}\textrm{Fe}$}

\newcommand{\figref}[1]{Figure~\ref{#1}}    
\newcommand{\figrefbra}[1]{Fig.~\ref{#1}}   
\newcommand{\Figref}[1]{Figure~\ref{#1}}    
\newcommand{\tabref}[1]{Table~\ref{#1}}      
\newcommand{\tabrefbra}[1]{Tab.~\ref{#1}}   
\newcommand{\Tabref}[1]{Table~\ref{#1}}     
\newcommand{\secref}[1]{Section~\ref{#1}}   
\newcommand{\secrefbra}[1]{Sec.~\ref{#1}}   
\newcommand{\eqnref}[1]{Eqn.~\eqref{#1}}      
\newcommand{\eqnrefbra}[1]{Eqn.~\eqref{#1}}   

\begin{document}
\section{Introduction}
\label{cmosPaper:sec:introduction}

Ingesting lead can have acute and chronic health effects and it is especially harmful to infants and children. There is no safe threshold for the onset of lead's negative effects on the human condition and damages are permanent -- \textit{e.g.} a child loses \SI{3}{IQ} points on average when it consumes as much lead as \SI{25}{\micro\gram\per\kilo\gram} body weight per week over a longer period \cite{WHO2011}. It is estimated that globally 26 million people in low- and middle-income countries are at risk of lead exposure \cite{PE2015} as \textit{e.g.} people living in rural areas in Mexico \cite{CARAVANOS2014269}. The main source of lead pollution is from improper recycling of lead-acid batteries. The tonnage of lead in production is rapidly increasing and has grown by more than an order of magnitude in the past decade \cite{PE2015}. The trend towards clean technologies, such as electric cars, will likely increase the demand on lead-acid battery recycling, therefore the corresponding pollution can be expected to increase.\\
Lead screening tests in the UK, for example, are usually carried out in both public and private water supplies \cite{NIwater}. In public water supplies these are done at water treatment facilities, service reservoirs, water supply points and customer taps in water supply zones. In private water supplies, samples are collected at the point of use. The laboratories that carry out the analysis of these samples are accredited by the United Kingdom Accreditation Service (UKAS) and the Drinking Water Testing Specification (DWTS). In low- and middle-income countries (LMICs), access to such testing is more limited and may be prohibitively expensive. Therefore, a low-cost sensor of lead in drinking water would allow a wider range of people access to on-demand assay methods. A broader programme of measurements enabled by a low-cost technology could have important impacts on mitigating lead intake through contaminated water.\\
Most people carry a CMOS sensor in their pocket -- the silicon chip in their mobile phones' cameras. Even lower cost CMOS sensors are available off-the-shelf. These silicon chips are in principle capable of measuring radiation as x-rays, $\gamma$-rays, and radiation from $\beta$- and $\alpha$-decays. Radioactive isotopes are found in trace amounts together with stable isotopes of lead \cite{Orrell:2015zca}. Detecting the decay radiation of these isotopes can potentially enable lead detection in food and drinking water. Thus, radio assay methods based on a cheap and already common sensor such as a CMOS chip can be of great help to mitigate lead ingestion, particularly in LMICs. A major challenge is to detect the small signal from trace contamination. This paper reports on a first step towards developing lead radio assay in CMOS by exploring the potential of a scientific CMOS, operated to minimize noise. In this paper we qualify how a scientific CMOS, built for optical light detection, performs as a radiation detector. Previous studies have shown that it is possible to use CMOS sensors to distinguish between $\alpha$-decays and other types of radiation, as well as counting events from different radiation dosages \cite{PEREZ2016}. It has also been shown that these sensors can provide good spatial resolution, since they allow for a geometrical confinement of the received signal, for either x-rays or for charged particles \cite{SERVOLI2010}. Spacial information can be used to distinguish electronic noise and background radiation from the actual signal of interest, although it is by it self not a pre-requisite for dosimetry applications.

\subsection{Lead-210 Measurement Applications}
\label{cmosPaper:sec:introduction:pb210}
One of the radio-isotopes occurring with stable lead is \pb{210}, which decays by $\beta^{-}$ decay ($Q$-value of \SI{63.5(5)}{\kilo\electronvolt}) to an excited state of $^{210}\text{Bi}$, which de-excites practically immediately while emitting a $\gamma$-ray with \SI{46.6}{\kilo\electronvolt} energy. This lead isotope can occur in trace amounts with $\textrm{Pb}$, because it is a precursor of the stable lead (\textit{i.e.} \pb{206} in this case) in the $^{238}\textrm{U}$ (or $^{222}\textrm{Rn}$) decay series. Measurements in \textit{e.g.} \cite{Orrell:2015zca} show a range of \SI{3.9E-8}{ppb} to \SI{2.4E-5}{ppb} for certain \pb{} samples. Assaying \pb{210} at trace levels is required for sediment layer dating in geology \cite{Appleby2008}, to assess pollution levels in environmental monitoring \cite{HYang2016}, and to identify trace radioactivity in materials for particle physics experiments (e.g. DM searches) \cite{DEAPDetectorPaper}.\\
The necessary radio-isotope assays for such studies are done with (high purity) germanium detectors reaching concentrations as low as $\sim\!\!\SI{1e-6}{ppb}$. Germanium detectors are also used for \textit{e.g} \pb{210} assays of materials for DM experiments \cite{Nantais:2013iwz,widorskiadevelopment} and many more applications in science and technology. To measure even lower concentrations of \pb{210} in a sample, ashing of that sample has been applied with success (plants: \cite{HYang2016}, acrylic: \cite{GMM1994}). Heavy (radio-)isotopes are retained in these processes, so the decay-rate per volume is enhanced after ashing. The lowest concentration measured reaches limits of down to \SI{1.1e-20}{\gram} \pb{210} per \si{\gram} of acrylic or \SI{1e-11}{ppb} \cite{Nantais:2013iwz}.\\ 
The measurement sensitivity goal for assay of lead in drinking water is the World Health Organization (WHO) upper limit. The WHO quotes a value of \SI{10}{ppb} \cite{WHO2011} as a guideline value for lead in water, although it states explicitly that there is no safe threshold for lead ingestion. In instances of drinking water contamination, levels from 100 to 1000 times this value have been measured \cite{flint}.

\subsection{Outline of this paper}

For this paper we study a CMOS sensor, which has been designed for imaging optical wavelengths. The paper is structured as follows: in \secref{cmosPaper:sec:neosCMOS} we give the technical details of the scientific CMOS sensor -- \textsc{Oxford instruments} Neo 5.5 scientific CMOS camera \cite{neosmos} (Neo sCMOS); in \secref{cmosPaper:sec:expSetUp} we outline the experimental set-up and our measurement procedures; \secref{cmosPaper:sec:anaProcedure} describes the analysis procedure -- the steps taken to go from camera exposure frames to the reconstruction of the energy deposited in the CMOS chip. Results from measurements with radioactive sources and a x-ray tube are presented in \secref{cmosPaper:sec:results}. We examine the general performance of the Neo sCMOS when exposed to x-rays and $\gamma$-rays.\footnote{Since the camera has a glass window before the sensor, we can not measure $\alpha$ particles emitted by the sources tested in \secref{cmosPaper:sec:results}.} The performance analysis includes a study of the calorimetric capabilities of the camera, \textit{i.e.} its capabilities to measure the energy of incident photons and its energy resolution as function of the incident energy (\secrefbra{cmosPaper:sec:results:caliometry}). Furthermore we measure the background rate without any source present, the minimal detectable rate with an \am{241} source and the camera's detection efficiency for x-rays and $\gamma$-rays (\secrefbra{cmosPaper:sec:results:caliometry:sensorEfficiency}). We use the measured efficiencies to make a rough estimate on the sensor thickness in \secref{cmosPaper:sec:results:subsec:toyMC}. To check our measured results for consistency we construct a simulation of different layers of the Neo sCMOS in Geant4 \cite{AGOSTINELLI}, which is described in \secref{cmosPaper:sec:genatsimulations} and we compare our measured spectra with these simulations in order to further understand the effects of the thickness of different layers in the chip. The paper concludes with a discussion of the results (\secrefbra{cmosPaper:sec:summary}) in which we estimate the sensitivity of the CMOS approach to measure lead concentration in water down to the WHO of \SI{10}{ppb} \cite{WHO2011}.

\section{CMOS camera specifications}
\label{cmosPaper:sec:neosCMOS}

An ideal silicon sensor for the measurement of x-rays and low energy $\gamma$-rays would have a thick conversion region to enhance the probability that photons are absorbed in the silicon, low noise, and ideally allows the photons to pass unhindered by \textit{e.g.} an entry windows to the chip. The DAMIC CCDs \cite{Aguilar-Arevalo:2016ndq} are a good example for scientific sensors designed for this purpose, as they search for DM by measuring energy deposits $\lesssim\SI{10}{\kilo\electronvolt}$ in their silicon. In this work we explore the potential of commercial CMOS cameras for dosimetry applications, with the aim of understanding the prospects for CMOS-based sensors in the field. The \textsc{Oxford instruments} Neo sCMOS camera \cite{neosmos} was chosen because it provides readout noise comparable with CCDs with relatively more widespread CMOS technology.\\
The Neo sCMOS features a chip with $2560 \times 2160$ active pixels, each with a height and width of \SI{6.5}{\micro\meter}. The active size of the sensor is thus $16.6 \times \SI{14.0}{\milli\meter\squared}$ (height $\times$ width). The spatial granularity of the pixels is relevant for the image background characterisation and event classification as discussed in \secref{cmosPaper:sec:anaProcedure:backgroundcorrection}, \secref{cmosPaper:sec:anaProcedure:clusterParameters} and \secref{cmosPaper:sec:anaProcedure:clusterSize}. Each pixel in the Neo sCMOS's chip has a typical well depth of $\SI{30e+3}{e^{-}}$ (electron) and is equipped with its own \textit{micro lens}. This micro lens array ensures that light arriving at the chip's surface is focused into the active region of the pixels. The Neo sCMOS has two acquisition modes: global shutter, where a full image frame is acquired and rolling shutter, where image data is acquired one row at a time. The detection limit for \pb{210} we are aiming to reach is small, and thus we chose operating conditions to minimize the noise. For rolling shutter with a readout frequency of \SI{200}{\mega\hertz}, the Neo sCMOS read noise specification is $\SI{1}{e^{-}}$ ($\SI{1.5}{e^{-}}$) mean (RMS) at $\SI{-30}{^{\circ}\textrm{C}}$. This temperature is enabled by the Neo sCMOS's cooling system, which allows for cooling the chip to $\SI{-30}{^{\circ}\textrm{C}}$ ($\SI{-40}{^{\circ}\textrm{C}}$) in a room temperature environment (with additional water cooling). For comparison, when a global shutter is used the mean read noise is $\SI{2.3}{e^{-}}$. Groups of hardware pixels ($2\times2$, $3\times3$, $4\times4$, $8\times8$) can be binned together, prior to readout, in order to reduce the overall contribution of readout noise and increase readout speed. These are then read together as one \textit{readout pixel}. Operating parameters were chosen to minimize the read noise, to optimize the SNR for the expected small signal. \Tabref{cmosPaper:sec:expSetUp:tab:standardSettings} lists the operating parameters for the CMOS during the measurements presented in this paper.\\
No information on the cross section of the camera chip, \textit{i.e.} its different layers and the thickness of the active silicon, is provided by the supplier. Information on the quantum efficiency is only available for radiation in the wavelength range from \SI{300}{\nano\meter} to \SI{1000}{\nano\meter} and not for short wavelengths (x-ray and $\gamma$-ray energies). Furthermore, it is not known how layers on top of the silicon (\textit{i.e.} the micro lenses) affect a measurement of x-rays / $\gamma$-rays. To understand the possible effect of these layers on our results, we performed Geant4 simulations for the different radiation sources, \textit{cf.} \secref{cmosPaper:sec:genatsimulations}. We furthermore attempt to assess the thickness of the sensor by comparing our measurements to toy Monte Carlo simulations treating the camera as only one silicon layer (\secrefbra{cmosPaper:sec:results:subsec:toyMC}).\\
Many of the Neo sCMOS features are not present for cheap, commercial CMOS sensors, especially the various features reducing the noise as the chip cooling or the general noise figures. On the other hand, the Neo sCMOS is a sensor optimised for optical wavelengths and the thickness of its conversion layer can be expected to be on the order of a few \si{\micro\meter} to $\sim\!\!\SI{10}{\micro\meter}$. The micro lense array on top of the actual pixels is an addtional layer as is \textit{e.g.} a \textit{Bayer filter} on top of the pixel matrix of a different camera chip. Furthermore incident radiation needs to passs through a window before it can reach the chip as is the case for commercial sensors enclosed in a housing, like the camera chip of a mobile phone. The camera has all these properties in common with cheap CMOS cameras.

\section{Experimental description}
\label{cmosPaper:sec:expSetUp}

\begin{table*}
\centering
\begin{tabular}{l|c}
  Setting & Value \\ \hline
  Readout binning & $4\times4$ \\
  Exposure time & \SI{10}{\second} \\
  Camera to source distance & \SI{1.75}{\centi\meter} \\
  Number of exposures / data taking run & 100 \\
  Pixel Readout Rate (inverse row time) & \SI{200}{\mega\hertz} \\
  Dynamic Range & 16-bit \\
  Mode & Low noise/ high well capacity \\
  Electronic Shuttering Mode & Rolling \\
\end{tabular}
\caption{\label{cmosPaper:sec:expSetUp:tab:standardSettings}Default settings of Neo sCMOS camera during all measurement runs. Runs with different settings are explicitly noted.}
\end{table*}
\begin{figure*}
\centering
\subfloat[]{\includegraphics[height=0.15\textheight, trim = 0 0 0 0, clip=true]{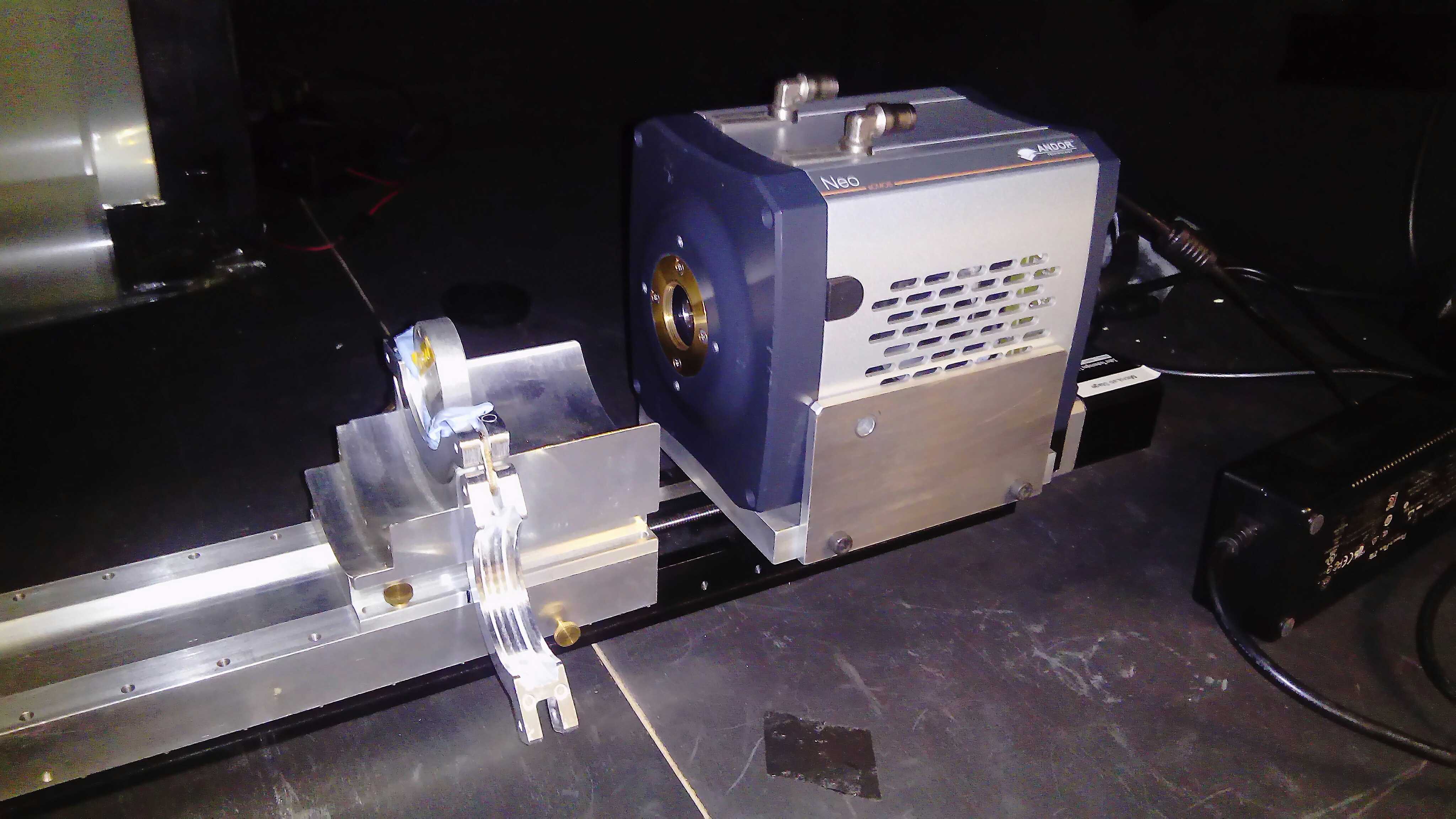}\label{cmosPaper:sec:expSetUp:fig:setupphoto}}
\subfloat[]{\includegraphics[height=0.15\textheight, trim = 300 1200 1700 200, clip=true]{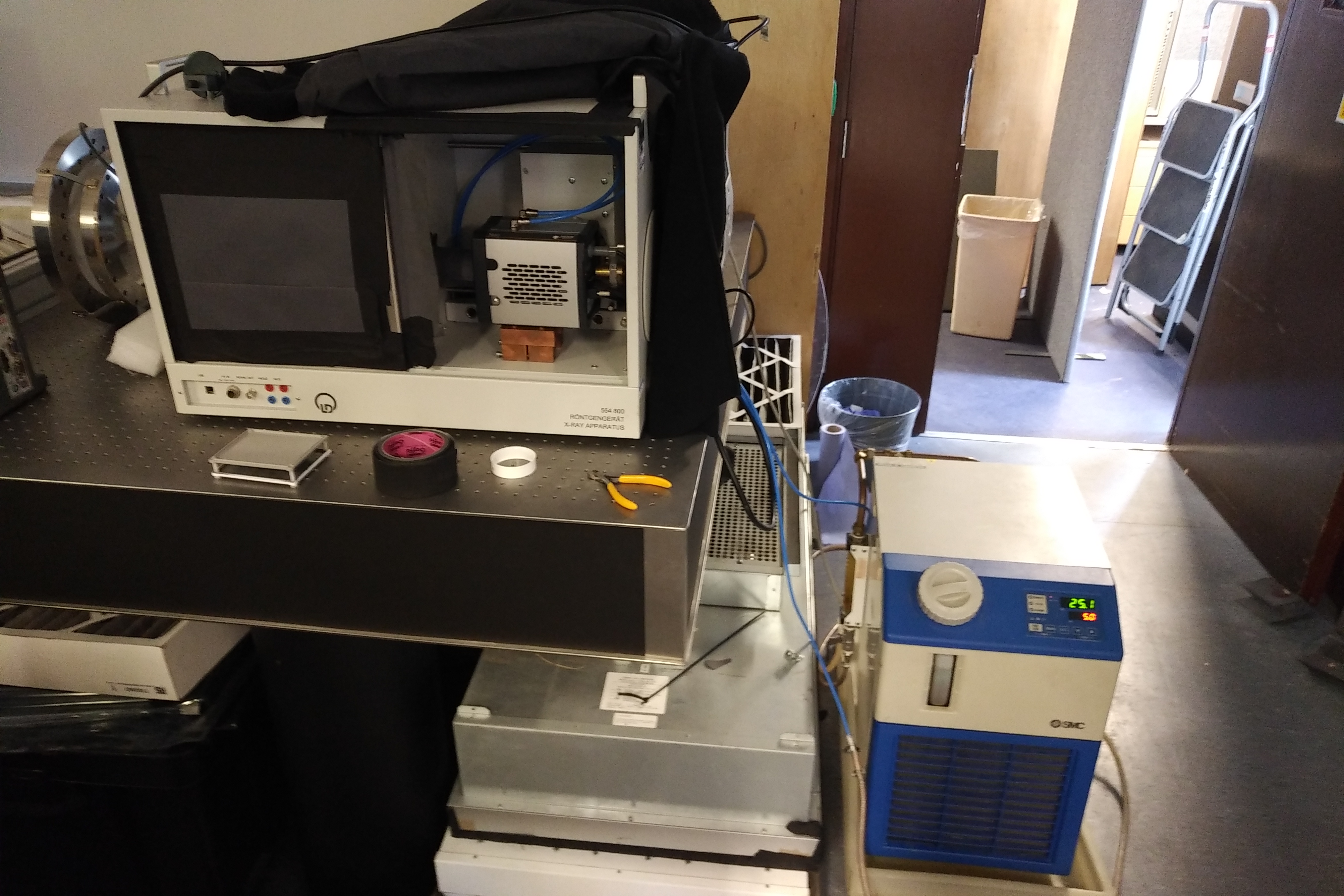}\label{cmosPaper:sec:expSetUp:fig:setupphoto:xray}}
\caption{\label{cmosPaper:sec:expSetUp:fig:setupphotos}\protect\subref{cmosPaper:sec:expSetUp:fig:setupphoto} Photograph of the source holder and the Neo sCMOS camera in the dark-box. \protect\subref{cmosPaper:sec:expSetUp:fig:setupphoto} The Neo sCMOS in the LD Didactic X-ray apparatus (554 800) -- the blue lines are for the water cooling circuit, which is not necessary for operation in the dark box.}
\end{figure*}
CMOS data for radioactive source measurements are acquired inside a dark box, with dimensions of $\SI{244}{\centi\meter}\times\SI{122}{\centi\meter}\times\SI{122}{\centi\meter}$ ($\text{L}\times\text{W}\times{H}$). The large size of the box allows the distance between the radioactive source and the camera to be increased up to $\sim\!\!\SI{2.5}{\meter}$. This allows for a measurement of the minimal detectable source activity, as described in \secref{cmosPaper:sec:results:subsec:minDetActivity}. The camera is positioned inside the box as shown in \figref{cmosPaper:sec:expSetUp:fig:setupphoto} and all data is taken whilst the box is closed. Several calibration sources are used to understand the camera's behaviour and to obtain different spectra for calorimetry. In particular, we take background data without any sources and also use \am{241}, \fe{55}, and \pb{210} sources. Data taking with the x-ray tube is also performed under dark-box conditions, in a different enclosure, shown in \figref{cmosPaper:sec:expSetUp:fig:setupphoto:xray} and described in \secref{cmosPaper:sec:expSetUp:subsec:xray}.\\ 
Prior to a data acquisition run the camera is cooled to $\SI{-30}{^{\circ}\textrm{C}}$ and the respective source is positioned in front of the camera, with the source supported such that it is aligned on the chip centre, as in \figref{cmosPaper:sec:expSetUp:fig:setupphoto}. The NEO sCMOS is controlled via a cable, which is fanned out of the dark-box and connects to a custom PCIe card hosted in the data acquisition computer. We used the \textit{Andor SOLIS for Imaging} software package for the data acquisition as well as to set camera's parameters. The operation settings in \tabref{cmosPaper:sec:expSetUp:tab:standardSettings} are chosen for several reasons: a readout binning of $4\times4$ is preferred over $1\times1$ due to limitations of the data transfer rate and also to optimise data processing; the number of exposures and exposure time are chosen to ensure enough energy deposits to result in a clear peak in the energy spectra; all the other settings are chosen to ensure a low readout noise. The analysis procedure developed to identify energy deposits in the camera chip relies on a low occupancy. For this reason, an even larger readout binning than $4\times4$ is not chosen, although a larger binning would be even more beneficial for data transfer and data processing. As an exposure time of \SI{10}{\second} is used, the readout time of the camera is not an issue.

\subsection{X-ray data taking}
\label{cmosPaper:sec:expSetUp:subsec:xray}

A LD Didactic X-ray apparatus (554 800) \cite{LD} is used for the X-ray data taking. \Figref{cmosPaper:sec:expSetUp:fig:setupphoto:xray} shows the camera inside the apparatus. The door to the compartment with the camera is closed before the data taking and the compartment is sealed light-tight. The x-ray tube and camera develop substantial heat, therefore the camera's water cooling is used to ensure stable operation at $-\SI{30}{^{\circ}C}$. The rate of x-rays of the apparatus is larger than the rate of any radioactive source we use in this paper, allowing much shorter exposure times than stated in \tabref{cmosPaper:sec:expSetUp:tab:standardSettings}: \SI{0.004}{\second} and \SI{0.025}{\second}.\\
The anode in the x-ray tube is made of $\text{Mo}$, with its characteristic $K_{\alpha}$ and $K_{\beta}$ lines at \SI{17.41}{\kilo\electronvolt} and \SI{19.61}{\kilo\electronvolt}, respectively. Data is also acquired with a $\text{Cu}$ or $\text{Zr}$ filter between the X-ray tube and the camera. The observation of absorption edges adds more energy measurements in addition to the two x-ray lines, which makes these tests valuable for the energy calibration of the sensor. In \secref{cmosPaper:sec:results:caliometry:energyMeasurement} our results with the x-ray source are discussed.

\section{Data analysis}
\label{cmosPaper:sec:anaProcedure}

\begin{figure*}
\centering
\subfloat[BKG]{
\label{cmosPaper:sec:anaProcedure:fig:2DaduDist:ADU:Bkg}
\includegraphics[width=0.45\textwidth,trim=0 0 0 31,clip=true]{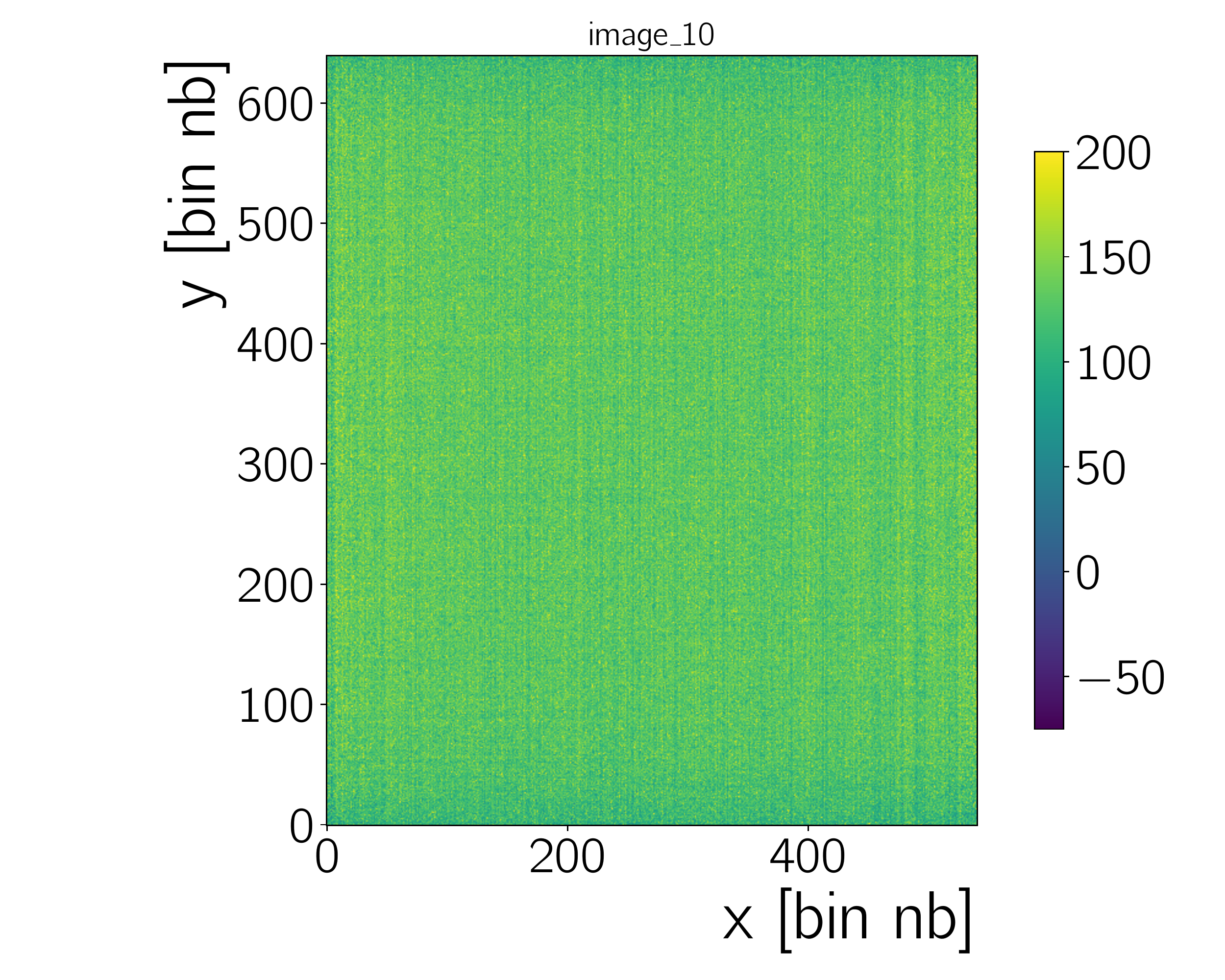}}
\subfloat[\am{241}]{
\label{cmosPaper:sec:anaProcedure:fig:2DaduDist:ADU:Sgl}
\includegraphics[width=0.45\textwidth,trim=0 0 0 31,clip=true]{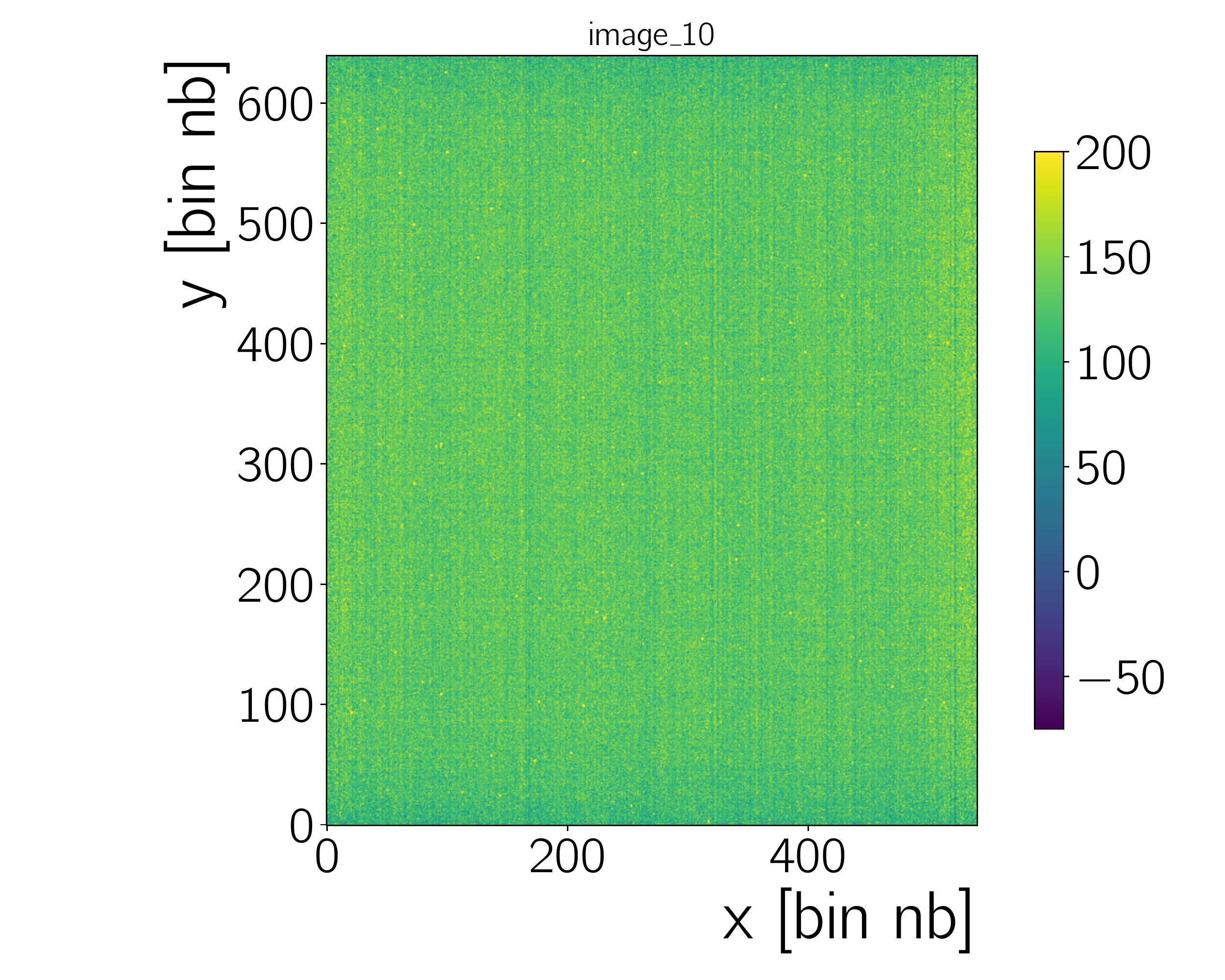}}
\caption{\label{cmosPaper:sec:anaProcedure:fig:2DaduDist:ADU:BkgAndSgl}Raw frames (without any correction) recorded when no source of radiation is present, \textit{i.e} background (BKG), and when the camera is irradiated with an \am{241} source. A zoom in $x$ and $y$ of Figure \protect\subref{cmosPaper:sec:anaProcedure:fig:2DaduDist:ADU:Sgl} is shown in \figref{cmosPaper:sec:anaProcedure:fig:2DaduDist:ADU:Sglzoom}. All images are zoomed on the intensity scale, visible in the colour-bar to the right of each image. (Which means values larger than 200 are displayed as 200.)}
\end{figure*}
The camera control software produces files in the FITS (Flexible Image Transport System) \cite{FITS} format. Each FITS file can contain several frames, \textit{i.e.} 2D arrays with one Analogue-to-Digital Unit (ADU) measurement for each camera pixel. For example, \figref{cmosPaper:sec:anaProcedure:fig:2DaduDist:ADU:BkgAndSgl} shows such a frame. After data taking, frames are processed by \textsc{python} code and CERN ROOT \cite{Brun:1997pa} routines. During normal data acquisition conditions (\tabref{cmosPaper:sec:expSetUp:tab:standardSettings}) we take several \textit{run}s of $N_{\textrm{f}}=100$ frames. The data analysis described below identifies clusters due to radiation in each frame. Ultimately all clusters found in the frames of several runs are combined into one set and further data analysis is done on this set (\secrefbra{cmosPaper:sec:results}).

\subsection{Frame background calculation}
\label{cmosPaper:sec:anaProcedure:backgroundcorrection}

The image processing analysis corrects for the pixel pedestal in two steps, described in \secref{cmosPaper:sec:anaProcedure:columns} and \secref{cmosPaper:sec:anaProcedure:timeSeries}, then finds clusters of signal pixels, described in \secref{cmosPaper:sec:anaProcedure:clusterFinding}, and measures the energy within each cluster. These clusters are identified by their difference to the remaining pedestal value, after the corrections.

\subsubsection{Column correction}
\label{cmosPaper:sec:anaProcedure:columns}

\begin{figure*}
\centering
\subfloat[]{
\includegraphics[width=0.45\textwidth,trim=0 0 0 0,clip=true]{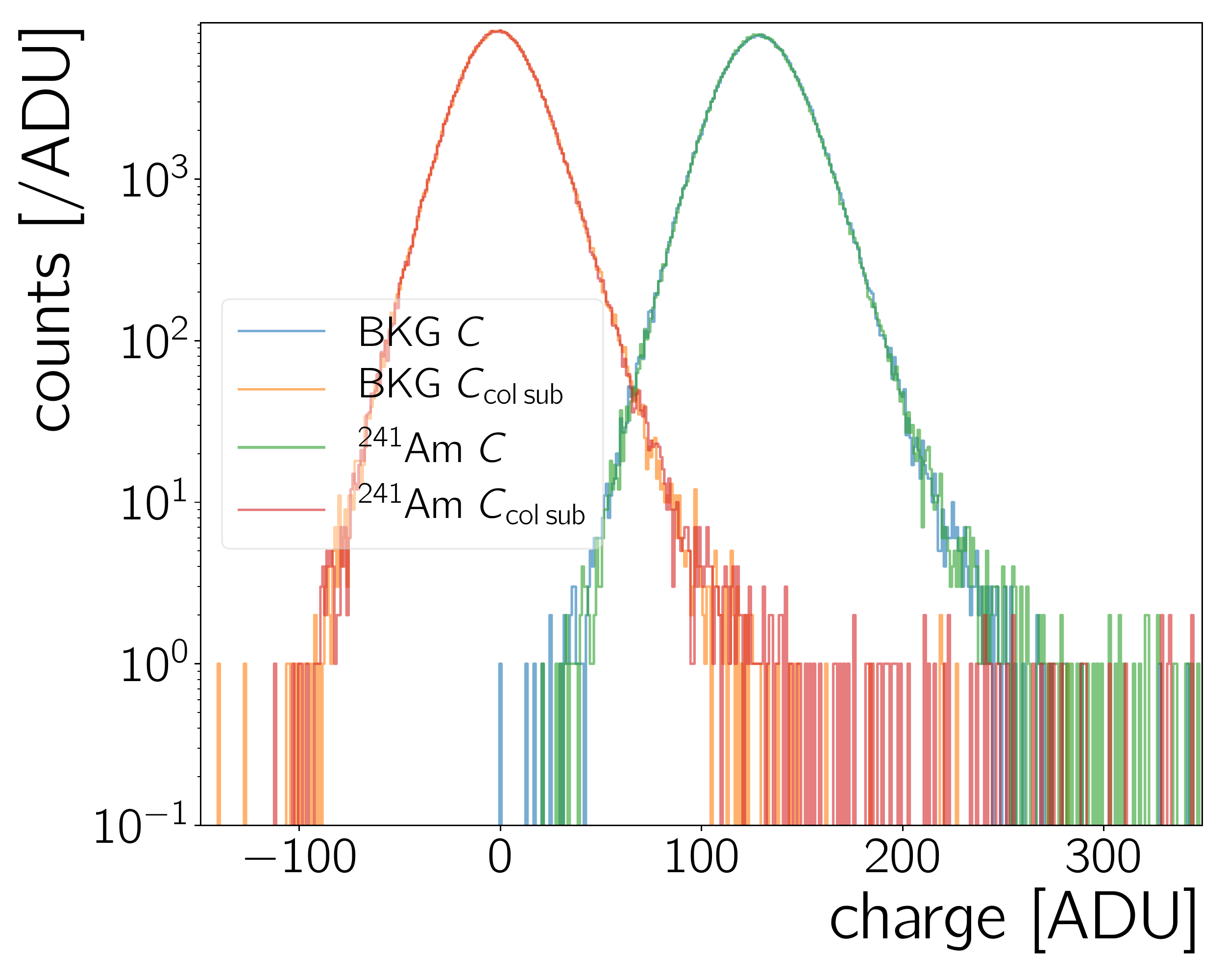}
\label{cmosPaper:sec:anaProcedure:fig:2DaduDist:ADUcolsub:1D}}
\subfloat[]{
\includegraphics[width=0.45\textwidth,trim=0 0 0 0,clip=true]{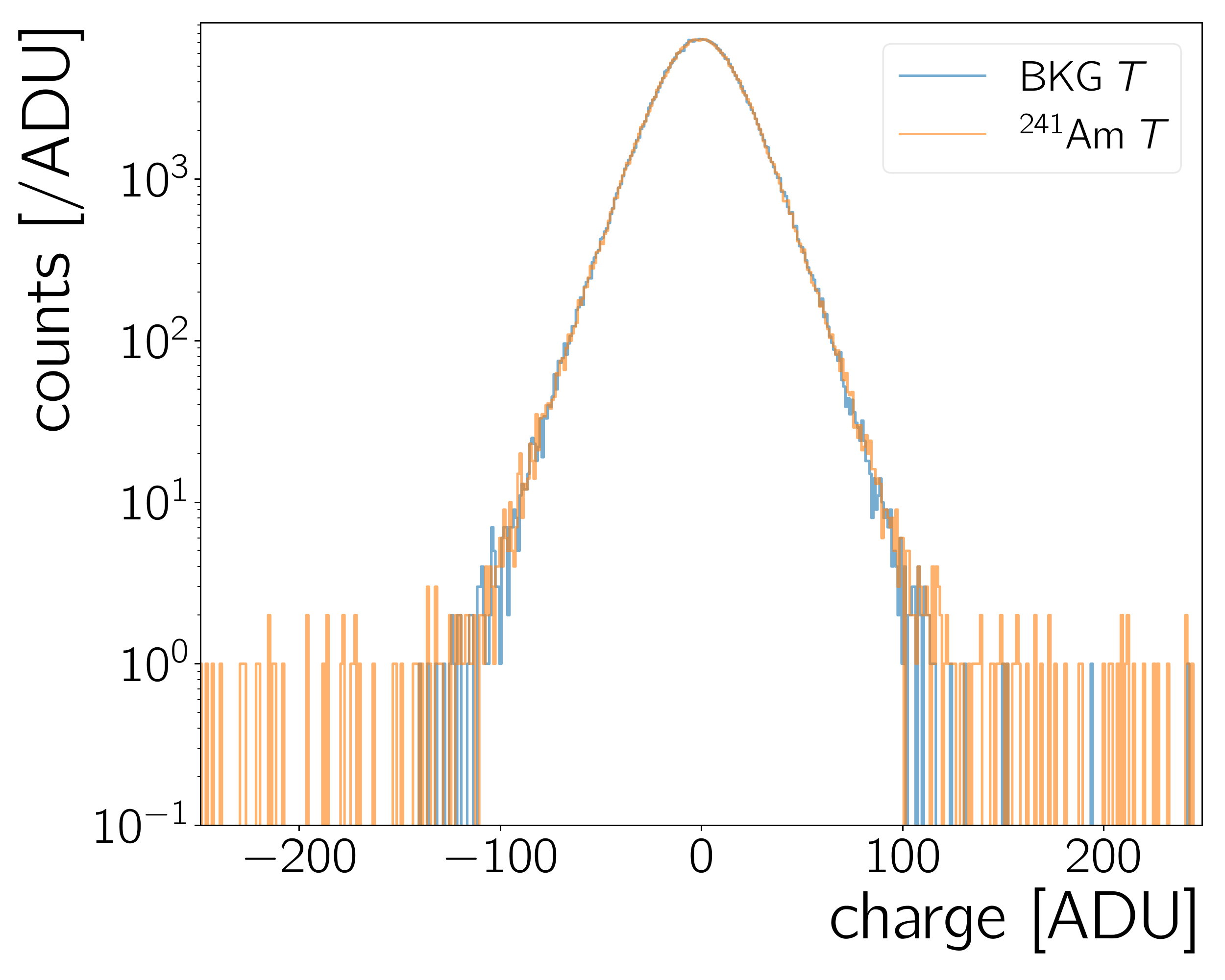}
\label{cmosPaper:sec:anaProcedure:fig:2DaduDist:ADUcolsub:1DTS}}
\caption{\label{cmosPaper:sec:anaProcedure:fig:2DaduDist:ADUcolsub:1DandTS}Histograms of the \protect\subref{cmosPaper:sec:anaProcedure:fig:2DaduDist:ADUcolsub:1D} raw pixel values $C$ of all pixels in the frame in \figref{cmosPaper:sec:anaProcedure:fig:2DaduDist:ADU:Bkg} (BKG $C$), \figref{cmosPaper:sec:anaProcedure:fig:2DaduDist:ADU:Sgl} ($^{241}\text{Am}$ $C$), and of the $C_{\text{col}\:\text{sub}}$ values calculated for the data in these frames using \eqnref{cmosPaper:sec:anaProcedure:columns:eq:columnsub}.
\protect\subref{cmosPaper:sec:anaProcedure:fig:2DaduDist:ADUcolsub:1DTS} Histogram of the $C_{\text{col}\:\text{sub}}$ values after time-series subtraction ($T$, form \eqnrefbra{cmosPaper:sec:anaProcedure:timeSeries:eq:ts}).}
\end{figure*}

The first step of the analysis is to correct for the raw image pedestal, which is defined as the background \si{ADU} measurement in each pixel in the absence of a source. We observe that the \si{ADU} values of each pixel in a given column are correlated with each other, giving rise to the distinct columns in \figref{cmosPaper:sec:anaProcedure:fig:2DaduDist:ADU:BkgAndSgl}. This correlation pattern is no fixed-pattern-noise as it changes form frame to frame. Therefore we employ the following approach to correct for it on a single frame basis. First, the mean column value $\left<\,C\,\right>_{\text{col}}\left({x}, {n_{\textrm{f}}}\right)$ and its standard deviation $\sigma_{{C}_{\text{col}}}\left(x,n_{\textrm{f}}\right)$ for each column is calculated: 
\begin{align}
\left<\,C\,\right>_{\text{col}}\left({x}, {n_{\textrm{f}}}\right) &= \frac{1}{N_{\textrm{y}}}\sum_{y=0}^{N_{\textrm{y}}} C({x},y,{n_{\textrm{f}}})
\label{cmosPaper:sec:anaProcedure:columns:eq:columnmean}\\
\sigma_{{C}_{\text{col}}}\left(x,n_{\textrm{f}}\right) &= \sqrt{\frac{1}{N_{\textrm{y}}-1}\sum_{y=0}^{N_{\textrm{y}}} (C({x},y,{n_{\textrm{f}}})-\left<\,C\,\right>_{\text{col}}\left({x}, {n_{\textrm{f}}}\right))^2}\ \ , \quad n_{\textrm{f}}=j,\ \ x=k
\label{cmosPaper:sec:anaProcedure:columns:eq:columnsigma}
\end{align}
where $C({x},y,{n_{\textrm{f}}})$ is the charge (in \si{ADU}) measured by a pixel at a given $x,y$ position in frame ${n_{\textrm{f}}}$. The column coordinate and the frame number are fixed ($x=k$, $n_{\textrm{f}}=j$) while the sum runs over the row coordinate ($y=0\dots N_{\textrm{y}}$). After a first calculation of the column mean and standard deviation using \eqnref{cmosPaper:sec:anaProcedure:columns:eq:columnmean} and \eqnref{cmosPaper:sec:anaProcedure:columns:eq:columnsigma}, all pixel values $C({k},y,{j})\not\in\left<\,C\,\right>_{\text{col}}\left({k}, {j}\right)\pm 5 \cdot \sigma_{{C}_{\text{col}}}\left(k,j\right)$ are excluded and $\left<\,C\,\right>_{\text{col}}\left({k}, {j}\right)$ and $\sigma_{{C}_{\text{col}}}\left(k,j\right)$ are calculated again until $\sigma_{{C}_{\text{col}}}\left(k,j\right)$ changes less than \SI{0.5}{\%} between two iterations. This iterative approach is necessary in order to exclude pixels with a high charge value as \textit{e.g.} hot pixels -- transient high values in a certain $x,y$ position -- or pixels with a higher charge value due to a signal induced by incident radiation. In the zoomed image in \figref{cmosPaper:sec:anaProcedure:fig:2DaduDist:ADU:Sglzoom} some pixels with a high charge value are visible with $C\geq200$. After the final mean and standard deviation is found, that mean is subtracted from each pixel value:
\begin{align}
C_{\text{col}\:\text{sub}}(x,y,n_{\textrm{f}}) = C(x,y,n_{\textrm{f}}) - \left<\,C\,\right>_{\text{col}}\left(x_{\textrm{k}}, n_{\textrm{f}}\right)
\label{cmosPaper:sec:anaProcedure:columns:eq:columnsub}
\end{align}
$x_{k}$ in $\left<\,C\,\right>$ indicates that the column mean is the same for all the $C(x,y,n_{\textrm{f}})$ along a column with $x=k$, \textit{i.e.} in $y$ direction.\\
The result of this column-pedestal correction procedure is shown in \figref{cmosPaper:sec:anaProcedure:fig:2DaduDist:ADUcolsub:1D}, for the raw data of \figref{cmosPaper:sec:anaProcedure:fig:2DaduDist:ADU:BkgAndSgl}. A raw frame recorded during data taking with no source and with $^{241}\text{Am}$ has a mean of \SI{129(20)}{ADU} and \SI{130(73)}{ADU}, respectively, where the uncertainty is chosen to be one standard deviation. After the column correction the mean moves to \SI{0(18)}{ADU} and \SI{0(73)}{ADU}, respectively.

\subsubsection{Time-series analysis}
\label{cmosPaper:sec:anaProcedure:timeSeries}

At this stage it is possible that there is still fixed-pattern-noise in the recorded frames, \textit{e.g.} pixels which have, in every frame, a $C$ value elevated over the neighbouring pixel's values. Such pixels may be hot pixels or pixels with charge values of only a few \SI{100}{ADU}. In order to correct for these we adopt a \textit{time-series} approach: all charge values $C_{\text{col}\:\text{sub}}(x,y,j)$ in the $n_{\textrm{f}}=j$ frame in a run, with $N_{\textrm{f}}$ frames in total, are subtracted from their corresponding values in the $n_{\textrm{f}}=j+1$ frame, $C_{\text{col}\:\text{sub}}(x,y,j+1)$. The result are $N_{\textrm{f}}-1$ frames with pixel intensities $T(x,y,n_{\textrm{f}})$ given by\footnote{Note that in Eq. \eqref{cmosPaper:sec:anaProcedure:timeSeries:eq:ts} the $n_\textrm{f}$ starts at zero -- hence the last index is $N_{\textrm{f}}-2$ for $N_{\textrm{f}}-1$ frames.}
\begin{align}
T(x,y,n_{\textrm{f}})=C_{\text{col}\:\text{sub}}(x,y,n_{\textrm{f}})-C_{\text{col}\:\text{sub}}(x,y,n_{\textrm{f}}+1)\ \ \quad n_\textrm{f}=0\dots N_{\textrm{f}}-2 \quad . 
\label{cmosPaper:sec:anaProcedure:timeSeries:eq:ts}
\end{align}
\begin{figure*}
\centering
\subfloat[BKG]{
\label{cmosPaper:sec:anaProcedure:fig:2DaduDist:ADUcolsub:BkgTS}
\includegraphics[width=0.45\textwidth,trim=0 0 0 31,clip=true]{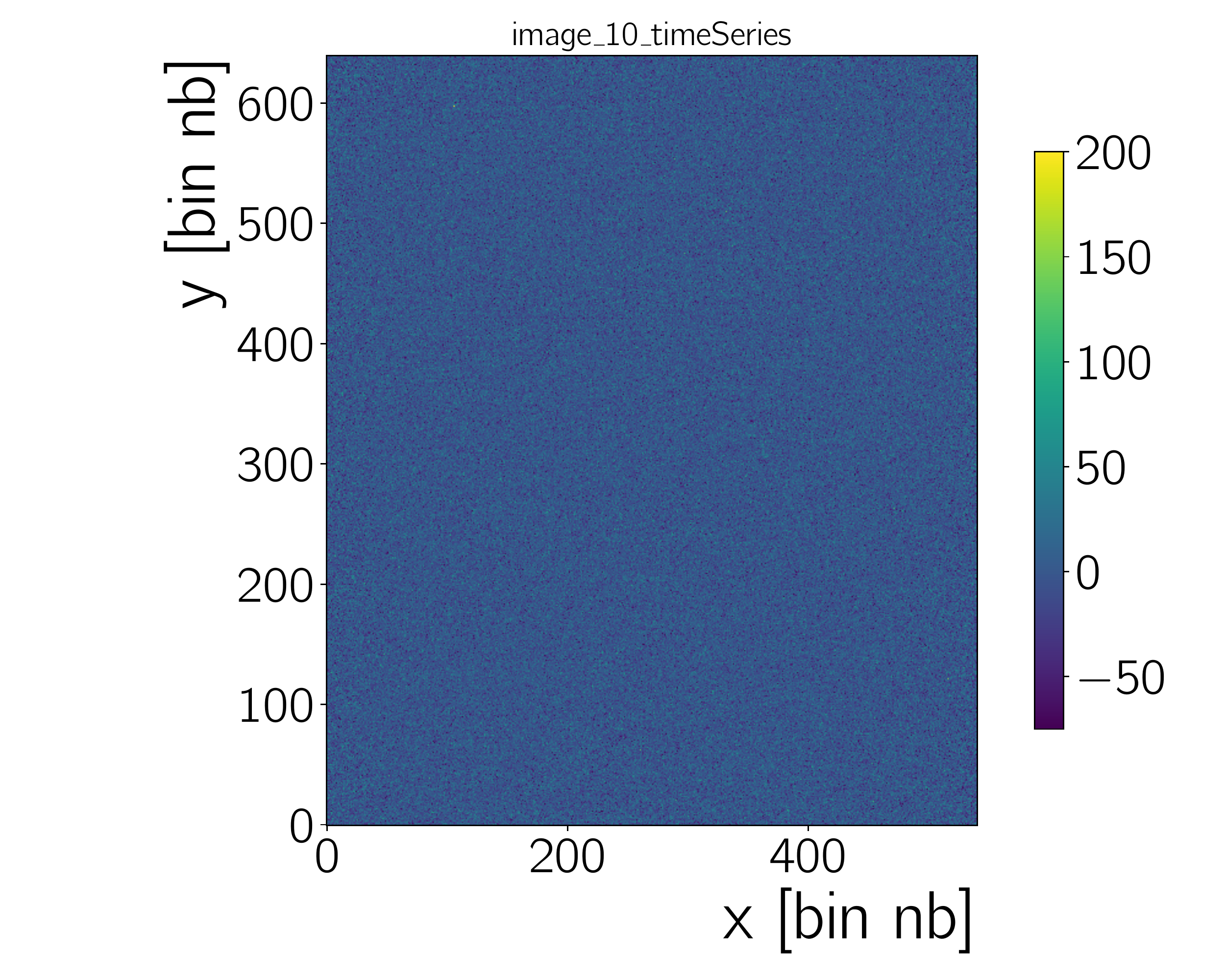}}
\subfloat[\am{241}]{
\label{cmosPaper:sec:anaProcedure:fig:2DaduDist:ADUcolsub:SglTS}
\includegraphics[width=0.45\textwidth,trim=0 0 0 31,clip=true]{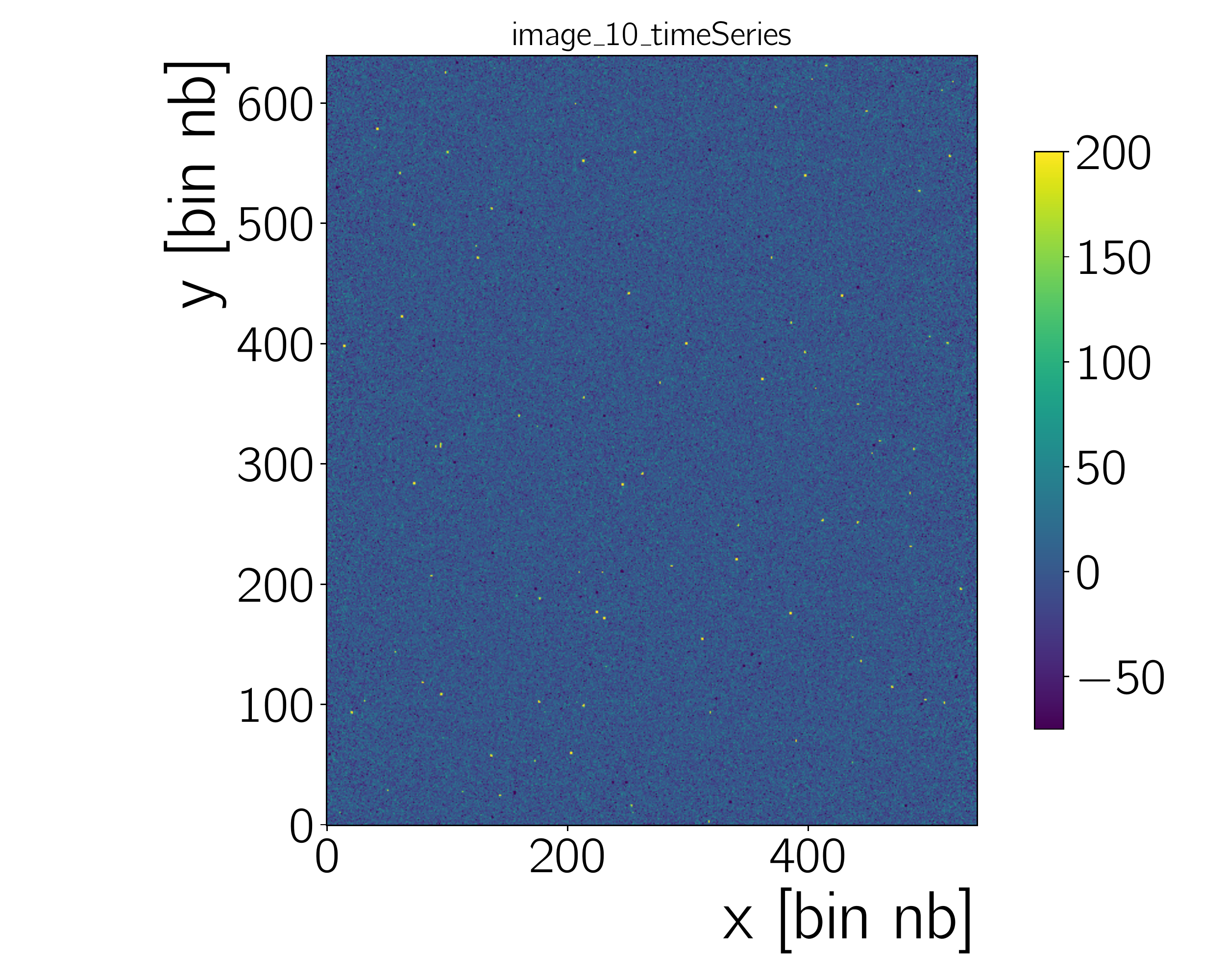}}
\caption{\label{cmosPaper:sec:anaProcedure:fig:2DaduDist:ADUcolsub:BkgAndSglTS}Pair wise subtracted frames (\textit{cf.} \secrefbra{cmosPaper:sec:anaProcedure:timeSeries}). The same frames as in \figref{cmosPaper:sec:anaProcedure:fig:2DaduDist:ADU:BkgAndSgl} are shown to illustrate this next step in the background rejection procedure. More high intensity points in the \am{241} frame than in the BKG frame are visble, when comparing the two plots.}
\end{figure*}
The time series pair-wise subtraction removes effects that are persistent in time, hence the name. Transient features, \textit{e.g.} radiation from a source, may create negative entries during this procedure, \textit{cf} \figref{cmosPaper:sec:anaProcedure:fig:2DaduDist:ADUcolsub:SglzoomTS}. This can be tolerated as long as the source rate is not too high such that transient features occur at the same $x,y$ coordinate in two subsequent frames. \Figref{cmosPaper:sec:anaProcedure:fig:2DaduDist:ADUcolsub:1DTS} shows the effect on the 1D charge distributions. The tails of the distributions change due to the subtraction of transient pixels with a high charge value. This leads to a mean of \SI{0(21)}{ADU} and \SI{0(112)}{ADU} for the time-series corrected frame with no source and $^{241}\text{Am}$, respectively. The standard deviation increases, since there are now more negative pixel values in the distribution, from the pairwise subtraction of transient features.

\subsection{Clustering}
\label{cmosPaper:sec:anaProcedure:clusterFinding}

The column corrected and time-series subtracted data, following \eqnref{cmosPaper:sec:anaProcedure:columns:eq:columnsub} and \eqnref{cmosPaper:sec:anaProcedure:timeSeries:eq:ts} are then searched for clusters. A cluster is defined as one or more spatially adjacent pixels which have a charge value larger than the remaining pedestal value.

\subsubsection{Threshold calculation}

The threshold value is constructed by a data driven method: We check all $N_{\textrm{f}}-1$ individual values a pixel at coordinates $x=m, y=i$ measures over the course of a run. In the notation introduced before, these values correspond to all $T(x,y,n_{\textrm{f}})$, where $x$ and $y$ are held constant and $n_{\textrm{f}}$ runs from 0 to $N_{\textrm{f}}-1$. From these charge values a run-averaged pixel pedestal value $p(x,y)$ and its standard deviation $\sigma_{\textrm{p}}(x,y)$ are calculated for all pixels at coordinates $x,y$.
\begin{align}
p\left(x,y\right) &= \left<\,T\,\right>\left({x}, y\right) = \frac{1}{N_{\textrm{f}}-1}\sum_{n_{\textrm{f}}=0}^{N_{\textrm{f}}-1} T({x},y,{n_{\textrm{f}}})
\label{cmosPaper:sec:anaProcedure:clusterFinding:eq:pedestalmean}\\
\sigma_{p}\left(x,y\right) &= \sqrt{\frac{1}{N_{\textrm{f}}-2}\sum_{n_{\textrm{f}}=0}^{N_{\textrm{f}}-1} (T({x},y,{n_{\textrm{f}}})-p\left({x}, {y}\right))^2}\ \ , \quad x=m, y=i
\label{cmosPaper:sec:anaProcedure:clusterFinding:eq:pedestalsigma}
\end{align}
This is again an iterative procedure, similar to what is done to calculate the column mean. From subsequent iterations all pixel values $T(x,y,n_{\textrm{f}})\not\in p\left(x, y\right)\pm 5 \cdot \sigma_{p}\left(x,y\right)$ are rejected when using \eqnref{cmosPaper:sec:anaProcedure:clusterFinding:eq:pedestalmean} and \eqnref{cmosPaper:sec:anaProcedure:clusterFinding:eq:pedestalsigma} to (re)calculate $p(x,y)$ and $\sigma_{p}\left(x,y\right)$. Both values are regarded as final when $\sigma_{p}\left(x,y\right)$ changes less than \SI{0.5}{\%} between two iterations. While $p(x,y)$ is by construction close to zero for column subtracted and time-series subtracted data, $\sigma_{p}\left(x,y\right)$ has a minimal value slightly above \SI{6}{ADU} and most probable value between \SI{14}{ADU} and \SI{15}{ADU}, skewed towards higher values.\footnote{Without time-series subtraction and column correction, the pedestal values and their standard deviation for every pixel should allow to discriminate between background and a charge signal, provided the fluctuations of the background are randomly distributed. However, the column mean of a specific column changes from exposure to exposure, motivating the approach described here.} The camera's manual states a read noise RMS value of $\SI{1.5}{\text{e}^{-}}$ and a dark current of $\SI{0.015}{\text{e}^{-}\per pixel\per\second}$. Combining these values in quadrature, whilst taking into account the exposure time (\SI{10}{\second}) and the readout binning ($4\times4$), and converting to \si{ADU}, we get $\sim\!\!\SI{4.22}{ADU}$, using the conversion factor of $\SI{0.67}{\text{e}^{-}\per ADU}$ as specified by the supplier. The modification of this RMS value by the before described column correction and the pair-wise subtraction have to be taken into account before comparing the RMS to $\sigma_{p}\left(x,y\right)$. While the corrections described in \secref{cmosPaper:sec:anaProcedure:columns} lead to a negligible reduction of the RMS, the pairwise subtraction (\secrefbra{cmosPaper:sec:anaProcedure:timeSeries}) increases the resulting RMS by a factor of $\sqrt{2}$. The smallest measured $\sigma_{p}\left(x,y\right)$ value of $\gtrsim\SI{6}{ADU}$ fits this expectation of \SI{5.97}{ADU}, indicating that the corrections applied here do remove most other noise contributions than the read- and dark-noise.

\subsubsection{Cluster finding}

The cluster finding algorithm employs two values, a \textit{seed} and a \textit{skirt} pixel threshold intensity. The \textit{seed} is a higher threshold value designed to quickly find the cluster's largest charge values. The \textit{skirt} is a lower threshold value designed to find potentially dimmer adjacent pixels to the \textit{seed} pixel associated with the cluster. The following threshold condition is used to discriminate whether a pixel value $T({x},y,{n_{\textrm{f}}})$ is part of the background or part of the charge deposit of a signal \textit{e.g.} by radiation incident on the chip:
\begin{align}
T({x},y,{n_{\textrm{f}}}) > p\left(x,y\right) + k \cdot \sigma_{p}\left(x,y\right)\ \quad k = k_{\textrm{seed}} \lor  k_{\textrm{skirt}}
\label{cmosPaper:sec:anaProcedure:clusterFinding:eq:threshold}
\end{align}
\begin{figure*}
\centering
\subfloat[raw]{
\label{cmosPaper:sec:anaProcedure:fig:2DaduDist:ADU:Sglzoom}
\includegraphics[width=0.32\textwidth,trim=0 0 0 31,clip=true]{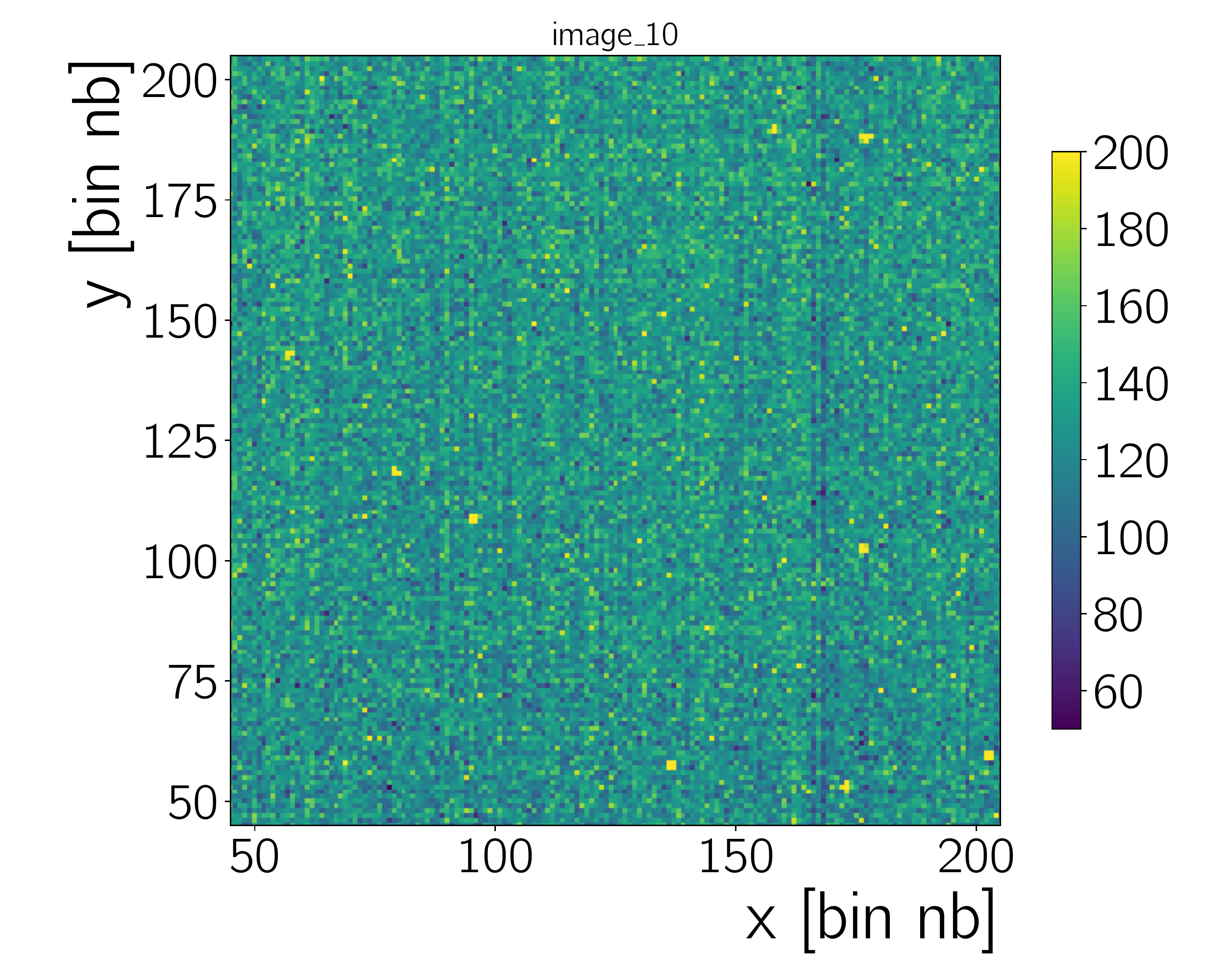}}
\subfloat[corrections applied]{
\label{cmosPaper:sec:anaProcedure:fig:2DaduDist:ADUcolsub:SglzoomTS}
\includegraphics[width=0.32\textwidth,trim=0 0 0 31,clip=true]{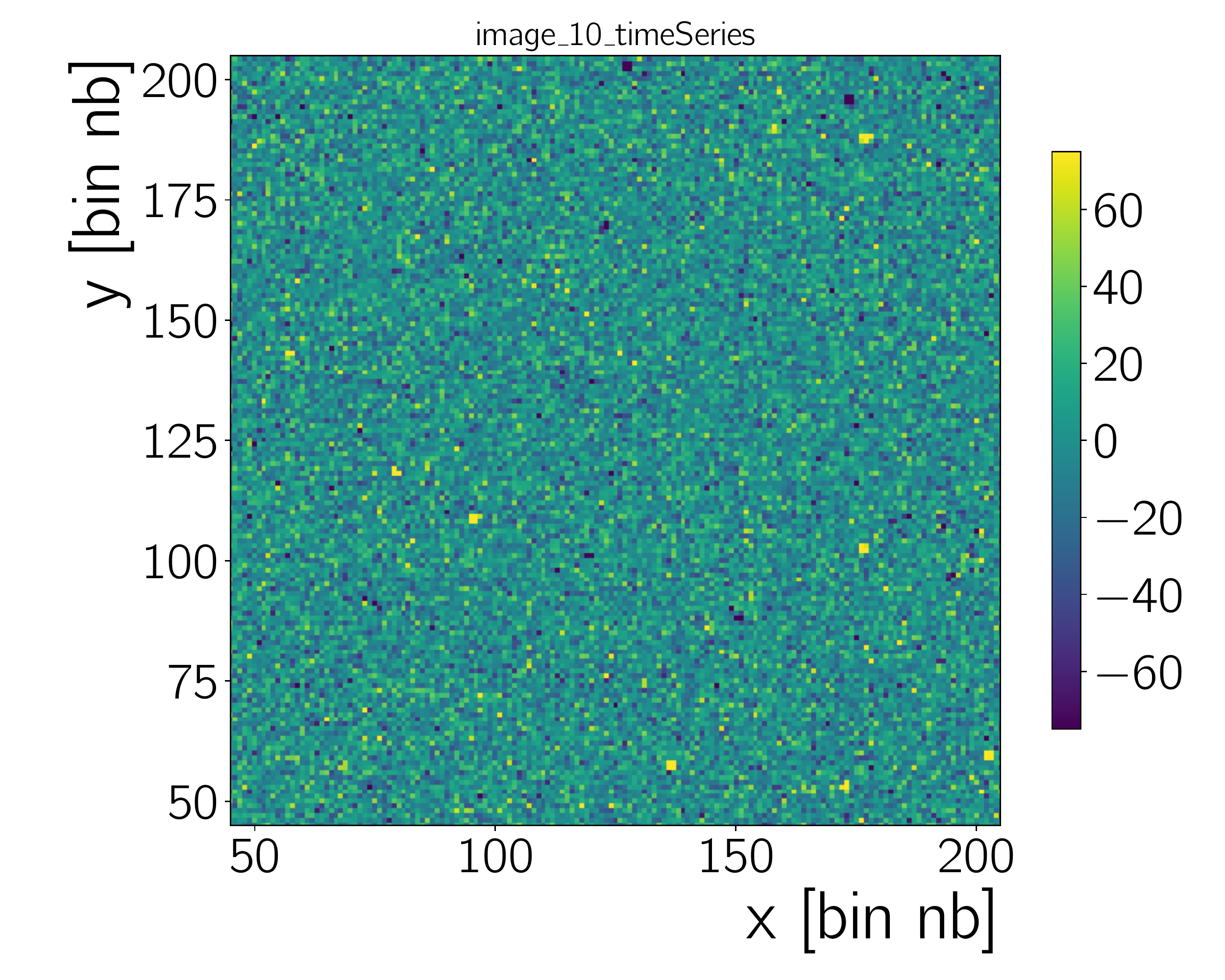}}
\subfloat[found clusters]{
\label{cmosPaper:sec:anaProcedure:fig:2DaduDist:ADUcolsub:SglClusterzoom}
\includegraphics[width=0.32\textwidth,trim=0 0 0 31,clip=true]{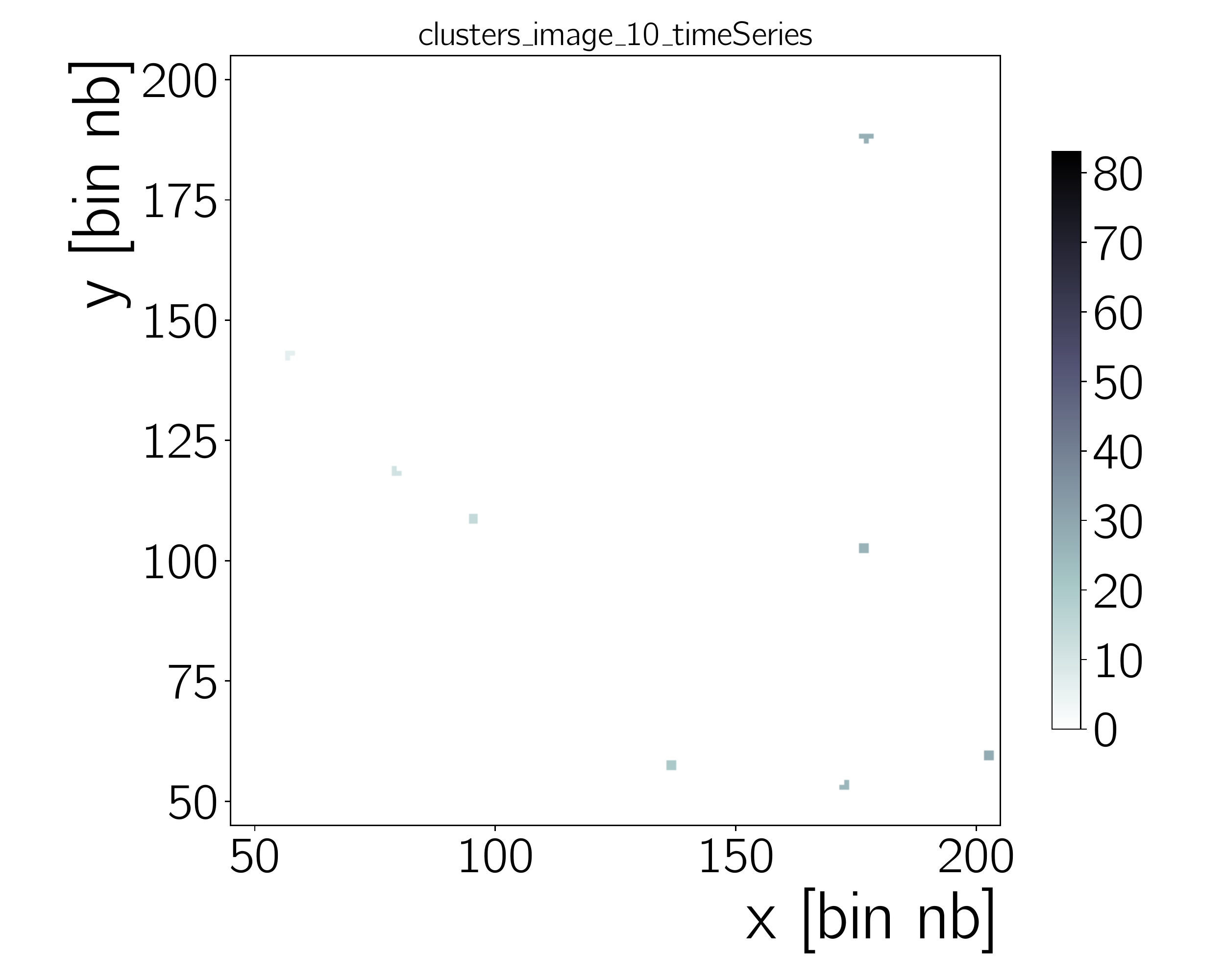}}
\caption{\label{cmosPaper:sec:anaProcedure:fig:2DaduDist:ADU:am241zooms}Detail view  \protect\subref{cmosPaper:sec:anaProcedure:fig:2DaduDist:ADU:Sglzoom} of the frame in \figref{cmosPaper:sec:anaProcedure:fig:2DaduDist:ADU:Sgl} -- raw frame without corrections applied -- and \protect\subref{cmosPaper:sec:anaProcedure:fig:2DaduDist:ADUcolsub:SglzoomTS} of the frame in \figref{cmosPaper:sec:anaProcedure:fig:2DaduDist:ADUcolsub:SglTS}, pair wise subtracted frame. \protect\subref{cmosPaper:sec:anaProcedure:fig:2DaduDist:ADUcolsub:SglClusterzoom} Identified clusters in the zoomed image shown in \protect\subref{cmosPaper:sec:anaProcedure:fig:2DaduDist:ADUcolsub:SglzoomTS} using $k_{\text{seed}}=10$ and $k_{\text{skirt}}=3$ as parameters for the cluster finding thresholds, \textit{cf}. \secref{cmosPaper:sec:anaProcedure:clusterFinding}, \eqnref{cmosPaper:sec:anaProcedure:clusterFinding:eq:threshold}. The scale to the right of the image shows the cluster number.}
\end{figure*}
We distinguish two cases ($k_{\textrm{seed}}$ or $k_{\textrm{skirt}}$) for the multiplier $k$: First, the factor to find the \textit{seed pixel} for a cluster ($k_{\textrm{seed}}$). After the seed pixel has been found we check in its vicinity for pixels fulfilling \eqnref{cmosPaper:sec:anaProcedure:clusterFinding:eq:threshold} with $k_{\textrm{skirt}}$, where $k_{\textrm{skirt}}\leq k_{\textrm{seed}}$. All contiguous pixels with charge values larger than the skirt threshold, as well as the seed pixel, constitute one cluster. For each cluster we store its defining properties such as size, charge, $x,y$ position, frame number, pedestal value and an identification number (\textit{cf.} \secrefbra{cmosPaper:sec:results}). \Figref{cmosPaper:sec:anaProcedure:fig:2DaduDist:ADUcolsub:SglClusterzoom} shows the clusters identified in the previously shown zoomed image of a frame taken with the $^{241}\text{Am}$ source.\footnote{The difference in the bare cluster counts and the shape of the cluster charge spectra (\textit{cf.} \figrefbra{cmosPaper:sec:anaProcedure:clusterAna:fig:clustercuts}) obtained during measurements with a radioactive source versus measurements without the presence of a source shows that the clusters selected in \figref{cmosPaper:sec:anaProcedure:fig:2DaduDist:ADUcolsub:SglClusterzoom} are not just noise, but due to the source radiation.}\\
The parameter $k_{\text{seed}}$ and $k_{\text{skirt}}$ are optimised using data obtained without the presence of a radioactive source (background data), while aiming for a low cluster count by recorded frame and a small cluster size. This optimisation is done on all the clusters found in our set of background run. Clusters in the background data will be due to cosmic radiation passing through the chip, due to noise fluctuations and created by radiation from natural radio-isotopes. For these sources we expect a cluster size to be small -- especially since we use $4\times4$ readout binning -- since $\gamma$- and x-ray photons should depose their energy localised and it is not likely that cosmic muons or $\beta$ particles pass exactly parallel through the chip. With $k_{\text{seed}}=10$ less than \SI{0.5}{clusters} per frame are found while a further increase to \textit{e.g.} $k_{\text{seed}}=20$ does not result in a further reduction. The cluster size decreases exponentially with $k_{\text{skirt}}$ and approaches a mean of $\sim\!\!\SI{3}{clusters}$ for all tested $k_{\text{seed}}$ values. The change is no longer significant for $k_{\text{skirt}}\geq3$. Therefore, we use  $k_{\text{seed}}=10$ and $k_{\text{skirt}}=3$ during our analysis. These values are to some extent arbitrary. However, using $k_{\text{seed}}=10$ and $k_{\text{skirt}}=3$ reduces the amount of background clusters found as stated before. When combined with additional cuts on cluster properties, as discussed in \secref{cmosPaper:sec:results}, these cluster finding settings allow to create almost background free samples.

\subsubsection{Cluster parameters}
\label{cmosPaper:sec:anaProcedure:clusterParameters}

\begin{figure*}
\centering
\subfloat[]{
\includegraphics[width=0.45\textwidth,trim=0 0 0 0,clip=true]{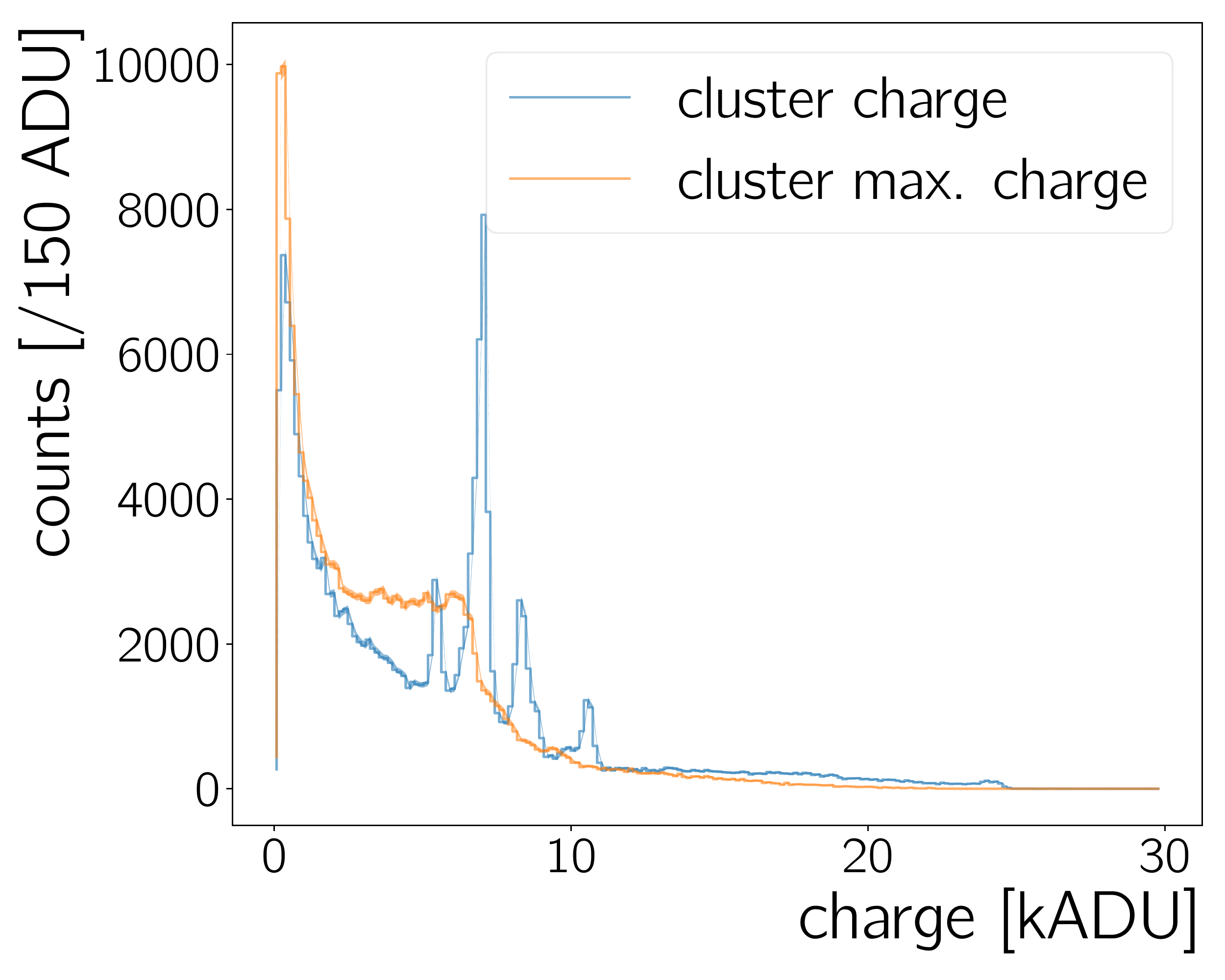}
\label{cmosPaper:sec:anaProcedure:clusterAna:fig:clustercuts:noCut}}
\subfloat[]{
\includegraphics[width=0.45\textwidth,trim=0 0 0 0,clip=true]{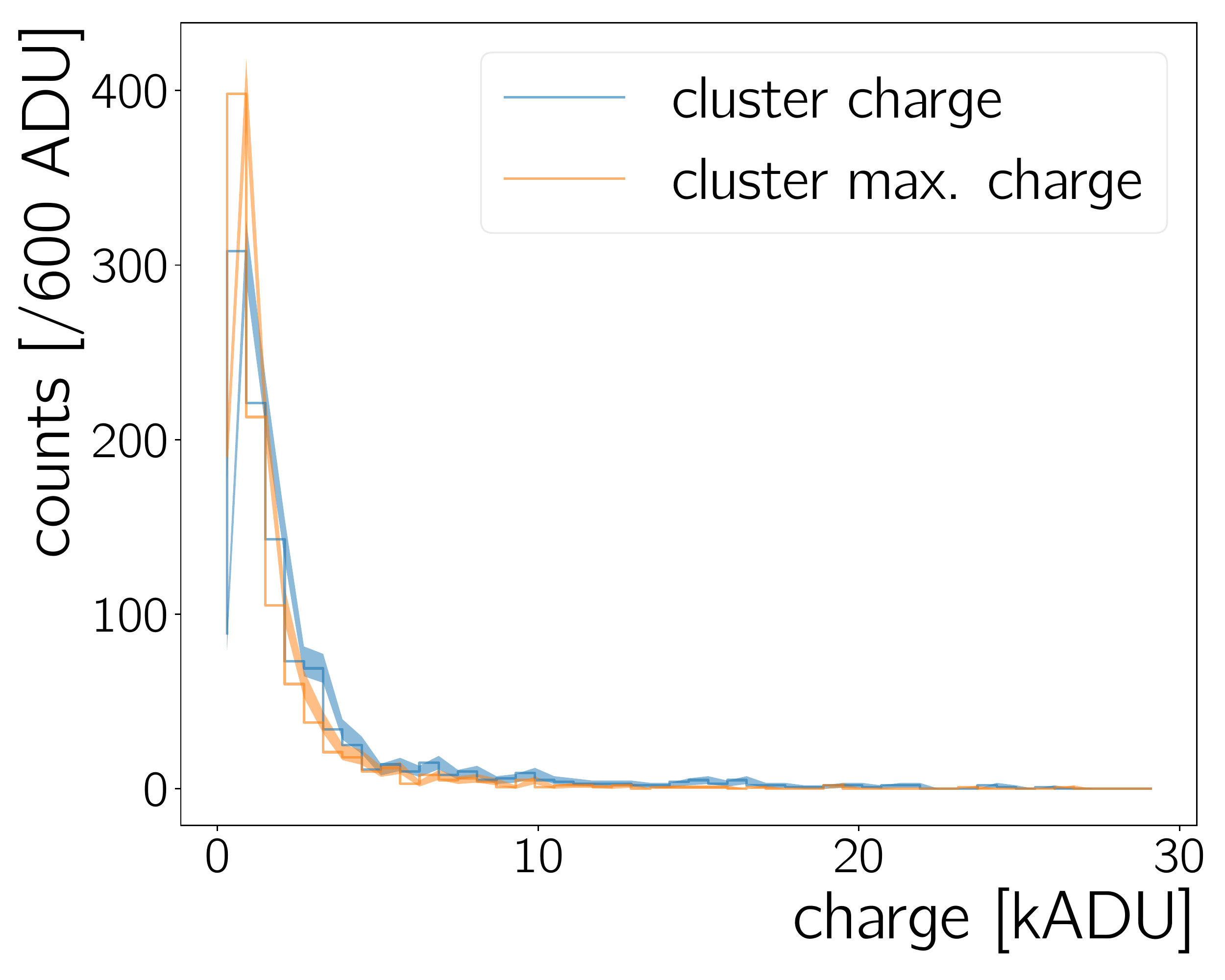}
\label{cmosPaper:sec:anaProcedure:clusterAna:fig:clustercuts:bkg:noCut}}\\
\subfloat[]{
\includegraphics[width=0.45\textwidth,trim=0 0 0 0,clip=true]{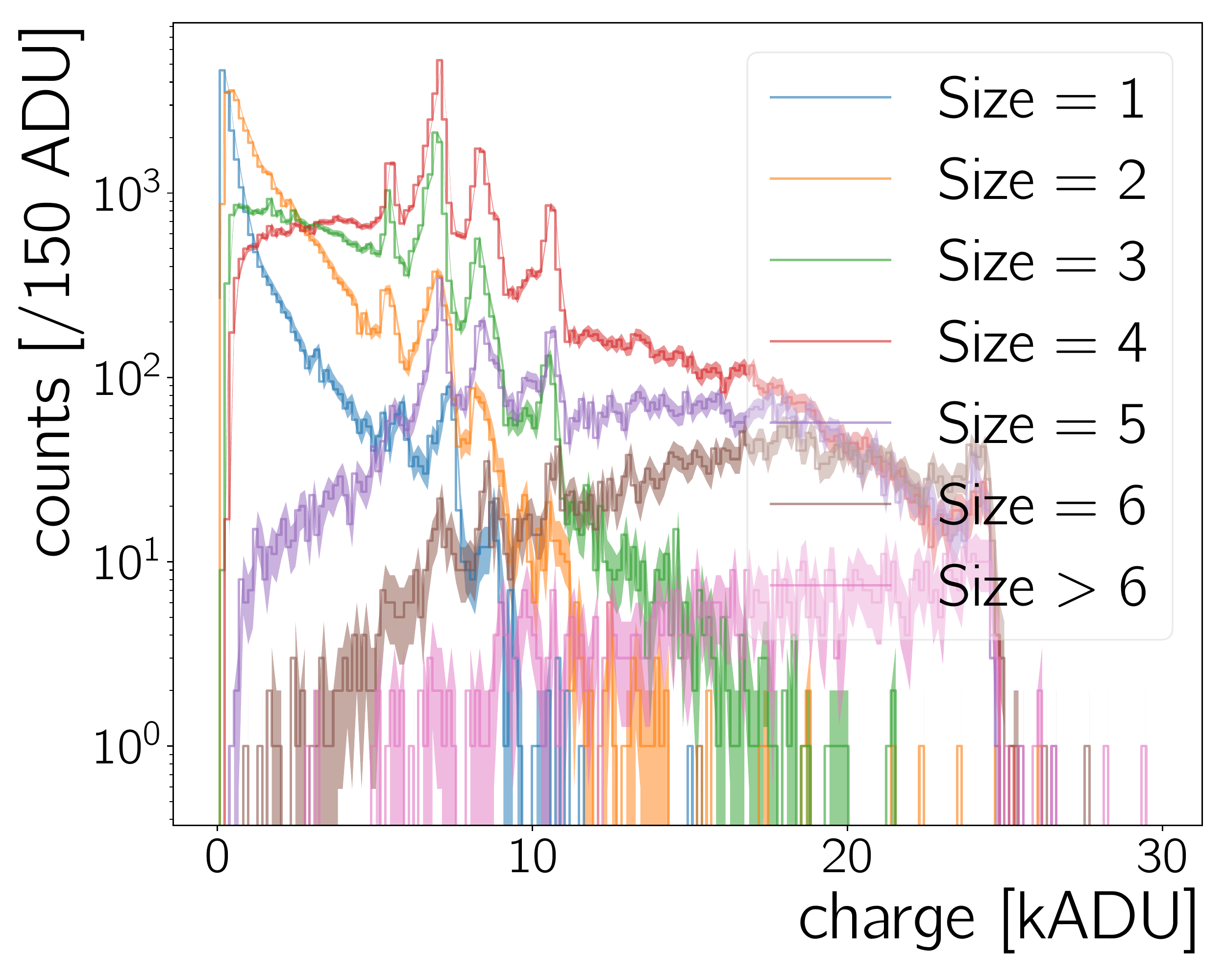}
\label{cmosPaper:sec:anaProcedure:clusterAna:fig:clustercuts:noCut:size}}
\subfloat[]{
\includegraphics[width=0.45\textwidth,trim=0 0 0 0,clip=true]{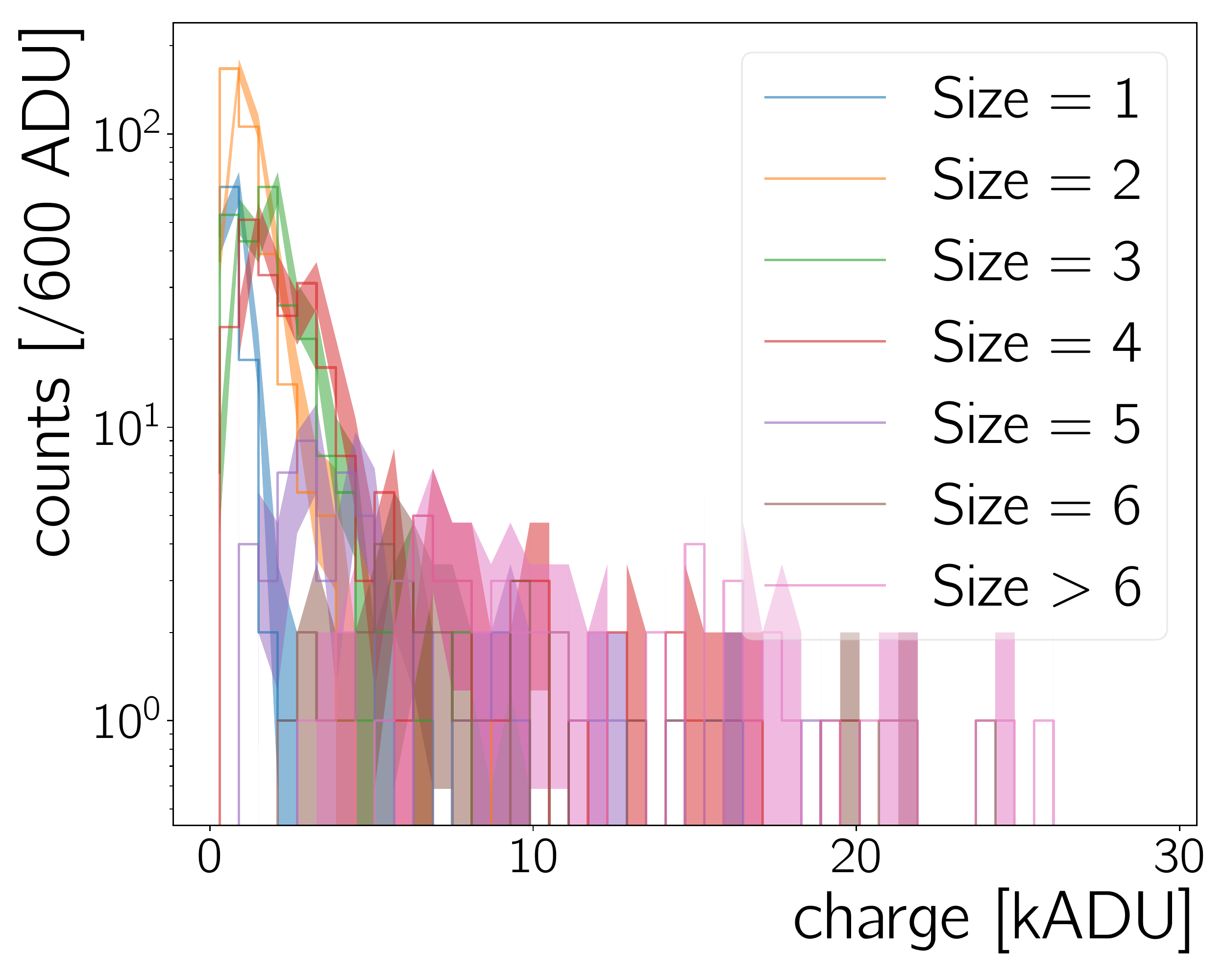}
\label{cmosPaper:sec:anaProcedure:clusterAna:fig:clustercuts:bkg:noCut:size}}\\
\subfloat[]{
\includegraphics[width=0.45\textwidth,trim=0 0 0 0,clip=true]{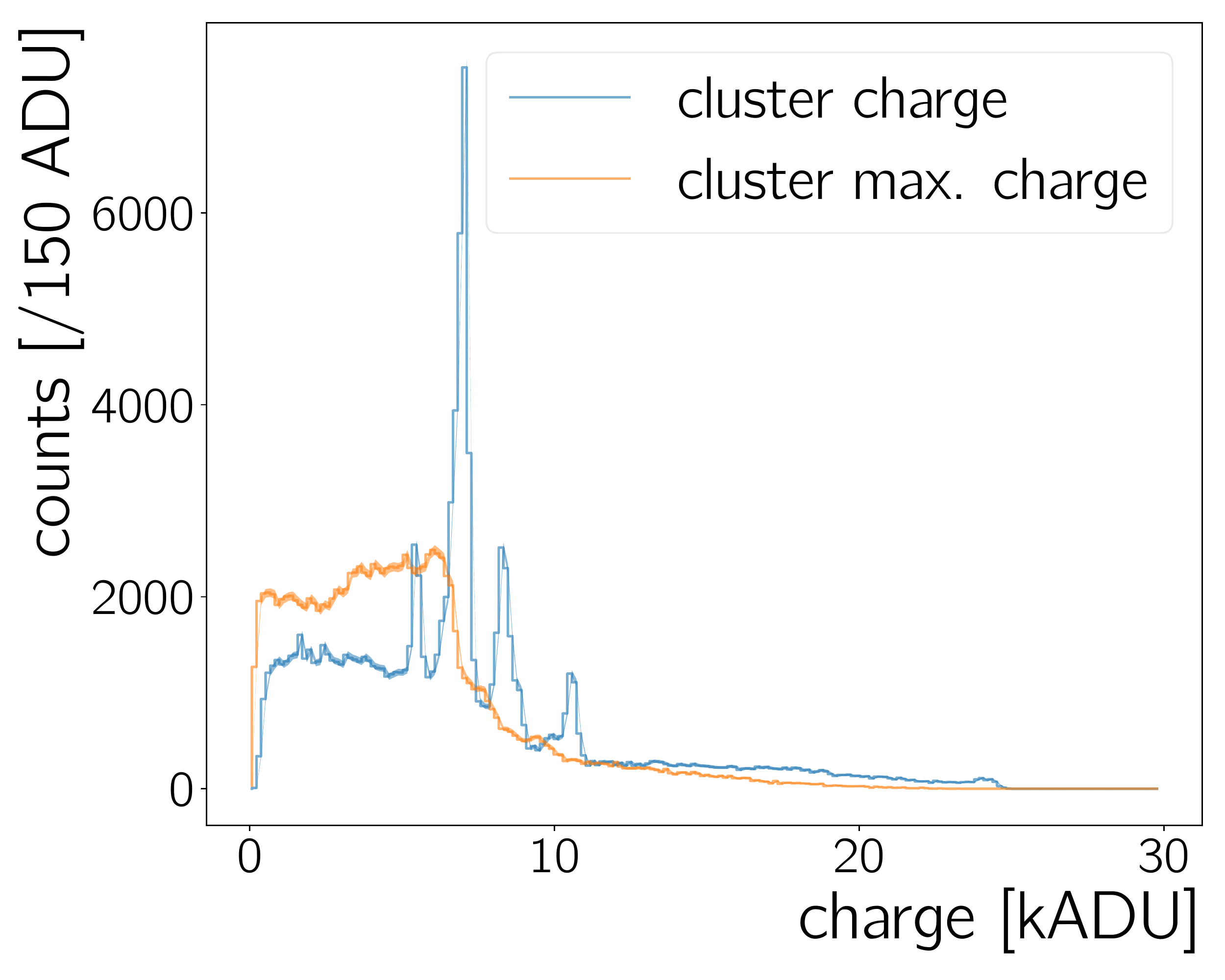}
\label{cmosPaper:sec:anaProcedure:clusterAna:fig:clustercuts:sizecut}}
\subfloat[]{
\includegraphics[width=0.45\textwidth,trim=0 0 0 0,clip=true]{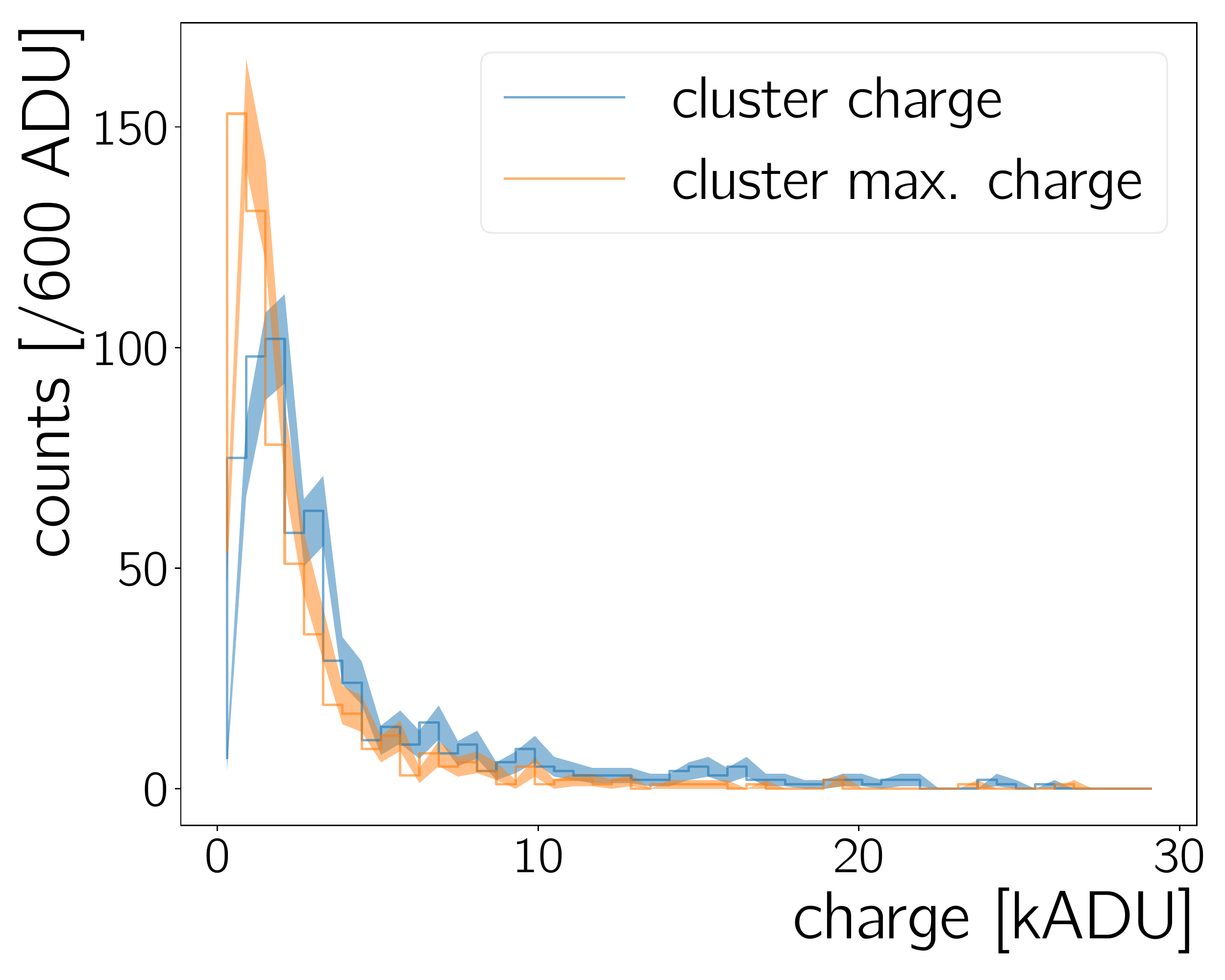}
\label{cmosPaper:sec:anaProcedure:clusterAna:fig:clustercuts:bkg:sizecut}}
\caption{\label{cmosPaper:sec:anaProcedure:clusterAna:fig:clustercuts}The plots show cluster charge and maximal charge spectra for $^{241}\text{Am}$ data in the left column (\protect\subref{cmosPaper:sec:anaProcedure:clusterAna:fig:clustercuts:noCut}, \protect\subref{cmosPaper:sec:anaProcedure:clusterAna:fig:clustercuts:noCut:size}, and \protect\subref{cmosPaper:sec:anaProcedure:clusterAna:fig:clustercuts:sizecut}), 
and for background data in the right column (\protect\subref{cmosPaper:sec:anaProcedure:clusterAna:fig:clustercuts:bkg:noCut}, \protect\subref{cmosPaper:sec:anaProcedure:clusterAna:fig:clustercuts:bkg:noCut:size} and \protect\subref{cmosPaper:sec:anaProcedure:clusterAna:fig:clustercuts:bkg:sizecut}). As energy unit kilo \si{ADU}, \textit{i.e} \si{\kilo ADU}, is used. The live-time of the camera during the $^{241}\text{Am}$ data taking is \SI{2470}{\second} where the camera is radiated with the corresponding source, while the live-time for the background data taking as \SI{2945}{\second}. The first row shows spectra containing all clusters found in the frames of the respective data taking runs, the second row shows cluster charge spectra grouped by cluster size in logarithmic scale while third row shows all data displayed in the first column, which passed a $\text{size}>2$ cut. Note the larger binning in the second column.}
\end{figure*}

For each cluster we store the following properties:
\begin{itemize} 
  \item the frame number
  \item the cluster number, which is a counter for all clusters in one frame
  \item the $x/y$ position, \textit{i.e} the coordinates of the seed pixel
  \item the cluster \textit{size}, \textit{i.e} how many pixels make up a cluster
  \item the \textit{cluster charge}, \textit{i.e} the integral over all $T({x},y,{n_{\textrm{f}}})$ in a cluster subtracted by the cluster pedestal\footnote{In case of the time-series approach $p\left(x,y\right)\sim\!\!0$. Without the time-series approach subtracting the cluster pedestal is essential since it is different from zero.}
  \item the \textit{cluster pedestal}, \textit{i.e} the integral over all the clusters pixels' $p\left(x,y\right)$ 
  \item the charge of the pixel with the highest charge in the cluster (\textit{maximal charge}).
\end{itemize}
Analysis based on clusters are done on the full set of clusters found in specific set of data taking runs. \Figref{cmosPaper:sec:anaProcedure:clusterAna:fig:clustercuts:noCut} and \figref{cmosPaper:sec:anaProcedure:clusterAna:fig:clustercuts:bkg:noCut} show the cluster charge as well as the maximal charge without additional cuts on cluster properties.\footnote{In some data taking runs the camera took a short time to reach a stable state. Therefore, as a precaution, we do not use clusters of the first five frames of each run. Occasional runs with no-stable camera conditions have  in general been rejected.}
Cluster charge and maximal charge spectra of the background data peak at a few \SI{100}{ADU} (\figrefbra{cmosPaper:sec:anaProcedure:clusterAna:fig:clustercuts:bkg:noCut}) and have a tail towards higher values. Their most probable cluster size is $\sim\!\!\SI{2}{pixel}$ (\figrefbra{cmosPaper:sec:anaProcedure:clusterAna:fig:clustercuts:bkg:noCut:size}). The shape of the spectra as well as the cluster size are compatible with the expectations that the background counts are created by noise fluctuations and cosmic radiation passing through the chip.\\
\begin{table*}
\centering
\begin{tabular}{c|c|c}
energy & source & intensity   \\ \hline
\SI{13.8}{\kilo\electronvolt} & $\text{Np}$: $L_{\alpha1}$, x-ray & \\
\SI{17.8}{\kilo\electronvolt} & $\text{Np}$: $L_{\beta1}$, x-ray  & \\
\SI{20.8}{\kilo\electronvolt} & $\text{Np}$: $L_{\gamma1}$, x-ray & \\
\SI{26.3}{\kilo\electronvolt} & $^{241}\text{Am}$: $\gamma$      & \SI{2.3}{\%}  \\ 
\SI{59.5}{\kilo\electronvolt} & $^{241}\text{Am}$: $\gamma$      & \SI{35.9}{\%} \\
\end{tabular}
\caption{\label{cmosPaper:sec:results:caliometry:tab:amenergies}Expected lines in the decay spectrum of $^{241}\text{Am}$ based on data from \cite{chong1997gamma,IAEA}. $\text{Np}$ x-rays in the energy region from \SI{11.87}{\kilo\electronvolt} to \SI{22.4}{\kilo\electronvolt} are expected to make up for another \SI{37}{\%} of intensity \cite{IAEA}. The x-ray energies are approximate and are composed of several overlapping lines. Therefore no intensities are given, since these require assumptions on the detectors energy resolution.}
\end{table*}
A clear peak-structure is observed in the case of the $^{241}\text{Am}$ data in \figref{cmosPaper:sec:anaProcedure:clusterAna:fig:clustercuts:noCut}. The charge of a cluster should be proportional to the energy deposited by the incident radiation. From $^{241}\text{Am}$-decay energy spectra in the literature there should be 5 prominent lines \cite{Demir_2013,chong1997gamma} at energies stated in \tabref{cmosPaper:sec:results:caliometry:tab:amenergies}. The spectra presented here show four prominent peaks (\figrefbra{cmosPaper:sec:anaProcedure:clusterAna:fig:clustercuts}, first column), which will be discussed in detail in \secref{cmosPaper:sec:results:caliometry}.

\subsubsection{Sizes of identified clusters}
\label{cmosPaper:sec:anaProcedure:clusterSize}

The peaks visible in \figref{cmosPaper:sec:anaProcedure:clusterAna:fig:clustercuts:noCut} and \figref{cmosPaper:sec:anaProcedure:clusterAna:fig:clustercuts:sizecut} sit on a floor which is itself related to the decay radiation of the source. Examining the cluster charge spectrum as a function of the cluster size (\figrefbra{cmosPaper:sec:anaProcedure:clusterAna:fig:clustercuts:noCut:size}) shows that this floor is mainly due to clusters with a size of 1 and \SI{2}{pixels}. The peaks are furthermore significantly less prominent for these cluster sizes than \textit{e.g.} for cluster sizes of \SI{3}{pixels} and \SI{4}{pixels}. For cluster sizes larger than four pixels the peak heights decrease again. It is interesting to note that there is a correlation between cluster size and cluster charge, \textit{i.e.} energy deposited in the sensor. For increasing cluster size the ratio of clusters with a large charge value to such with a low charge value increases. Gamma radiation and x-rays are expected to interact in the CMOS sensor and to release their energy locally. Therefore, the observed cluster sizes are larger than expected, even more so, given the readout binning of $4\times4$, resulting side length per readout pixel of $4\times\SI{6.5}{\micro\meter}=\SI{26}{\micro\meter}$ each. As stated in \secref{cmosPaper:sec:expSetUp} the exact layout of the actual CMOS is not known -- its different layers may lead to a spread of the charge which reaches a few \SI{10}{\micro\meter}. Incident radiation can \textit{e.g.} be absorbed in a non-active layer of the chip and then diffuse towards the collection zones. Another possible explanation is that a substantial fraction of the incoming $\gamma$ energy gets transferred to a few $\delta$-electrons which can then travel more than a pixel length in the sensor, while they produce further ionisation. It can be excluded that the cluster size gets inflated by pixels accidentally assigned to the respective cluster. Comparing the spectrum where only the most energetic pixel per cluster is shown (maximal charge) with the cluster charge spectrum shows that the information from the lower energy pixels in a cluster is needed to measure a spectrum with distinguishable peaks (\figrefbra{cmosPaper:sec:anaProcedure:clusterAna:fig:clustercuts:noCut} and \figrefbra{cmosPaper:sec:anaProcedure:clusterAna:fig:clustercuts:sizecut}). \\
For the analysis of the Neo sCMOS' calorimetric response we require the cluster size to be larger than \SI{2}{pixels} in order to improve the quality of the peak spectrum. \Figref{cmosPaper:sec:anaProcedure:clusterAna:fig:clustercuts:sizecut} shows the data displayed in \figref{cmosPaper:sec:anaProcedure:clusterAna:fig:clustercuts:noCut} but with a $\text{size}>\SI{2}{pixels}$ cut. The floor below the peaks is substantially reduced. Approximately half of the entries in the background spectrum are removed when this cut is applied to the background data (\figrefbra{cmosPaper:sec:anaProcedure:clusterAna:fig:clustercuts:bkg:sizecut} vs \figrefbra{cmosPaper:sec:anaProcedure:clusterAna:fig:clustercuts:bkg:noCut}).

\section{Performance of the CMOS as radiation detector}
\label{cmosPaper:sec:results}

\subsection{Calorimetric capabilities of the Neo sCMOS}
\label{cmosPaper:sec:results:caliometry}

\begin{figure*}
\centering
\subfloat[Combined plot]{
\includegraphics[width=0.45\textwidth,trim=0 0 0 0,clip=true]{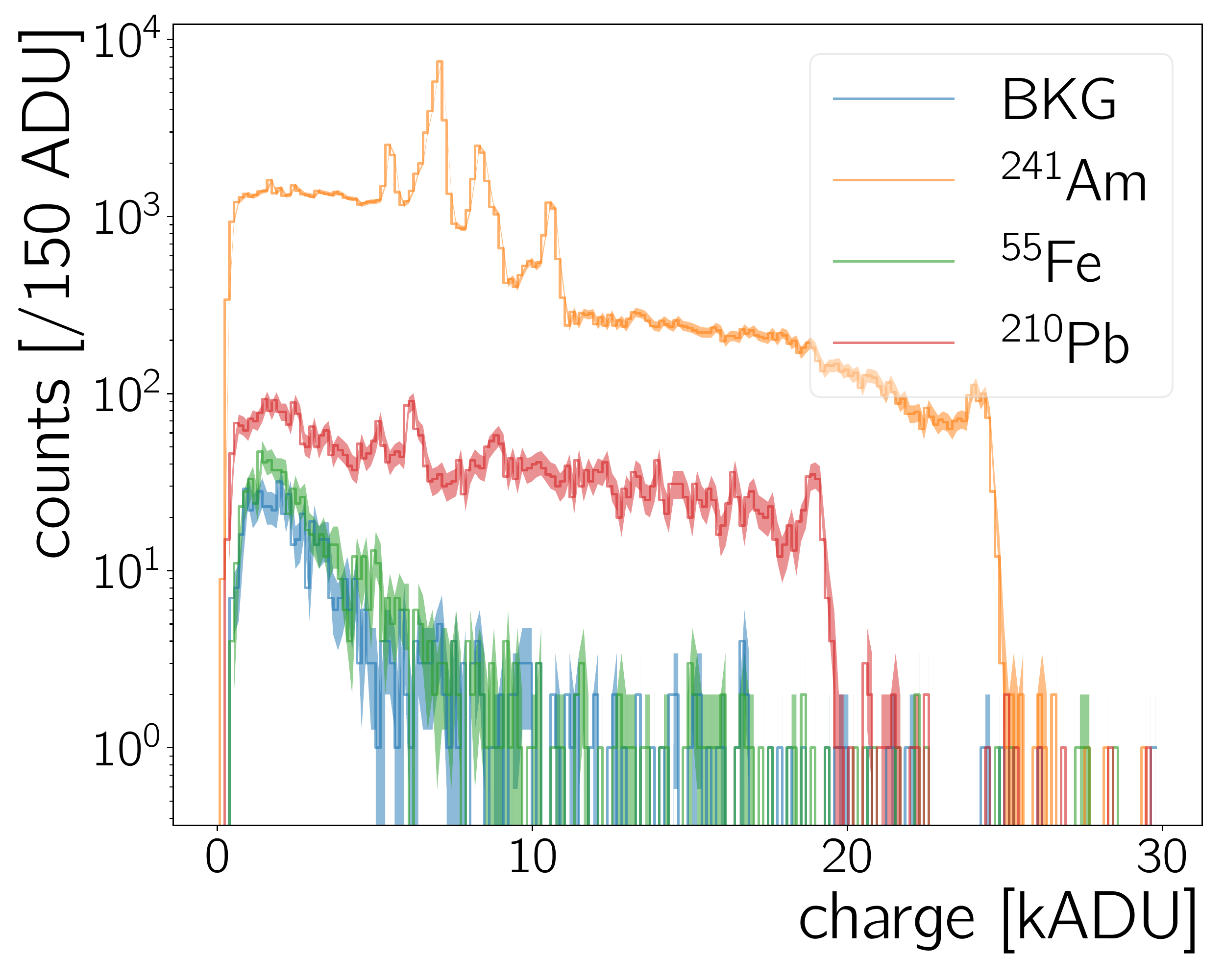}
\label{cmosPaper:sec:results:caliometry:fig:allSourcesLog}}
\subfloat[\am{241} spectrum, counts scaled]{
\includegraphics[width=0.45\textwidth,trim=0 0 0 0,clip=true]{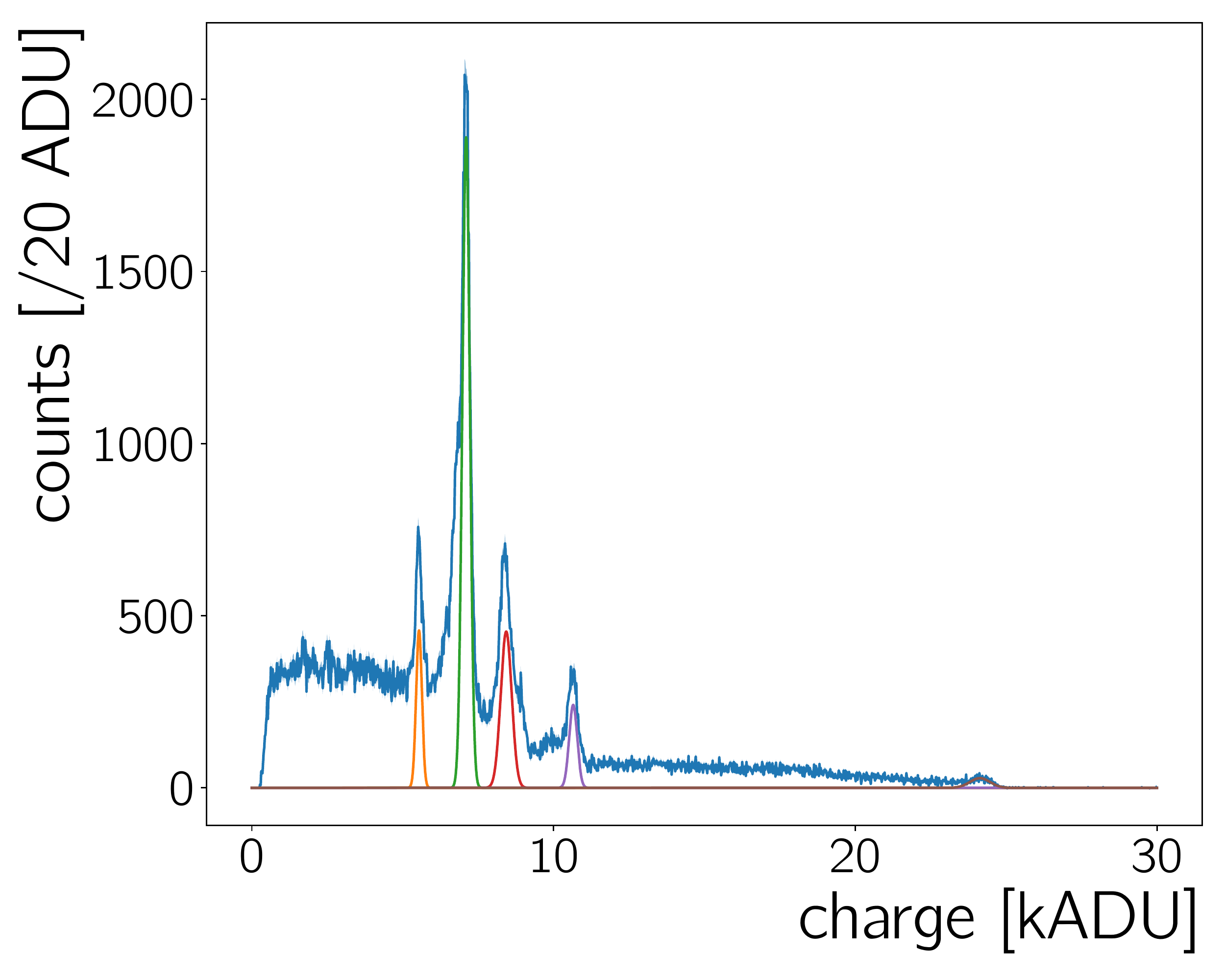}
\label{cmosPaper:sec:results:caliometry:fig:am241scaledMinusBkg}\label{fig:conversion2}}\\
\subfloat[\fe{55} spectrum, counts scaled]{
\includegraphics[width=0.45\textwidth,trim=0 0 0 0,clip=true]{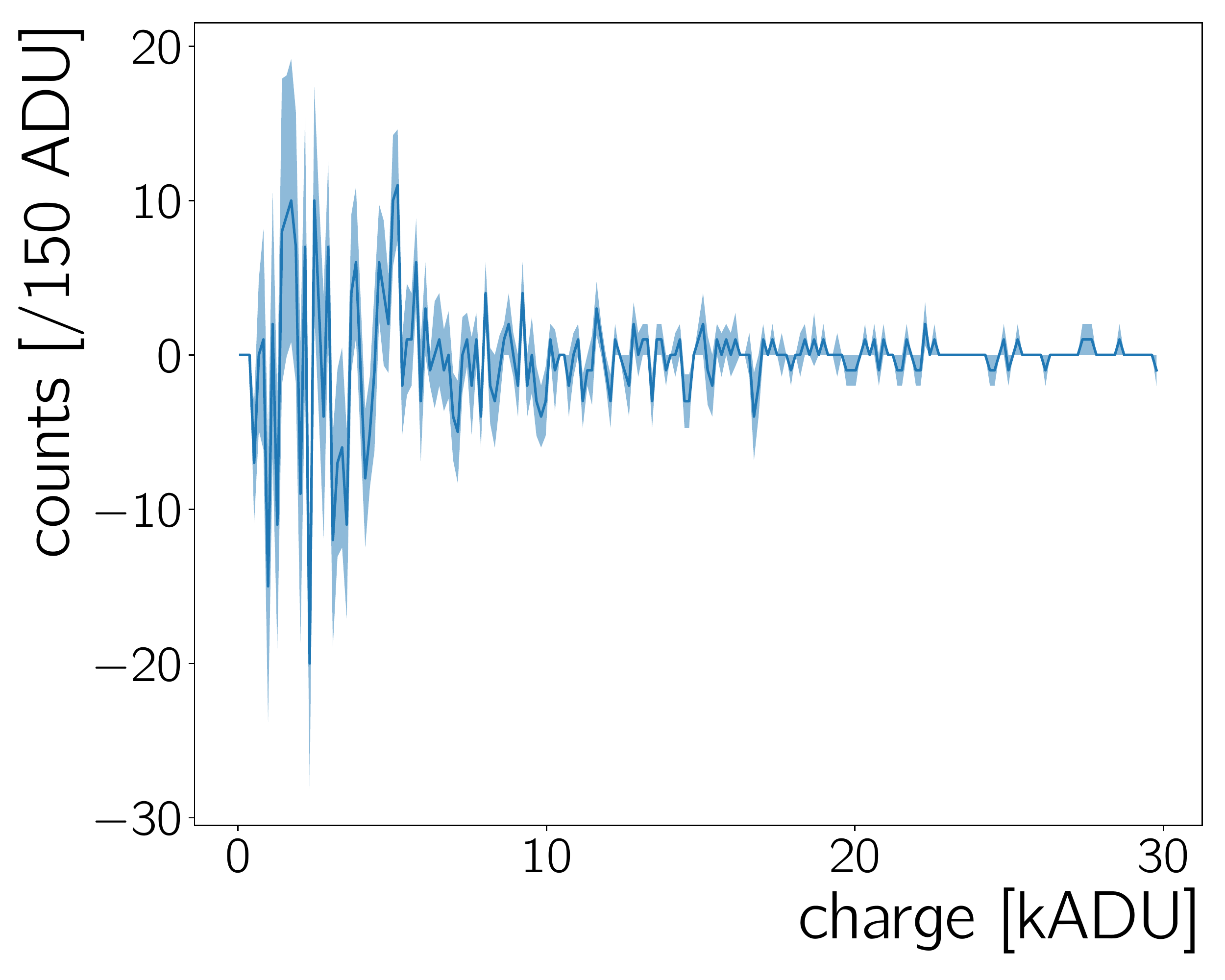}
\label{cmosPaper:sec:results:caliometry:fig:fe55scaledMinusBkg}}
\subfloat[\pb{210} spectrum, counts scaled]{
\includegraphics[width=0.45\textwidth,trim=0 0 0 0,clip=true]{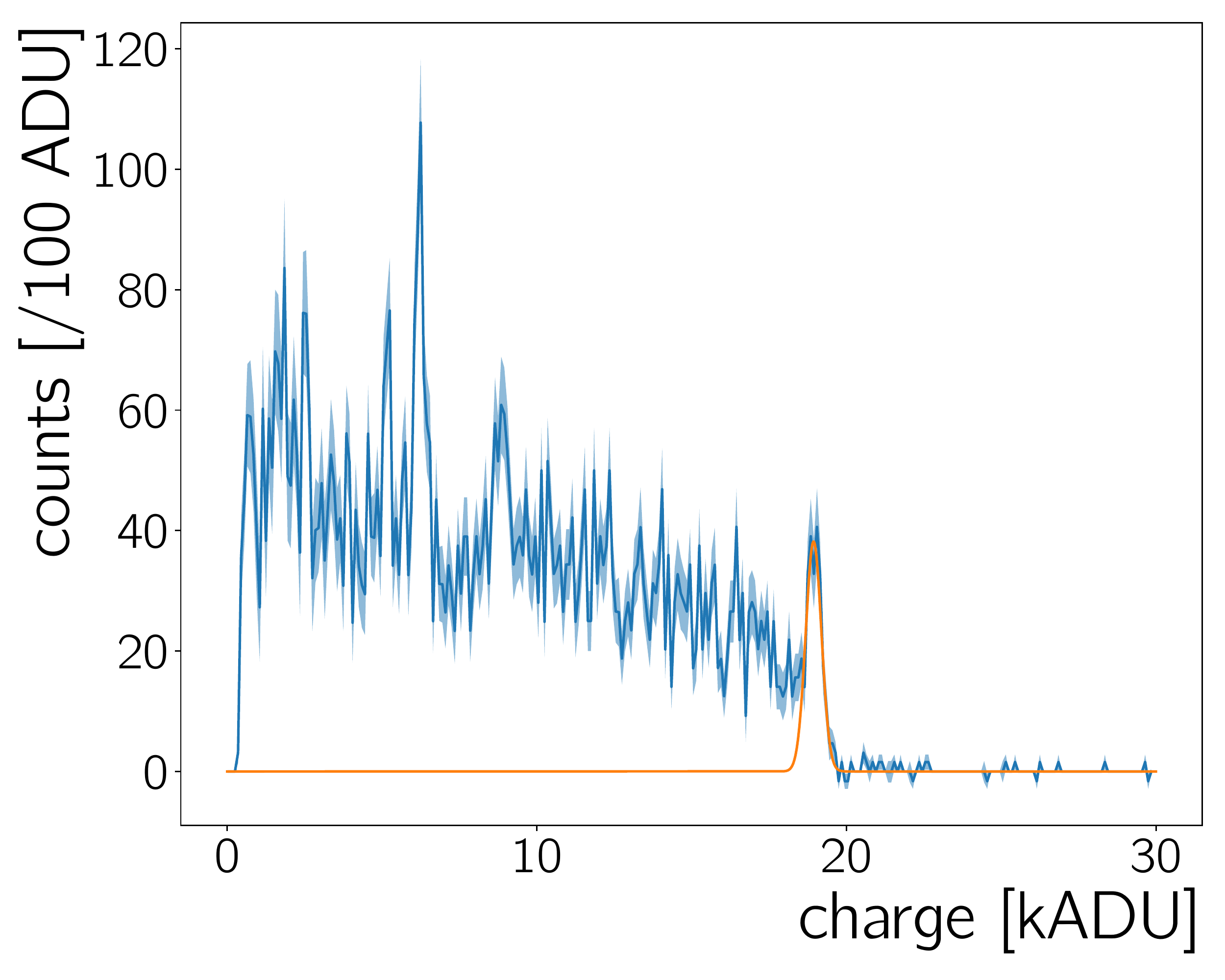}
\label{cmosPaper:sec:results:caliometry:fig:pb210scaledMinusBkg}\label{fig:conversion3}}
\caption{\label{cmosPaper:sec:results:caliometry:allamfepb}\protect\subref{cmosPaper:sec:results:caliometry:fig:allSourcesLog} Combined plot showing an overview of the different data sets acquired with the Neo sCMOS. The live-time of the background (\am{241}, \fe{55} and \pb{210}, respectively) measurement is \SI{29450}{\second} (\SI{24700}{\second}, \SI{41800}{\second} and \SI{30400}{\second}, respectively). The spectra are shown on a log scale for better visibility since the rates of the sources vary as does the observed event rate. \protect\subref{cmosPaper:sec:results:caliometry:fig:am241scaledMinusBkg}, \protect\subref{cmosPaper:sec:results:caliometry:fig:fe55scaledMinusBkg}, \protect\subref{cmosPaper:sec:results:caliometry:fig:pb210scaledMinusBkg} shows spectra for the respective sources. These spectra have been scaled to a live-time of \SI{47500}{\second} and are subtracted with the background spectrum scaled to the same live-time. Shaded regions represent the statistical error. For all plots a cluster size $>\SI{2}{pixels}$ is required. The spectra in \protect\subref{fig:conversion2} and \protect\subref{fig:conversion3} illustrate furthermore the peaks fitted with Gau{\ss}ian functions, \textit{cf}. \secref{cmosPaper:sec:results:caliometry:energyMeasurement}. (Note that only the Gau{\ss}ians are plotted, and not the additionally fitted backgrounds.)}
\end{figure*}

The camera response to \fe{55}, \pb{210} and \am{241} source radiation is examined in order to determine the Neo sCMOS' calorimetric measurement capabilities. Background data obtained with no source present is used as well for this analysis. \Figref{cmosPaper:sec:results:caliometry:allamfepb} shows cluster charge spectra measured for these sources. To create these, the analysis procedures detailed in \secref{cmosPaper:sec:anaProcedure} are applied to the raw frames and a cluster size larger than two pixels is required for all entries in the plots.\\
The contribution of the source radiation to the spectra has to be disentangled from the contribution of the background radiation. To this end, spectra obtained with radioactive sources and the background spectrum are normalised to the same live-time and then the background spectrum is subtracted from the source spectra. The results are shown in \figref{fig:conversion2} to \figref{cmosPaper:sec:results:caliometry:fig:pb210scaledMinusBkg}.
\paragraph{\am{241}:}The cleanest spectrum is obtained with the \am{241} source, which has an activity of \SI{344(17)}{\kilo\becquerel} as of at the time the measurement.\footnote{The uncertainty on the initial source activity is not known, therefore a \SI{5}{\%} error is assumed.} Americium-241 decays via an $\alpha$-decay to $^{237}\text{Np}$. There are many possible $\alpha$-decays with different $Q$ values from \SI{5000}{\kilo\electronvolt} to \SI{5500}{\kilo\electronvolt} \cite{IAEA}, where the most probable (\SI{85}{\%}) decay has an energy of \SI{5485}{\kilo\electronvolt}. These $\alpha$-decays occur together with $\gamma$-ray emission and x-ray emission by the $^{237}\text{Np}$ atom \cite{IAEA}. \Tabref{cmosPaper:sec:results:caliometry:tab:amenergies} lists the two $\gamma$ energies with the largest yield per decay as well as x-ray lines measured in \am{241} spectra elsewhere. The CMOS chip of the Neo sCMOS camera is housed behind a glass window, with an assumed thickness of \SI{1}{\milli\meter} -- therefore the $\alpha$-particles will not reach the sensor, since the range of $\alpha$s of this energy is less than \SI{100}{\micro\meter} \cite{ASTAR}. The energy deposits measured with the \am{241} source are thus for the most part due to $\gamma$- and x-rays. Attenuation lengths for different $\gamma$- and x-ray energies are given in \tabref{proposedResearch:fig:attenuationEnergies}. 
\paragraph{\fe{55}:}Iron-55 decays via electron capture to $^{55}\text{Mn}$ \cite{IAEA}. After the decay, the electron shell re-arranges to match the levels of $^{55}\text{Mn}$ and to fill the hole from the electron capture. By doing so, Auger-Meitner electrons with an energy of up to \SI{6}{\kilo\electronvolt} are released as well as x-rays of \SI{5.9}{\kilo\electronvolt} and \SI{6.5}{\kilo\electronvolt}. For these x-ray energies the yield per decay is \SI{16.6}{\%} and \SI{7}{\%}, respectively. Although the source used has a rate of $\sim\!\!\SI{100}{\kilo\becquerel}$, the background subtracted spectrum in \figref{cmosPaper:sec:results:caliometry:fig:fe55scaledMinusBkg} is compatible with zero. For low charge values, \textit{i.e} low energy deposits, the spectrum is more erratic -- however, no clear peak can be identified. For photon energies $\leq\SI{10}{\kilo\electronvolt}$ we estimate a lower limit for the photon absorption in glass with the data from \cite{xrayAbs}, assuming \SI{10}{\kilo\electronvolt} photon energy and a glass density of \SI{2.23}{\gram\per\centi\meter\cubed}. For a window of \SI{1}{\milli\meter} and \SI{2}{\milli\meter}, at least \SI{97}{\%} and \SI{99.04}{\%} of the x-rays are absorbed in the glass before they reach the chip, respectively. Therefore, the non observation of any clear peak is most likely due to the x-ray absorption in the Neo sCMOS window.
\begin{table*}
\centering
\hfill
\begin{tabular}[t]{c|c|c}
Energy $\ $                               & \multicolumn{2}{c}{$\ $Attenuation length/$\ $ } \\
 $\left[ \si{\kilo\electronvolt} \right]$ & \multicolumn{2}{c}{$\ $Range in Silicon$\ $}     \\
                                          & $\gamma$-/X-ray & $\ \beta^{-}$ \\ \hline
10         & $\ $\SI{111}{\micro\meter} $\ $ & \SI{1.2}{\micro\meter}       \\
15         & $\ $\SI{365}{\micro\meter} $\ $ & \SI{2.5}{\micro\meter}       \\
45         & $\ $\SI{7}{\milli\meter}   $\ $ & \SI{16}{\micro\meter}        \\
\end{tabular}
\hfill
\begin{tabular}[t]{c|c|c}
Energy $\ $                               & \multicolumn{2}{c}{$\ $Attenuation length/$\ $ } \\
 $\left[ \si{\kilo\electronvolt} \right]$ & \multicolumn{2}{c}{$\ $Range in Silicon$\ $}     \\
                                          & $\gamma$-/X-ray & $\ \beta^{-}$     \\ \hline
60         & $\ $\SI{12}{\milli\meter} $\ $ & \SI{28}{\micro\meter}             \\
100        & $\ $\SI{20}{\milli\meter} $\ $ & \SI{66}{\micro\meter}             \\
1000       & $\ $\SI{59}{\milli\meter} $\ $ & \SI{2}{\milli\meter}              \\
\end{tabular}
\hfill\null
\caption{\label{proposedResearch:fig:attenuationEnergies}Approximate ranges of $\gamma$-rays and $\beta^{-}$s (electrons) for typical decay energies in Silicon. For the $\gamma$-rays the attenuation length is calculated from the attenuation cross section given in \cite{XCOM} using the density of silicon-dioxide. The same density is used to calculate the electron range from the CSDA range for electrons given in \cite{ESTAR}.}
\end{table*}
\paragraph{\pb{210}:}Lead-210 decays via $\beta^{-}$ decay to $^{210}\text{Bi}$ as mentioned in \secref{cmosPaper:sec:introduction:pb210}. The most probable $\beta^{-}$ decay (\SI{84}{\%}) results in an excited state of $^{210}\text{Bi}$, whilst emitting an electron with an average decay energy of \SI{4.16}{\kilo\electronvolt}. The nucleus de-excites by emitting a $\gamma$ of \SI{46.5}{\kilo\electronvolt} with a \SI{4}{\%} yield per decay. The de-excitation is accompanied by the emission of x-rays from approximately \SI{9}{\kilo\electronvolt} to \SI{16}{\kilo\electronvolt} with a yield of \SI{22}{\%} per decay. The second most probable decay (\SI{16}{\%}) is a $\beta^{-}$ decay with an electron mean energy of \SI{16.2}{\kilo\electronvolt} to $^{210}\text{Bi}$ in the ground state \cite{IAEA}. The \pb{210} source used has a rate of $\sim\!\!\SI{185}{\kilo\becquerel}$. It holds the lead diluted in nitric acid in a small glass vial. It is not likely that any of the low energy $\beta$-radiation is detected by the Neo sCMOS, given that the decay electrons have to traverse the liquid, the glass of the vial and of the Neo sCMOS before it can be detected by the CMOS chip. Therefore, similarly to the \am{241} source, only the x-rays and $\gamma$-rays are measured.\\
The \pb{210} spectrum in \figref{cmosPaper:sec:results:caliometry:fig:pb210scaledMinusBkg} contains fewer counts than the \am{241} spectrum (\figrefbra{cmosPaper:sec:results:caliometry:fig:am241scaledMinusBkg}). There are several factors contributing to this: First, the activity of the \pb{210} source is a factor of 1.85 lower than the activity of the \am{241} source. The latter source has also a significantly smaller extent -- compared to the CMOS sensor it can be considered as a point source, while the lead source extends over a vial of more than \SI{1}{\centi\meter} length and \SI{0.5}{\centi\meter} diameter. Next, the $\gamma$ yield for the two sources differs greatly -- comparing $\sim\!\!\SI{4}{\%}$ to $\sim\!\!\SI{36}{\%}$ for the \pb{210} \SI{46.5}{\kilo\electronvolt} $\gamma$-ray and the \am{241} \SI{59.5}{\kilo\electronvolt} $\gamma$-ray. In order to establish whether the \am{241} and \pb{210} spectra are consistent with each other, we first need to establish the overall energy scale and compare peaks at a known energy directly.

\subsubsection{Energy response calibration}
\label{cmosPaper:sec:results:caliometry:energyMeasurement}

\begin{figure*}
\centering
\subfloat[]{
\includegraphics[width=0.45\textwidth,trim=0 0 0 0,clip=true]{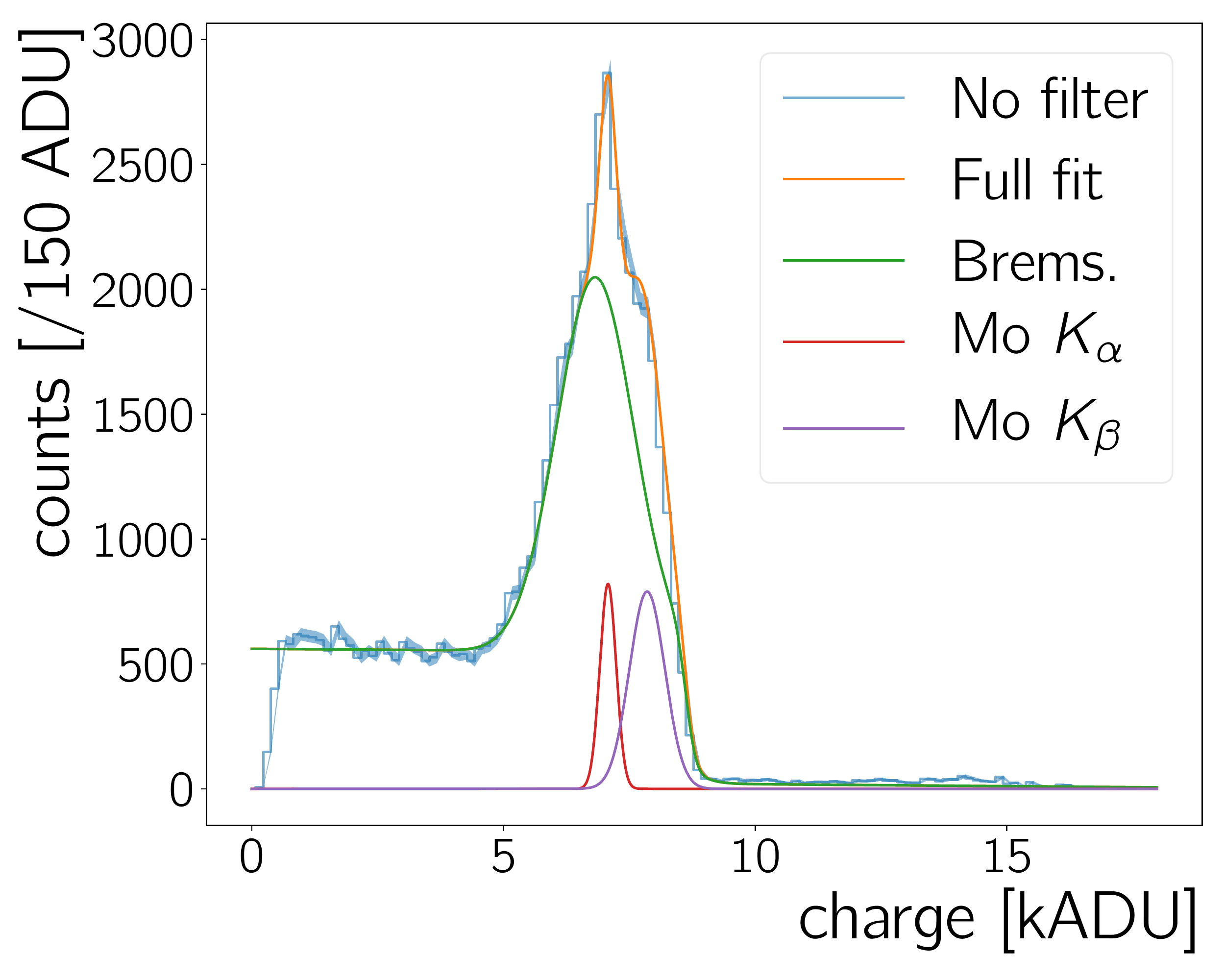}
\label{cmosPaper:sec:results:caliometry:fig:xrayMoFit}}
\subfloat[]{
\includegraphics[width=0.45\textwidth,trim=0 0 0 0,clip=true]{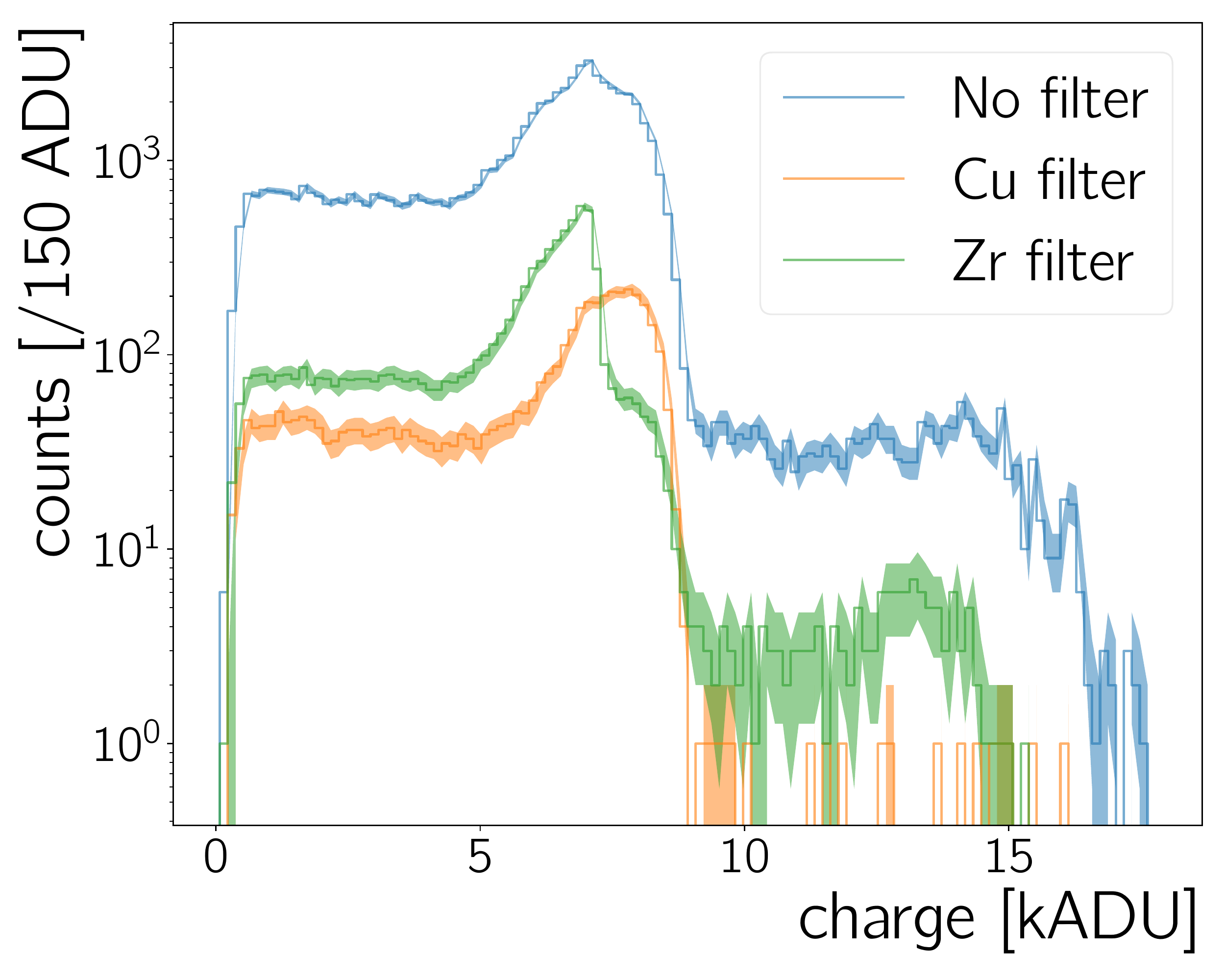}
\label{cmosPaper:sec:results:caliometry:fig:xrayWithFilter}}
\caption{\label{cmosPaper:sec:results:caliometry:fig:xrayData}\protect\subref{cmosPaper:sec:results:caliometry:fig:xrayMoFit} Spectrum recorded with the molybdenum x-ray tube (\textit{cf.} \secrefbra{cmosPaper:sec:expSetUp:subsec:xray}) as well as fitted curves to establish the position of the $K_{\alpha}$ and $K_{\beta}$ peak. \protect\subref{cmosPaper:sec:results:caliometry:fig:xrayWithFilter} The same data as in \protect\subref{cmosPaper:sec:results:caliometry:fig:xrayMoFit} is shown together with data recorded when a $\text{Zr}$ or a $\text{Cu}$ filter is placed between the x-ray tube with $\text{Mo}$ target and the Neo sCMOS. For this plot all data has been normalised to a live-time of \SI{50}{\milli\second} and a cluster size $>\SI{2}{pixels}$ is required. Shaded regions indicate the statistical error.}
\end{figure*}

All the spectra presented so far are shown with \textit{analogue-to-digital} units as unit of the deposited energy in the detector. The Neo sCMOS's manuals consulted during this work do not state a conversion factor from \si{ADU} to energy in \si{\electronvolt}. However, the report \cite{gabriella} specifies a gain of either $\SI{0.59}{\text{e}^{-}\per ADU}$ or $\SI{0.67}{\text{e}^{-}\per ADU}$ according to the supplier. These gain values translate to a conversion factor of either \SI{2.154}{\electronvolt\per ADU} or \SI{2.446}{\electronvolt\per ADU}, respectively, accounting for the $W$ factor in $\text{Si}$ of \SI{3.65}{\electronvolt} to create an electron-hole-pair \cite{kolanoski2016teilchendetektoren}. In order to establish the exact energy scale, the known energies of radioactive sources from literature are matched to the \si{ADU} values at which peaks are observed. All large peaks in the cluster charge spectra are fitted with a Gau{\ss}ian curve and their mean energy, $\varepsilon_{\text{peak}}$, and $\sigma$ is extracted. \Tabref{cmosPaper:sec:results:caliometry:tab:allEnergyResults} lists all peaks used for this analysis and the result of the fits. The Gau{\ss}ians are furthermore plotted in \figref{fig:conversion2}, \figref{fig:conversion3} and \figref{cmosPaper:sec:results:caliometry:fig:xrayMoFit}. The fits are done locally -- in the ranges from  $\varepsilon_{\text{min}}$ to $\varepsilon_{\text{max}}$ as specified in \tabref{cmosPaper:sec:results:caliometry:tab:allEnergyResults} -- and where necessary a polynomial of order one is added to the Gau{\ss}ian curve to account for the floor due to other radiation. For the peaks at the high energy end of the \am{241} and \pb{210} spectrum an error-function is used instead of a polynomial.\\
\begin{table*}
\centering
\begin{tabular}{ll||c|c|c|c|c||c||c|c|c}
\multicolumn{2}{l||}{Source} & \multicolumn{5}{c||}{\am{241}} & \pb{210} & \multicolumn{3}{c}{$\text{Mo}$ x-ray tube} \\ 
\multicolumn{2}{l||}{Radiation} & \multicolumn{2}{c|}{$\gamma$} & \multicolumn{3}{c||}{x-ray ($\text{Np}$)} & $\gamma$ & \multicolumn{2}{c|}{x-ray} & $\text{Zr}$ edge\\ \hline \hline
\multicolumn{2}{l||}{Expected energy $[\si{\kilo\electronvolt}]$}                     
& 26.3  & 59.5  & 13.8 & 17.8 & 20.8 & 46.5  & 17.4 & 19.6 & 18.0 \\ \hline
\multirow{2}{*}{Fit range}  & $\varepsilon_{\text{min}}$ $[\si{\kilo ADU}]$ &  9.87  & 23.49 & 4.99 & 6.01 & 7.78  & 18.37 & \multicolumn{2}{c|}{1} & n.a. \\
                            & $\varepsilon_{\text{max}}$ $[\si{\kilo ADU}]$ & 11.45  & 24.5  & 6.12 & 7.9  & 9.99 & 19.64 & \multicolumn{2}{c|}{15} & n.a. \\ \hline
\multirow{3}{*}{Fit result} & $\varepsilon_{\text{peak}}$ $[\si{\kilo ADU}]$ 
& 10.65 & 24.13 & 5.54 & 7.1  & 8.37 & 18.94 & 7.08 & 7.85 & 7.3  \\
                            & $\sigma_{\text{peak}}$ $[\si{ADU}]$ 
& 133   & 276   & 95   & 124  & 199  & 239   & 163 & 353  & n.a. \\
                           & $\chi^2/N_{\text{dof}}$    
& 1.22  & 1.05  & 1.61 & 2.2  & 1.32 & 0.77  & \multicolumn{2}{c|}{4.62} & n.a. \\
\end{tabular}
\caption{\label{cmosPaper:sec:results:caliometry:tab:allEnergyResults}The table lists the fit results (peak position $\varepsilon_{\text{peak}}$ and standard deviation $\sigma$) of Gau{\ss}ian peak fits in \figref{fig:conversion2}, \figref{fig:conversion3}, and \figref{cmosPaper:sec:results:caliometry:fig:xrayMoFit} as well as the absorption edge read off from \figref{cmosPaper:sec:results:caliometry:fig:xrayWithFilter}. (Since the properties of the absorption edge are not determined by a fit, the corresponding values in the ``$\text{Zr}$ edge'' column are labelled \textit{n.a.} for non applicable.). The \textit{expected energies} have been extracted from \cite{IAEA} and the $\text{Zr}$ absorption edge form \cite{XCOM}.}
\end{table*}
The second, independent, dataset to obtain the energy scale calibration uses measurements where the CMOS is irradiated by an x-ray tube, described in \secref{cmosPaper:sec:expSetUp}. For the spectra obtained with the x-ray tube using only Gau{\ss}ian fits with a local background is not sufficient (\figrefbra{cmosPaper:sec:results:caliometry:fig:xrayMoFit}): The two characteristic peaks of the molybdenum x-ray tube are expected to be located on top of the \textit{bremsstrahlung} spectrum of the tube. For low x-ray energies the camera has negligible calorimetric capabilities as seen in the measurements with the \fe{55} source (previous section, \figrefbra{cmosPaper:sec:results:caliometry:fig:fe55scaledMinusBkg}). Hence, the Neo sCMOS should become efficient for x-rays of the molybdenum x-ray tube somewhere after $\sim\!\!\SI{6}{\kilo\electronvolt}$ -- from that point onwards there should be an increasing number of counts due to \textit{bremsstrahlung} and eventually the molybdenum $K_{\alpha}$ and $K_{\beta}$ peaks at \SI{17.4}{\kilo\electronvolt} and \SI{19.6}{\kilo\electronvolt}, respectively. \Figref{cmosPaper:sec:results:caliometry:fig:xrayMoFit} shows the spectrum, the fit to the spectrum, and the fit's components. An onset of counts is observed at $\sim\!\!\SI{5}{\kilo ADU}$ however, no clear double peak structure is observed. As there is no clear expected functional shape for the bremsstrahlung contribution, we model it as the minimal functional addition ($Brems(\varepsilon)$) needed so $Brems(\varepsilon) + Gau\textit{\ss}(\varepsilon)_{K_{\alpha}} + Gau\textit{\ss}(\varepsilon)_{K_{\beta}}$ fits the data well.
\begin{align}
Brems\left(\varepsilon\right) = p_{0} \cdot \text{exp}\left(-\frac{1}{2}\left(\frac{\varepsilon - p_{1}}{\sigma}\right)^2\right) + p_{2} \cdot \left(1-\text{erf}\Big(p_3\cdot \left(\varepsilon - p_{4}\right)\right)\Big) + p_5 \cdot \varepsilon + p_6 \label{cmosPaper:sec:results:caliometry:eq:xrayBkg}\\
Gau\textit{\ss}(\varepsilon)_{K_{j}} = p_{0}^{j} \cdot \text{exp}\left(-\frac{1}{2}\left(\frac{\varepsilon - \varepsilon_{\text{peak}}^{j}}{\sigma^{j}}\right)^2\right) \label{cmosPaper:sec:results:caliometry:eq:gauss} \quad j = \alpha \lor \beta
\end{align}
In these equations $\varepsilon$ is the cluster charge (or energy deposited in the chip) in \si{ADU}. The parametrisation \eqref{cmosPaper:sec:results:caliometry:eq:xrayBkg} for the bremsstrahlung contribution yields the lowest $\chi^2/N_{\text{dof}}$ of 4.62 for the total fit of $Brems(\varepsilon) + Gau\textit{\ss}(\varepsilon)_{K_{\alpha}} + Gau\textit{\ss}(\varepsilon)_{K_{\beta}}$ to the data, whilst all fit parameters are free. The extracted parameters of the two $K$ lines ($\varepsilon_{\text{peak}}$, $\sigma_{\text{peak}}$) are listed in \tabref{cmosPaper:sec:results:caliometry:tab:allEnergyResults}.\\
Using a $\text{Cu}$ or a $\text{Zr}$ foil to filter the molybdenum x-rays results in the spectra shown in \figref{cmosPaper:sec:results:caliometry:fig:xrayWithFilter}. The absorption edges of those two elements for energies higher than $\sim\!\!\SI{6}{\kilo\electronvolt}$ are at \SI{8.98}{\kilo\electronvolt} ($\text{Cu}$) and at \SI{18}{\kilo\electronvolt} ($\text{Zr}$). The shape of the spectrum recorded with the $\text{Cu}$ filter does not feature a drop which can be identified with an absorption edge -- the edge is thus placed in the energy range where the Neo sCMOS is not sensitive to allow x-ray calorimetry. There is a larger reduction of counts for energies $\lesssim\SI{7}{\kilo ADU}$, relative to the not filtered spectrum and the one with the $\text{Zr}$ filter. After, the number of counts increases again. Starting from low energies, the shape of spectrum with the zirconium filter matches the un-filtered spectrum, until the edge at $\sim\!\!\SI{7.5}{\kilo ADU}$. This drop is identified with the $\text{Zr}$ absorption edge. The \si{ADU} value at which the amplitude reaches the \SI{50}{\%} value between maximal peak height and the floor in the spectrum is taken as its energy position  (\tabref{cmosPaper:sec:results:caliometry:tab:allEnergyResults}). The uncertainty on the edge's position is taken to be half of the \si{ADU} range between the edge's \SI{10}{\%} and \SI{90}{\%} value.\\
\begin{figure*}
\centering
\subfloat[]{
\includegraphics[width=0.45\textwidth,trim=0 0 0 0,clip=true]{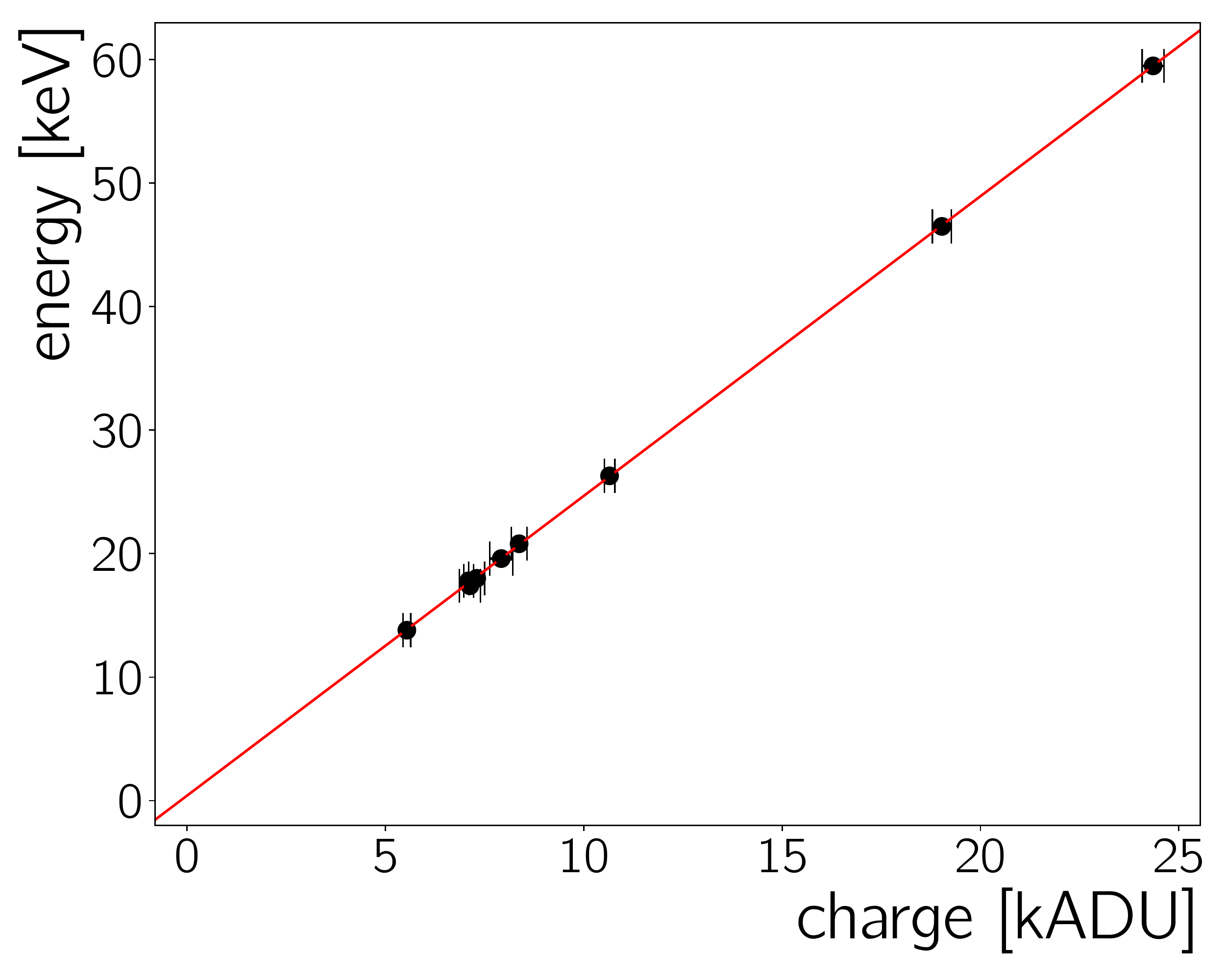}
\label{fig:conversion1}}
\subfloat[]{
\includegraphics[width=0.45\textwidth,trim=0 0 0 0,clip=true]{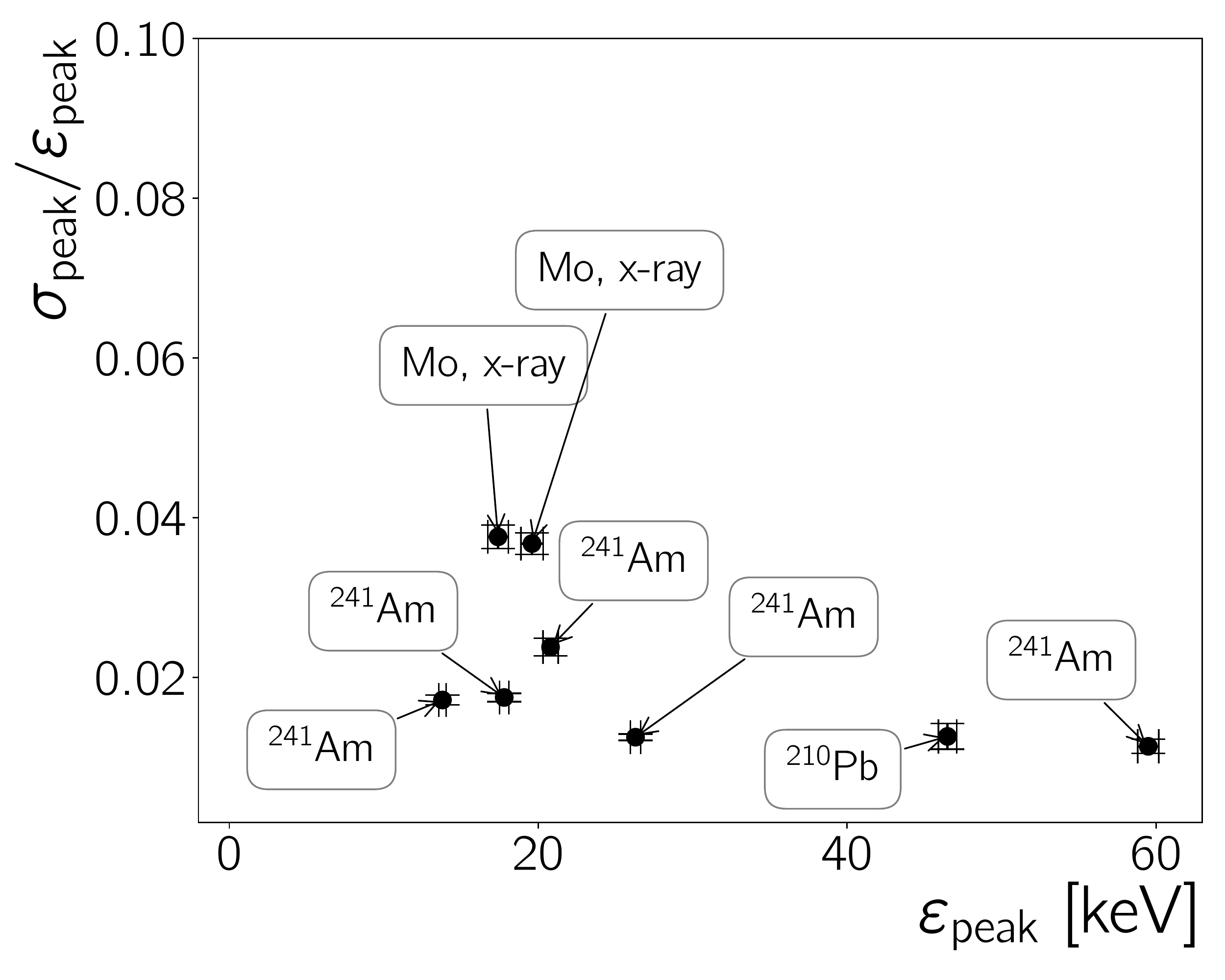}
\label{cmosPaper:sec:results:caliometry:fig:allEnergyResults:energyResolution}}
\caption{\label{cmosPaper:sec:results:caliometry:fig:allEnergyResults}\protect\subref{fig:conversion1} Comparison between the expected and the measured peak and edge energies ($\varepsilon_{\text{peak}}$) in \tabref{cmosPaper:sec:results:caliometry:tab:allEnergyResults}. One $\sigma_{\text{peak}}$ of the peak is used as uncertainty for $\varepsilon_{\text{peak}}$ and the red line through the points is a fit without an additional axis intercept. \protect\subref{cmosPaper:sec:results:caliometry:fig:allEnergyResults:energyResolution} Measured energy resolution ($\sigma_{\text{peak}}$/$\varepsilon_{\text{peak}}$) as function of the measured peak position. The boxes indicate to what spectrum a given peak belongs to.}
\end{figure*}
\Figref{fig:conversion1} displays the measured charge values plotted against their expected energies for all peaks and edges in \tabref{cmosPaper:sec:results:caliometry:tab:allEnergyResults}. Fitting a linear function without an axis intercept to these points yields the conversion factor from \si{\electronvolt} to \si{ADU} (and vice versa) to be \SI{0.405(1)}{ADU\per\electronvolt} (\SI{2.467(7)}{\electronvolt\per ADU}). For this fit $\chi^2/N_{\text{dof}}$ is $1.64$ while using a function with an intercept results in a  $\chi^2/N_{\text{dof}}$ of 0.56, an intercept of -\SI{0.36(9)}{\kilo\electronvolt}, and a slope of \SI{2.434(9)}{\electronvolt\per ADU} (\SI{0.41(1)}{ADU\per\electronvolt}). These values are compatible with the conversion factor mentioned before, although slightly different from an intercept of zero. In the next sections the conversion factor without an intercept is favoured over the conversion with an intercept, since the low $\chi^2/N_{\text{dof}}$ in the latter case indicates over-fitting.\\
The measured conversion factors matches well with the higher of the two gain values discussed before, \textit{i.e.} \SI{2.446}{ADU\per\electronvolt} which is located between the two different fit values. This agreement is taken as another reason to use in the following the conversion factor determined without an intercept, given the supplier does not specify an offset.

\subsubsection{Energy resolution}
\label{cmosPaper:sec:results:caliometry:energyResolution}

In \figref{cmosPaper:sec:results:caliometry:fig:allEnergyResults:energyResolution} the energy resolution is shown as $\sigma_{\text{peak}}$ divided by the peak positions $\varepsilon_{\text{peak}}$ (\tabrefbra{cmosPaper:sec:results:caliometry:tab:allEnergyResults}). As the uncertainty on $\sigma$ the uncertainty of the fit is used while $\sigma_{\text{peak}}$ itself is used as the uncertainty of the peak position $\varepsilon_{\text{peak}}$. The uncertainty on the energy resolution, $\Delta\left(\sigma_{\text{peak}}/\varepsilon_{\text{peak}}\right)$ includes both of these contributions. For the most part the resolution is better than \SI{2}{\%}. Outliers from this trend are the two molybdenum x-ray lines and the $\text{Np}$, $L_{\gamma1}$ line (at $\sim\!\!\SI{20}{\kilo\electronvolt}$). The two x-ray lines are extracted from a more complicated fit with the worst $\chi^2/N_{\text{dof}}$ and the uncertainty on their $\sigma_{\text{peak}}$ values is likely to be underestimated.\\
The energy resolution is determined by several factors. The full containment of all electrons produced during the photon conversion and their subsequent readout will play an important role. Furthermore their can be pixel-to-pixel variations of each pixels' amplifier gain. An increasing trend in cluster size with increasing energy has been observed, as stated in \secref{cmosPaper:sec:anaProcedure:clusterSize}, but the good linearity of the \si{\electronvolt} to \si{ADU} relation suggests that all electrons produced by a photon interaction are read out. The pixel-to-pixel amplifier gain variations for the Neo sCMOS are not known. Typical values are in the few \si{\%} range -- \textit{e.g.} \cite{4215175} shows a variation of about $\lesssim\SI{2}{\%}$ \cite{4215175}. Such a variation would be consistent with the energy resolution shown here. Note that the energy measurements in \figref{cmosPaper:sec:results:caliometry:fig:allEnergyResults:energyResolution} are the result of summing the energy measured in each pixel of a cluster. If the per-pixel amplifier variation would be the dominating factor for the energy resolution, the Neo sCMOS would have variations slightly larger than $\sim\!\!\SI{2}{\%}$.

\subsection{Radiation detection efficiency}
\label{cmosPaper:sec:results:caliometry:sensorEfficiency}

In this section we quantify the minimum detectable radioactivity using the Neo sCMOS, and measure the efficiency of the sensor as a detector for $\gamma$- and x-rays. Both are done using the \am{241} source, since this source has a well suited activity and an energy spectrum with clear peaks.

\subsubsection{Geometric acceptance of the experimental set-up}

\begin{figure*}
\centering
\includegraphics[width=0.45\textwidth, trim=0 0 0 0, clip=true]{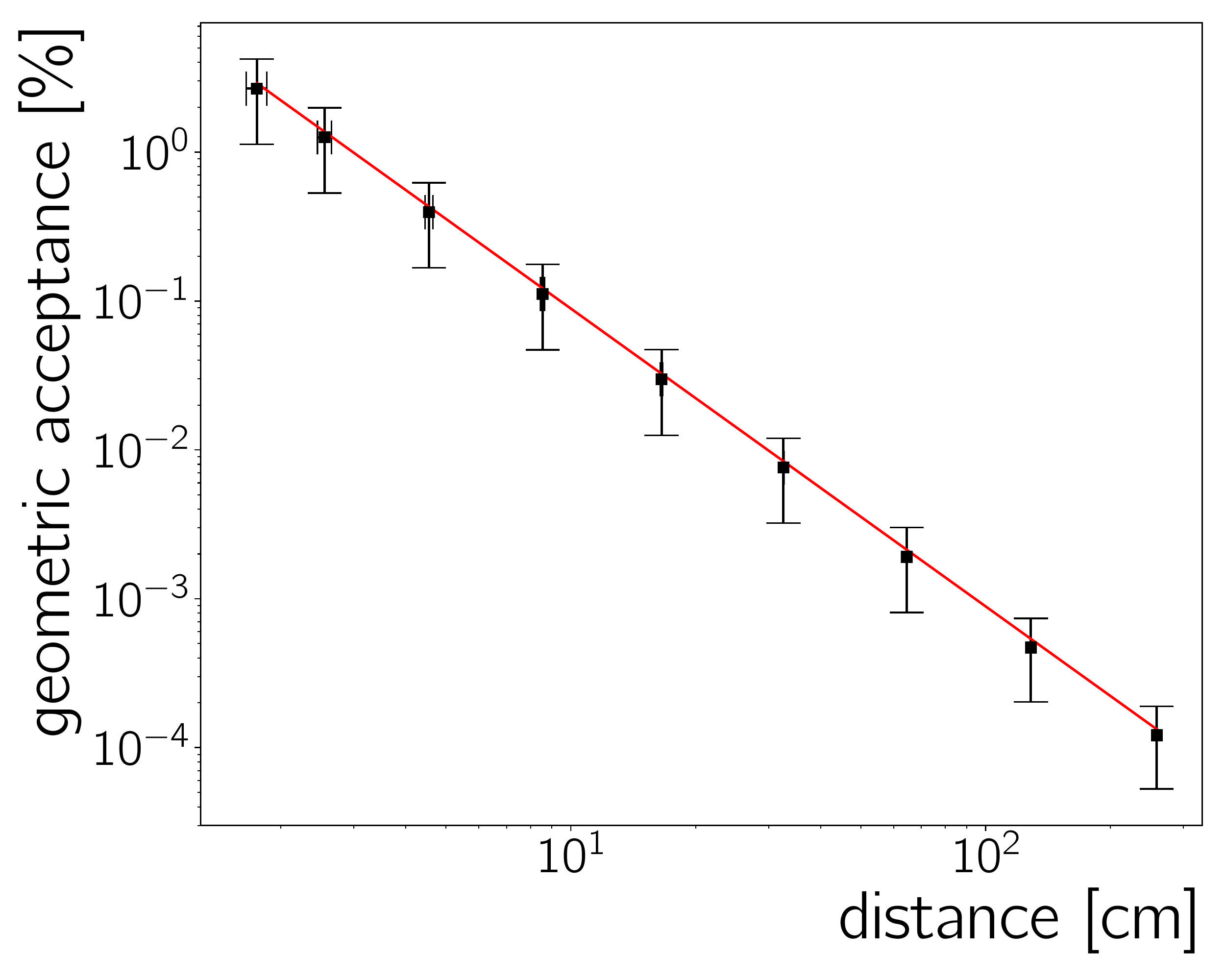}
\caption{\label{fig:ga1} Geometric acceptance of the experimental set-up with the Neo sCMOS: Shown is the analytical function for a point source in \eqnref{geoaccet:eq:1} and values from a toy Monte Carlo resembling the actual source geometry.}
\end{figure*}

The fraction of the \am{241} activity detected by the sensor depends on the source-detector distance. The geometric acceptance ($\epsilon_{\text{G}}$) is calculated assuming the \am{241} emits radiation as a point source, as
\begin{equation}
  \begin{split}
  \epsilon_{\text{G}} & =\frac{A_{\text{spherical}\ \text{cap}}}{A_{\text{sphere}}} \frac{A_{\text{camera}}}{A_{\ocircle\ \text{camera}\ \text{plane}}}\\
    & = \frac{\SI{0.0889}{\centi\meter\squared}}{r^{2}} \quad\quad (r\ \text{in}\ \si{\centi\meter})
  \end{split} 
  \label{geoaccet:eq:1}
\end{equation}
The calculation of this expression exploits the sphere into which the source emits radiation and its intersection with the plane of the camera chip. Therefore, $A_{\text{sphere}}$ is surface area of that sphere, $A_{\text{spherical}\ \text{cap}}$ is the surface area of the base of the spherical cone covering the camera chip, $A_{\ocircle\ \text{camera}\ \text{plane}}$ is the corresponding surface area of an otherwise similar cone with a flat base, $A_{\text{camera}}$ is the surface area of the camera and $r$ is the camera to source distance. \Figref{fig:ga1} compares the analytical estimate of Eqn. \eqref{geoaccet:eq:1} with a toy Monte Carlo simulation, using the actual source geometry, showing good consistency.
\begin{figure*}
\centering
\subfloat[]{
\includegraphics[width=0.45\textwidth, trim=0 0 0 0, clip=true]{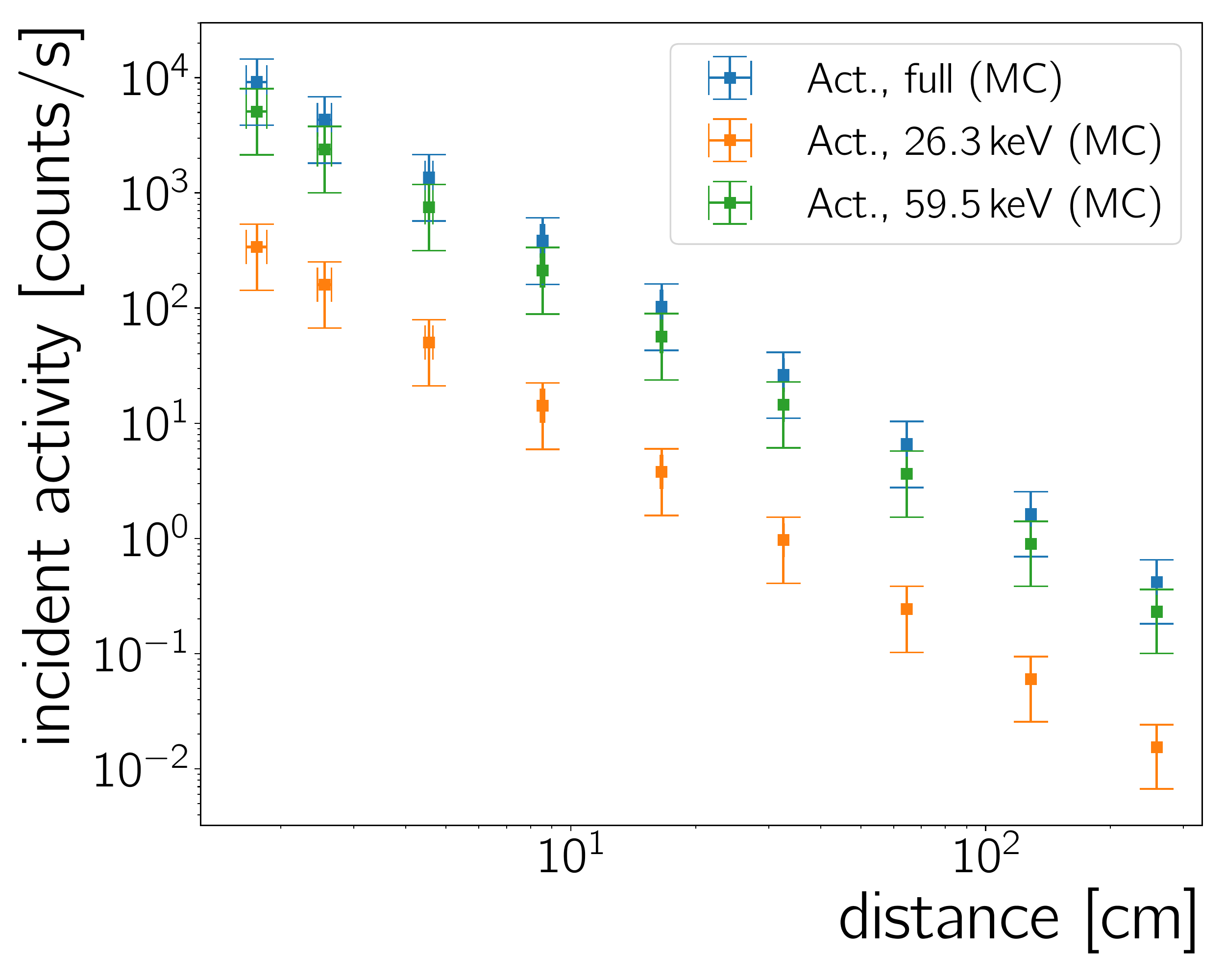}
\label{fig:minact1MC}}
\subfloat[]{
\includegraphics[width=0.45\textwidth, trim=0 0 0 0, clip=true]{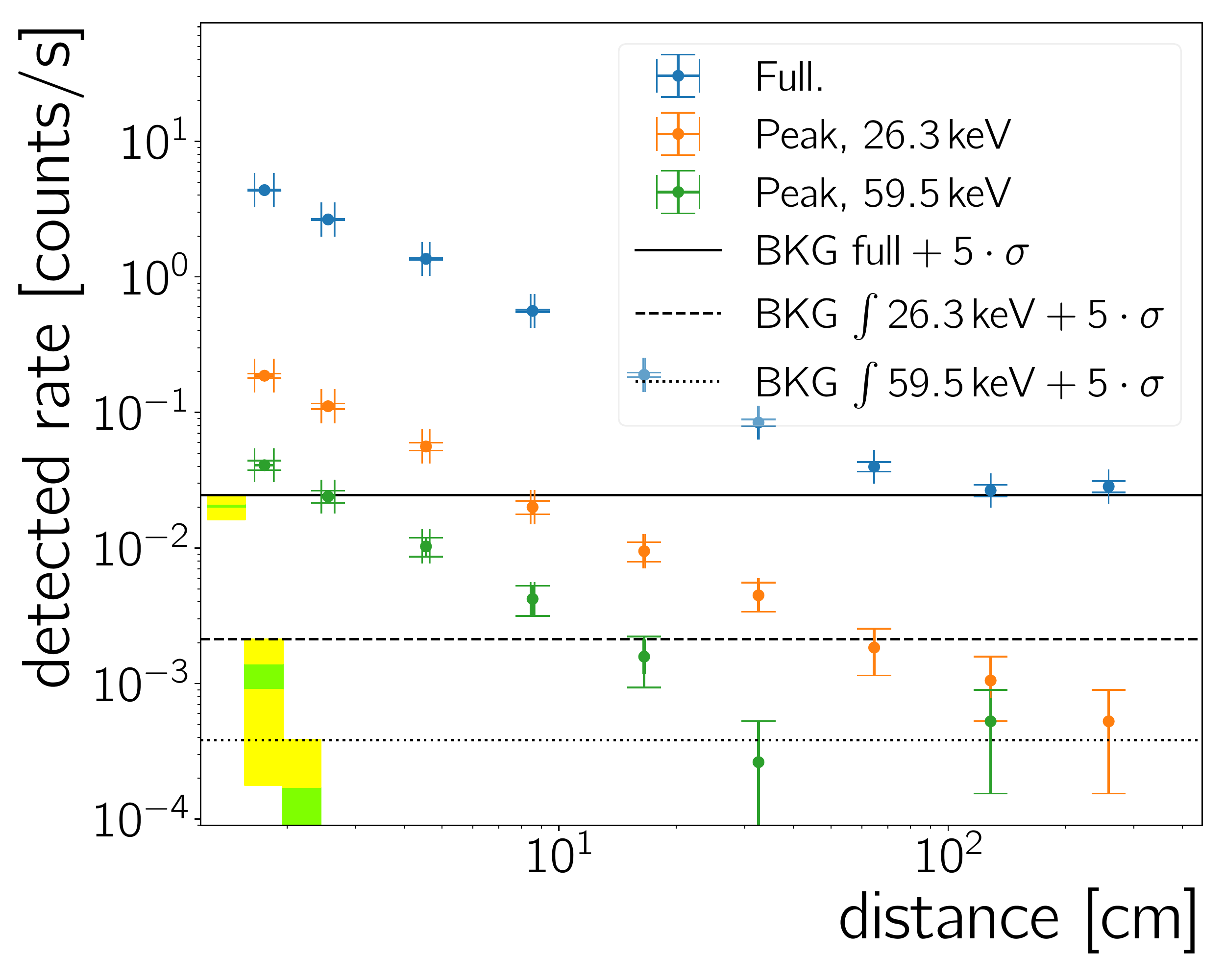}
\label{fig:minact1}}\\
\subfloat[]{
\includegraphics[width=0.45\textwidth, trim=0 0 0 0, clip=true]{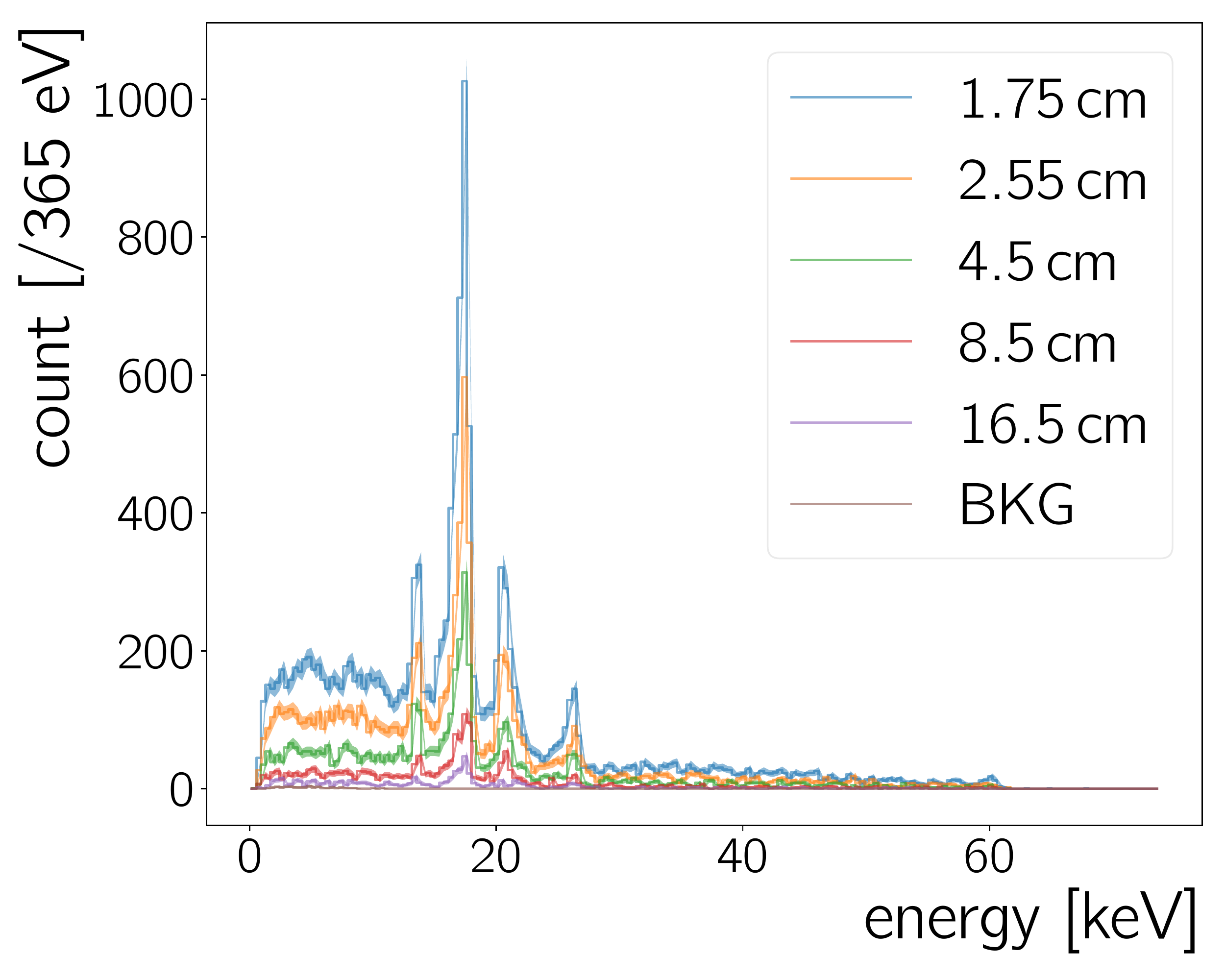}
\label{fig:detact1Near}}
\subfloat[]{
\includegraphics[width=0.45\textwidth, trim=0 0 0 0, clip=true]{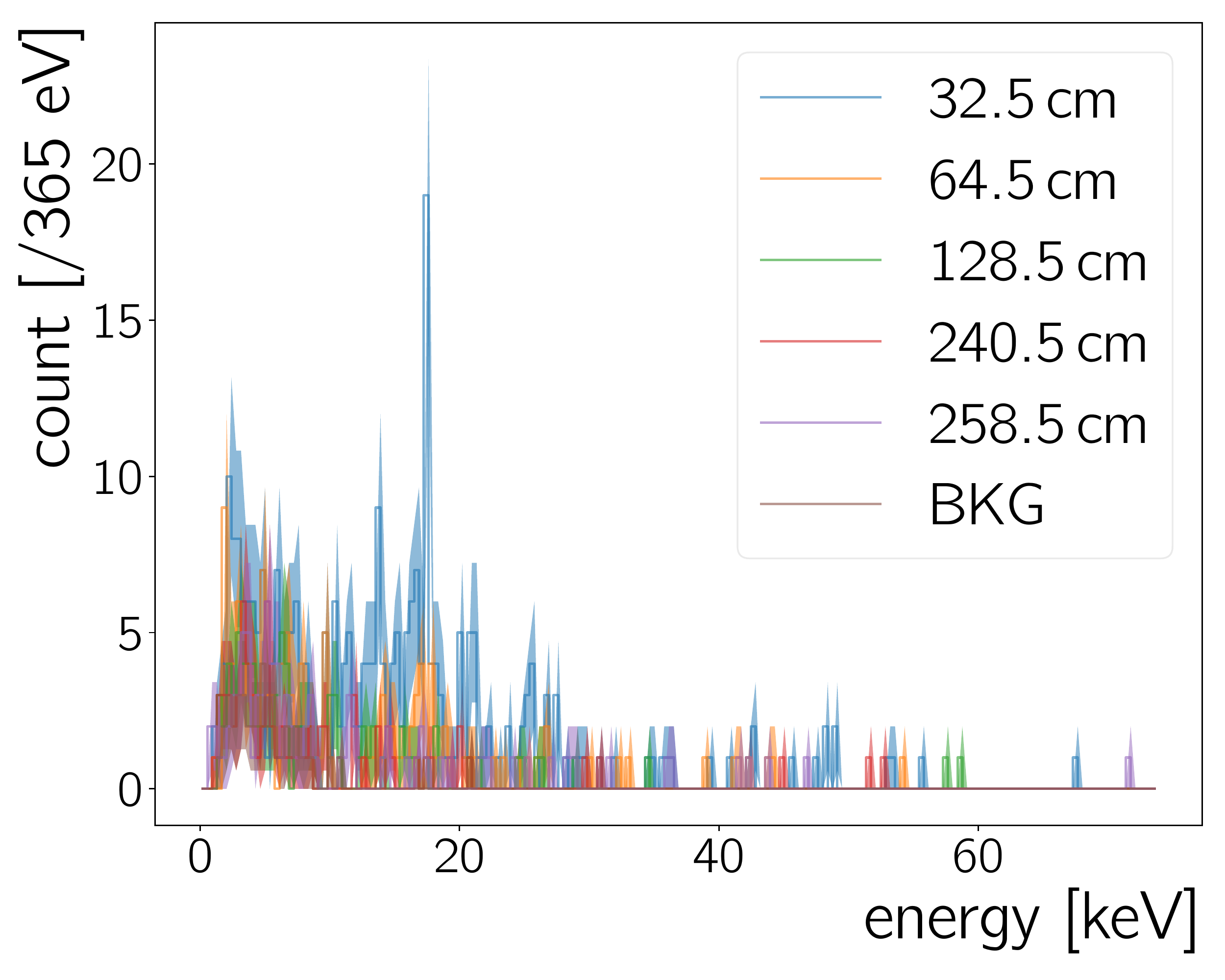}
\label{fig:detact1Far}}
\caption{\label{cmosPaper:sec:results:caliometry:sensorEfficiency:fig:acceptance}\protect\subref{fig:minact1MC} Simulated incident activity on the camera chip for different camera-to-source distances for the full source activity, and for the fractional activities corresponding to the \SI{26.3}{\kilo\electronvolt} and \SI{59.5}{\kilo\electronvolt} $\gamma$-lines. \protect\subref{fig:minact1} The measured rates are calculated using \eqnref{eq:2} and integrating over the full \am{241} spectra in Figures \protect\subref{fig:detact1Near} and \protect\subref{fig:detact1Far}, or integrating only over the energy region of the \SI{26.3}{\kilo\electronvolt} and \SI{59.5}{\kilo\electronvolt} peaks in the same spectra. The measured background rate is established by integrating over the full spectrum recorded in absence of any source and in $\pm5\sigma_{\text{peak}}$ windows around the mentioned peak energies. The shaded areas in the left half of the plot below the lines of the respective background rates show the five and 1.28 standard deviation ($\sigma$) regions around the background rate in yellow and green, respectively. \protect\subref{fig:detact1Near} and \protect\subref{fig:detact1Far} Measured \am{241} spectra at different distances with a live time of \SI{3800}{\second}. The data has to pass the cluster size $>\SI{2}{pixels}$ cut and is not background subtracted. The background data sample in both plots is the same as it has been shown previously (\textit{e.g.} \figrefbra{cmosPaper:sec:anaProcedure:clusterAna:fig:clustercuts:bkg:sizecut}) and is normalised to the same live-time.}
\end{figure*}

\subsubsection{Minimum detectable radioactivity}
\label{cmosPaper:sec:results:subsec:minIncActivity}

Data is acquired for \SI{3800}{\second} each at distances from \SI{1.8}{\centi\meter} up to \SI{258.6}{\centi\meter} between source and sensor.\footnote{From now on we omit ``between camera chip and radioactive source'' when referring to the distance between these two components.} The detected spectra, the inferred incident activity using the geometric acceptance from Eqn. \eqref{geoaccet:eq:1} and the known source activity are shown in \figref{cmosPaper:sec:results:caliometry:sensorEfficiency:fig:acceptance}. At a distance of $\SI{258.6 \pm 0.1}{\centi\metre}$ this activity is as low as $\SI{0.42(24)}{\becquerel}$. It is also of interest to determine the incident activity considering only certain $\gamma$-ray energies emitted by the \am{241} source, as opposed to the full source activity. These are obtained by multiplying the incident activity for all \am{241} decays by the respective peak yields which are $\SI{2.40(4)}{\%}$ and $\SI{35.9(4)}{\%}$ for the \SI{26.3}{\kilo\electronvolt} and $\SI{59.5}{\kilo\electronvolt}$ energy peaks, respectively \cite{IAEA}. These incident activities can be seen in \figref{fig:minact1MC}. To obtain the uncertainty on the incident activities in this figure we propagate the distance uncertainty through the calculations in \eqnref{geoaccet:eq:1}, as well as the uncertainty of the initial activity of the \am{241} source and the uncertainty of the emission probabilities. In the case of the toy Monte Carlo simulation, the statistical uncertainty on the counts obtained is used.\\
\label{cmosPaper:sec:results:subsec:minDetActivity}
In order to estimate the minimum detectable radioactivity and eventually the Neo sCMOS' detection efficiency, the detected rates in \figref{fig:minact1} are determined as
\begin{equation}
  rate = \frac{\int{spectrum}\ \text{d}time}{time}\left[\frac{\text{counts}}{\si{\second}}\right] \quad .
  \label{eq:2}
\end{equation}
The measured rate without any source present is calculated as well using the spectrum in \figref{cmosPaper:sec:anaProcedure:clusterAna:fig:clustercuts:bkg:sizecut}. During \SI{30400}{\second} of data taking 619 clusters with a size $>\SI{2}{pixels}$ are recorded, yielding a background rate of \SI{20.4(8)}{\milli\hertz}. In windows corresponding to $5\sigma$ of the integration limits around the \SI{26.3}{\kilo\electronvolt} and the \SI{59.5}{\kilo\electronvolt} $\gamma$-peak energies (\tabrefbra{cmosPaper:sec:results:caliometry:tab:allEnergyResults}) the background rate measured with no source present is \SI{1.1(2)}{\milli\hertz} (\SI{0.10(6)}{\milli\hertz}) for the \SI{26.3}{\kilo\electronvolt} (\SI{59.5}{\kilo\electronvolt}) peak region. The exposure time per frame and the number of considered frames are known with great certainty, hence the statistical uncertainty on the cluster count is the only contribution to the uncertainty of the quoted rates. \Figref{fig:minact1} shows the background rates added with five times their uncertainty as yellow shaded regions and as horizontal lines. The green shaded regions correspond to \SI{90}{\%} Confidence Level (CL), \textit{i.e.} $1.28\sigma$, around the measured background rate, where we use the background uncertainty as standard deviation, \textit{i.e.} $\sigma$.\footnote{Thus count rates higher than \SI{21.41}{\milli\hertz}, \SI{1.4}{\milli\hertz} and \SI{0.17}{\milli\hertz} for the full background rate, the background rate in the \SI{26.3}{\kilo\electronvolt} peak window, and the background rate in the \SI{59.5}{\kilo\electronvolt} peak window, respectively, are larger than  the background at \SI{90}{\%} CL.} Overall, our results are for the most case the same when when using $5\sigma$ or a \SI{90}{\%} CL criterion to differentiate between the measured background rate and the source rate. \Tabref{cmosPaper:sec:results:subsec:minIncActivity:tab:results} shows all result for both cases, in the following we refer to the results obtained with the $5\sigma$ criterion. The detected rate measured for a distance of \SI{128.5}{\centi\meter} is the first to be compatible with the background rate added with five standard deviations (\SI{27(3)}{\milli\hertz}), while the rate measured at \SI{64.5}{\centi\meter} distance (\SI{40(3)}{\milli\hertz}) is significantly larger than the background rate. The incident source activity at \SI{64.5}{\centi\meter} and \SI{128.5}{\centi\meter} distance is \SI{7(4)}{\becquerel} and \SI{2(1)}{\becquerel}, respectively.\\
\begin{table}
\centering
\begin{tabular}{c||c|c||c|c}
                                 & \multicolumn{2}{c||}{incident activity}           & \multicolumn{2}{c}{measured rate}    \\  
                                 &     $5\cdot\sigma$      &   \SI{90}{\%} CL        &     $5\cdot\sigma$        &   \SI{90}{\%} CL \\ \hline \hline
full spectrum                    & \SI{7(4)}{\becquerel}   &  \SI{2(1)}{\becquerel}  & \SI{40(3)}{\milli\hertz}  & \SI{27(3)}{\milli\hertz} \\
\SI{26.3}{\kilo\electronvolt}-peak & \SI{1.0(6)}{\becquerel} & \SI{1.0(6)}{\becquerel} & \SI{4(1)}{\milli\hertz}   & \SI{4(1)}{\milli\hertz}  \\
\SI{59.5}{\kilo\electronvolt}-peak & \SI{57(33)}{\becquerel} & \SI{57(33)}{\becquerel} & \SI{1.5(1)}{\milli\hertz} & \SI{1.5(1)}{\milli\hertz} \\
\end{tabular}
\caption{\label{cmosPaper:sec:results:subsec:minIncActivity:tab:results}The lowest incident activities and the corresponding measured rates obtained with the Neo sCMOS. See \secref{cmosPaper:sec:results:subsec:minIncActivity} and \figref{fig:minact1}. Values for two cases -- either a $5\sigma$ or a \SI{90}{\%} Confidence Level (CL) condition -- are quoted.} 
\end{table}
Performing the same analysis for the two most intense \am{241} $\gamma$-lines allows to establish whether this limit can be improved taking calorimetric information into account. For the \SI{26.3}{\kilo\electronvolt} line a falling trend for the detected rate is observed over the full distance range (\figrefbra{fig:minact1}). Comparing to the rate measured when no source is present the point at \SI{64.5}{\centi\meter} is already compatible within five standard deviations. For \SI{32.5}{\centi\meter} distance the detected activity is \SI{4(1)}{\milli\hertz}, corresponding to an incident activity of \SI{1.0(6)}{\becquerel}. In the case of the \SI{59.5}{\kilo\electronvolt} $\gamma$ line, the \SI{32.5}{\centi\meter} point is already compatible with the background rate for this energy window within five standard deviations. The detected rate at \SI{16.5}{\centi\meter} of \SI{1.5(1)}{\milli\hertz}, corresponding to an incident activity of \SI{57(33)}{\becquerel}, is thus the minimal detected rate, different from the background rate in this energy window.

\subsubsection{Detection efficiency of the Neo sCMOS}
\label{cmosPaper:sec:results:subsec:intrinsicAndAbsoluteEfficency}

\begin{figure*}
\centering
\subfloat[]{
\includegraphics[width=0.45\textwidth,trim=0 0 0 0,clip=true]{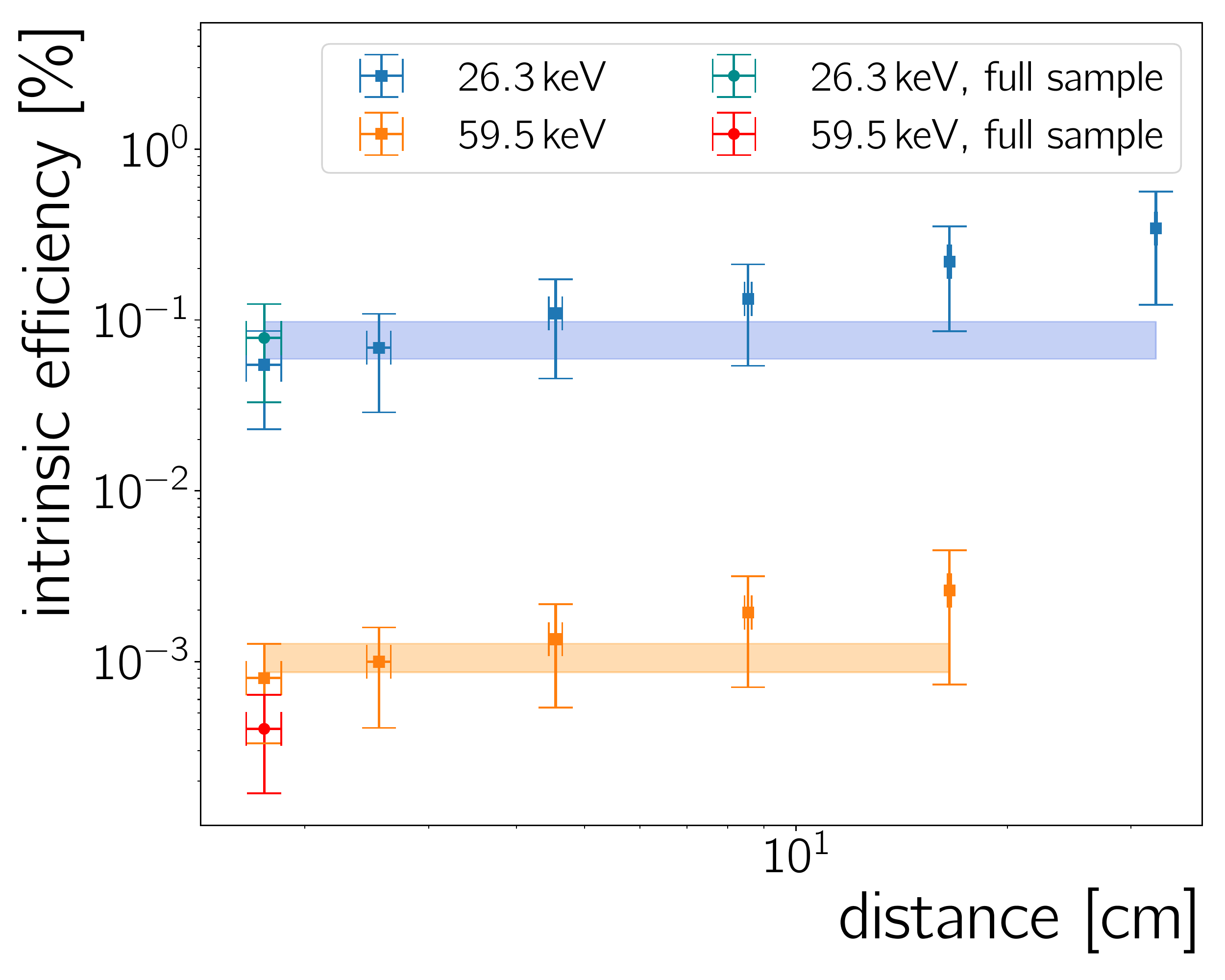}
\label{fig:eff_int_peaks}}
\subfloat[]{
\includegraphics[width=0.45\textwidth,trim=0 0 0 0,clip=true]{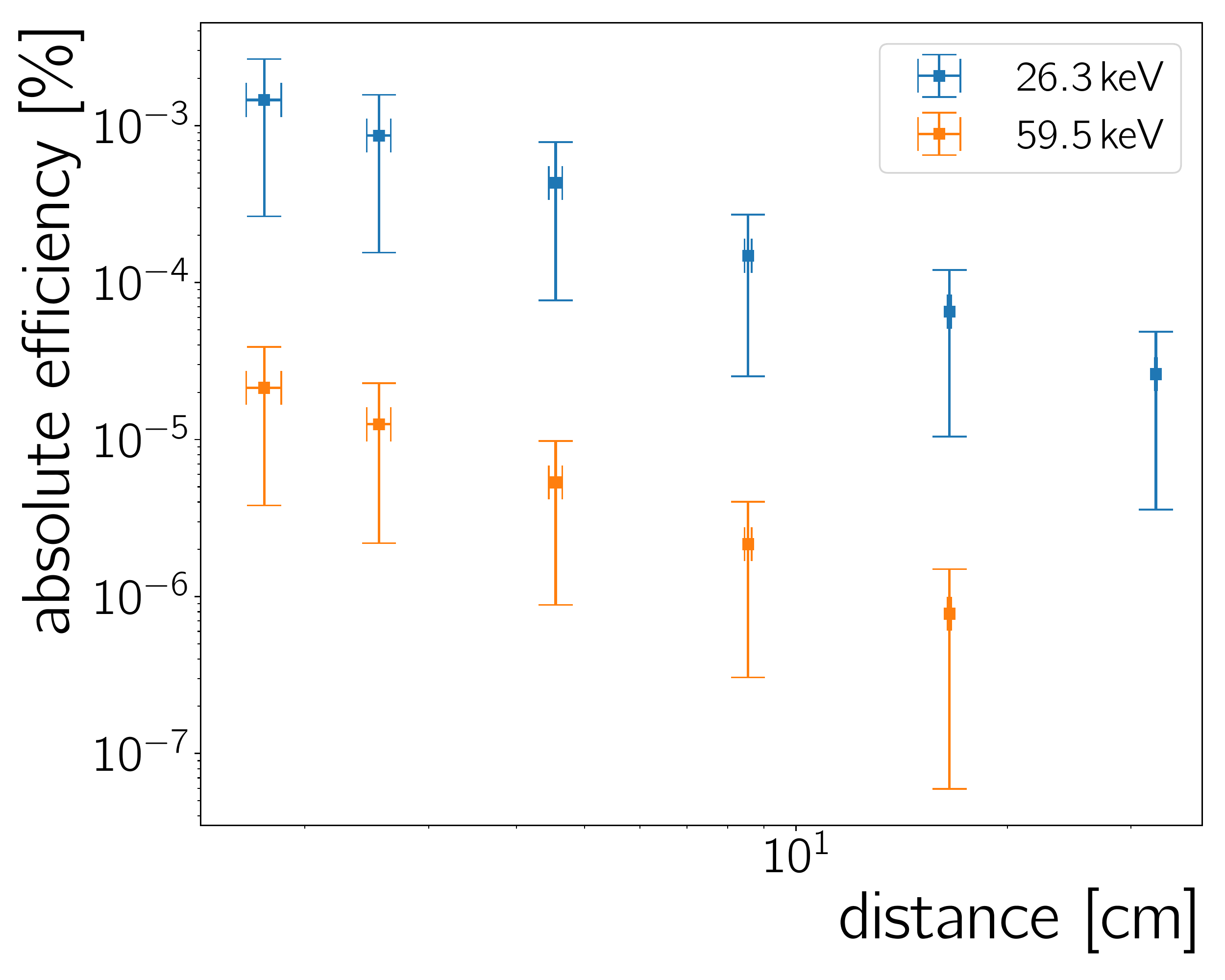}
\label{fig:eff_abs_peaks}}
\caption{\label{fig:eff_peaks}\protect\subref{fig:eff_int_peaks} Intrinsic and \protect\subref{fig:eff_abs_peaks} absolute efficiency for the \SI{26.3}{\kilo\electronvolt} and the \SI{59.5}{\kilo\electronvolt} peaks. The bands in \protect\subref{fig:eff_int_peaks} correspond to the result of fitting a constant to the data as discussed in \secref{cmosPaper:sec:results:subsec:intrinsicAndAbsoluteEfficency} with its error bars. The two \textit{full sample} points are extracted from data in \figref{cmosPaper:sec:results:caliometry:fig:am241scaledMinusBkg} in the same manner as all the other points are extracted from the data in \figref{cmosPaper:sec:results:caliometry:sensorEfficiency:fig:acceptance}.}
\end{figure*}

A comparison between \figref{fig:minact1MC} and \figref{fig:minact1} allows to calculate both the intrinsic and absolute efficiency of the detector. These are defined as
\begin{align}
  \epsilon_{\text{intrinsic}} &= \frac{\text{number\ of\ particles\ recorded}}{\text{number\ of\ particles\ incident\ on\ the\ detector}}
  \label{cmosPaper:sec:results:subsec:intrinsicAndAbsoluteEfficency:eq:int}\\
  \epsilon_{\text{absolute}} &= \frac{\text{number\ of\ particles\ recorded}}{\text{number\ of\ particles\ emitted\ by\ source}} = \epsilon_{\text{intrinsic}} \cdot \epsilon_{\text{G}} \quad,
  \label{cmosPaper:sec:results:subsec:intrinsicAndAbsoluteEfficency:eq:abs}
\end{align}
where $\epsilon_{\text{G}}$ is the geometric acceptance, which is given by \eqnref{geoaccet:eq:1}. The ratio of the recorded rate (\figrefbra{fig:minact1}) and the incident activity (\figref{fig:minact1MC}) yield the intrinsic efficiency of the Neo sCMOS camera for $\gamma$-rays of an \am{241} source. This is shown in \figref{fig:eff_int_peaks} for the \SI{26.3}{\kilo\electronvolt} and \SI{59.5}{\kilo\electronvolt} peaks as a function of the camera-to-source distance. For completeness, the absolute efficiencies for these two $\gamma$-rays are shown in \figref{fig:eff_abs_peaks}.\\
By definition $\epsilon_{\text{intrinsic}}$ has to be independent of the distance. The results of a constant fit to the data are shown in \figref{fig:eff_int_peaks}. However, a weak distance dependence of $\epsilon_{\text{intrinsic}}$ is observed; this is likely caused by a slight discrepancy between the calculated geometric acceptance and the actual source geometry. However, all points are within error-bars compatible with a constant as expected from \eqnref{cmosPaper:sec:results:subsec:intrinsicAndAbsoluteEfficency:eq:int}. The absolute efficiency's dependence on the distance in \figref{fig:eff_abs_peaks} is expected, due its dependence on the geometric efficiency (Eqn. \eqref{cmosPaper:sec:results:subsec:intrinsicAndAbsoluteEfficency:eq:abs}). The corresponding intrinsic efficiencies for the \SI{26.3}{\kilo\electronvolt} and the \SI{59.5}{\kilo\electronvolt} peaks are \SI{0.08(2)}{\%} and \SI{0.0011(2)}{\%}, respectively. Thus, the efficiency to detect a \SI{26.3}{\kilo\electronvolt} $\gamma$-ray is about a factor of 80 higher than the efficiency to detect a $\gamma$-ray with an energy of \SI{59.5}{\kilo\electronvolt}. To check this ratio for consistency the material composition and thickness of the sensor would need to be known, which is however not the case. \Figref{fig:sim4} in the next section shows photon absorption efficiencies calculated from the attenuation coefficients in \cite{NIST}. Depending on the CMOS thickness, the difference in photo-absorption for \SI{59.5}{\kilo\electronvolt} and \SI{26.3}{\kilo\electronvolt} can easily reach a factor 80. However, the efficiency at the \SI{59.5}{\kilo\electronvolt} $\gamma$-line appears lower than expected from photo-absorption cross sections in $\text{Si}$.

\subsubsection{CMOS sensor thickness}
\label{cmosPaper:sec:results:subsec:toyMC}

\begin{figure*}
\centering
\subfloat[]{
\includegraphics[width=0.45\textwidth,trim=0 0 0 0,clip=true]{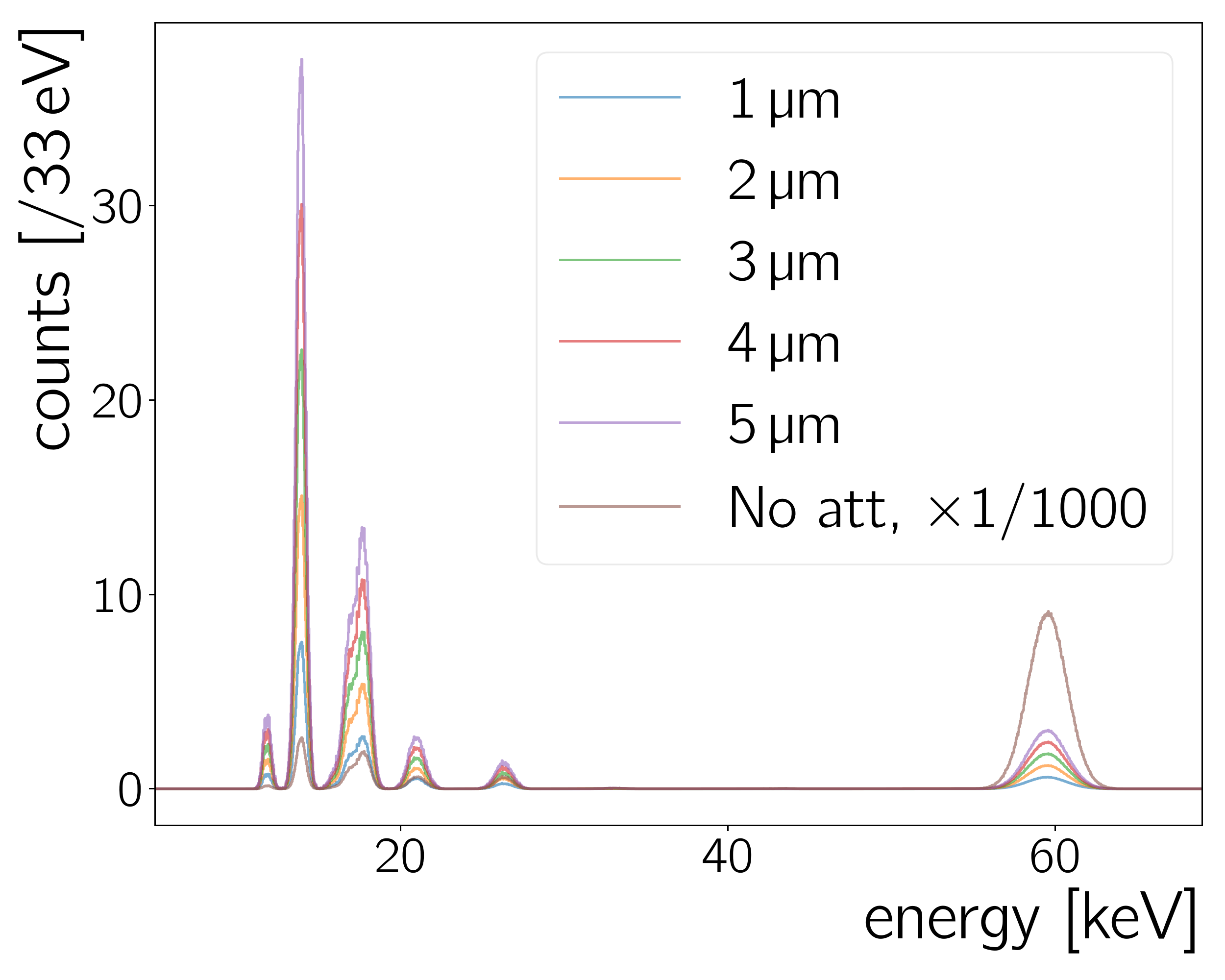}
\label{fig:sim1}}
\subfloat[]{
\includegraphics[width=0.45\textwidth,trim=0 0 0 0,clip=true]{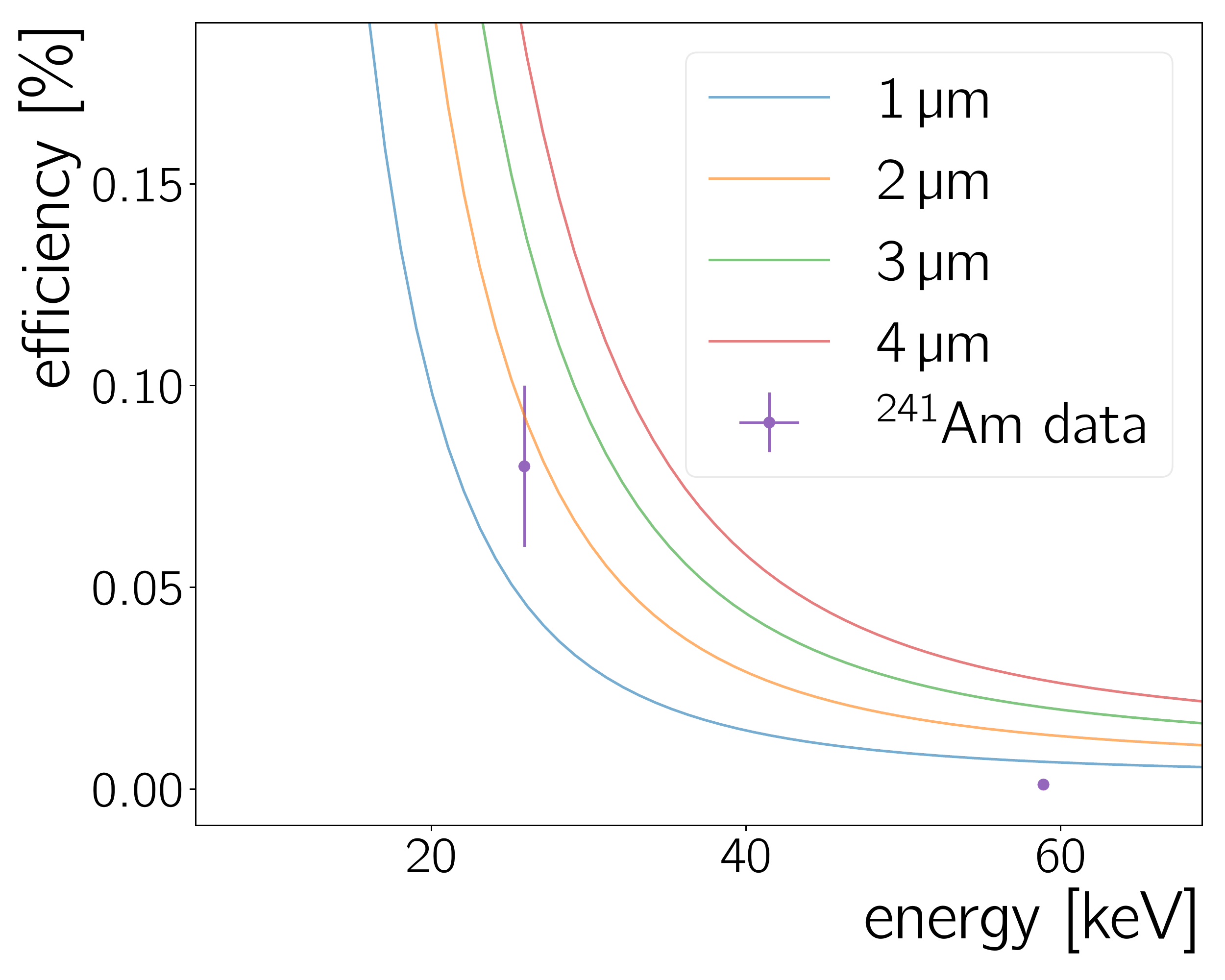}
\label{fig:sim4}}
\caption{\label{cmosPaper:sec:results:subsec:toyMC:simResults}\protect\subref{fig:sim1}\am{241} $\gamma$-ray and x-ray simulation for $1\,\text{M}$ decays. The spectrum with no material effects taken into account (\textit{No att.}) is scaled by a factor of $1/1000$ to improve the readability of the plot. The other spectra show the photons which are absorbed by a silicon layer of given thickness. \protect\subref{fig:sim4} Photon absorption efficiencies in silicon for different silicon layer thickness as well as the measured intrinsic efficiency of the Neo sCMOS at the two \am{241} $\gamma$-ray energies.}
\end{figure*}

Based on the efficiencies determined in the previous section one can estimate the thickness of the CMOS chip used in the Neo sCMOS camera. A toy Monte Carlo (MC) simulation of the \am{241} spectrum is used together with the photon attenuation coefficient from \cite{NIST} to calculate, for each photon, the energy-dependent photon absorption probability in silicon. This step is repeated for different silicon thicknesses. For each thickness we create a spectrum of the photons absorbed in $\text{Si}$ -- these photons are the ones which would be measured by a chip as in the Neo sCMOS. The ratio of the unattenuated \am{241} toy MC spectrum divided by the spectra of absorbed photons is constructed. Finally, these ratios are compared to the ratio of the measured count rate at the $\gamma$ peak energies divided by the incident activity for the respective $\gamma$ line.\\
For the toy MC, all possible $\gamma$-rays and x-rays of \am{241} decays as listed in \cite{ND1999} are taken into account. These are used to create a probability density function for photons emitted during \am{241} decays. Each $\gamma$- and x-ray line is represented by a Gau{\ss}ian peak where the amplitude is proportional to the yield per decay. The $\varepsilon_{\text{peak}}$ is set to the respective x-ray or $\gamma$-ray energy and the $\sigma_{\text{peak}}$ is set to \SI{2}{\%} of the peak energy to approximate the measurement for the Neo sCMOS in \secref{cmosPaper:sec:results:caliometry:energyResolution}. Creating $1\,\text{M}$ decays from this probability density function results in the \textit{No att.} cluster charge spectrum in \figref{fig:sim1}. This unattenuated spectrum does not take material effects into account.\\
\Figref{fig:sim1} shows also spectra of photons absorbed in silicon layers with a thickness from \SI{1}{\micro\meter} to \SI{5}{\micro\meter}. Toy MC spectra for absorbed photons and the measured spectrum (\textit{e.g.} \figrefbra{fig:conversion2}) are similar. The ordering of the different peaks' height is the same except for the fact that the largest measured peak is at \SI{18}{\kilo\electronvolt}, while the peak with the most counts in the simulation is the peak at \SI{14}{\kilo\electronvolt}. In \secref{cmosPaper:sec:results:caliometry:energyMeasurement} the detection efficiency is observed to drop at lower energies -- most notably making it impossible to detect a peak below $\sim\!\!\SI{10}{\kilo\electronvolt}$, \textit{cf}. \figref{cmosPaper:sec:results:caliometry:fig:fe55scaledMinusBkg} and \figref{cmosPaper:sec:results:caliometry:fig:xrayData}. The efficiency turn-on responsible for this behaviour may also affect the peak height of the \SI{14}{\kilo\electronvolt} peak and lead to the non-observation of the small peak visible at $\sim\!\!\SI{11}{\kilo\electronvolt}$ in \figref{fig:sim1}.\\
\Figref{fig:sim4} shows photon absorption efficiency curves calculated from the coefficients in \cite{NIST} for \SI{1}{\micro\meter}, \SI{2}{\micro\meter} and \SI{4}{\micro\meter} silicon layer thickness. Furthermore, the intrinsic efficiency values calculated in \secref{cmosPaper:sec:results:subsec:intrinsicAndAbsoluteEfficency} are shown (\figrefbra{fig:eff_int_peaks}). The two points follow roughly the trend expected by photon absorption in silicon, although the efficiency measured for the \SI{59.5}{\kilo\electronvolt} peak is lower than expected from any absorption efficiency curve. The actual peak height of any $\gamma$-line is given by the absorption probability as well as by the capability of the active material to contain the full energy deposit. The latter is not contained in the toy MC simulation and this may explain that the \SI{59.5}{\kilo\electronvolt} does not fit with the displayed curves.The \SI{26.3}{\kilo\electronvolt} favours a silicon layer thickness of \SI{2}{\micro\meter}. This scale of a few \si{\micro\meter} is on the same order of magnitude as the pixel dimension and seems reasonable for commercial CMOS chips, with a typical silicon-dioxide layer thickness of less than \SI{10}{\micro\meter} \cite{SERVOLI2010,PEREZ2016}. Since the silicon thickness is not the only contribution to the Neo sCMOS intrinsic efficiency, the estimate in this section can only be seen as a lower limit to the actual sensor thickness.

\subsection{Geant4 simulations of the Neo sCMOS detector}
\label{cmosPaper:sec:genatsimulations}

To extend the limited knowledge of the sensor geometry beyond the estimations based on the toy MC simulations, a Geant4 MC simulation study is carried out to estimate the impact of the window, the micro lens array and sensor thickness on the $\gamma$- and x-ray absorption.\\
Geant4 \cite{AGOSTINELLI,ALLISON2016186,1610988} version 10.5 patch 1 is used to simulate the primary particles and the production of the resulting secondary particles, for the tracking of all particles through the detector geometry, and to assess the energy deposition in the sensitive detector parts. This work employs physics lists which follow mainly the \textit{Shielding physics list} of the above mentioned Geant4 version with one physics modification; namely $\mathrm{G4EmStandardPhysics\_option4}$. $\mathrm{G4EmStandardPhysics\_option4}$ is used instead of $\mathrm{G4EmStandardPhysics}$, because the former is more accurate for low-energy electromagnetic interactions.

\subsubsection{Detector geometry and simulated particles}
\label{cmosPaper:sec:genatsimulations:subsec:gemoetry}

The sensor is modelled as a silicon CMOS layer, behind a glass layer to represent the entrance window, behind an acrylic layer to represent the micro lens array. For this simulation, the material of the window is chosen to be $\mathrm{SiO_{2}}$. The camera specifications indicate an organic material for the micro lens. Thus, for the material of this volume element, acrylic $\mathrm{C_{5}H_{8}O_{2}}$ is chosen. For the thicknesses of each volume element, values in the range of $\mathcal{O}(1-\SI{1000}{\micro\meter})$ are used. This study varies the thicknesses of the silicon and glass window. As discussed in \secref{cmosPaper:sec:results:caliometry}, \pb{210} decays either by a combined $\gamma$- and $\beta$-decay (\SI{84(3)}{\%}), where the $\gamma$-ray has an energy of \SI{46.5}{\kilo\electronvolt} and an average $\beta$ energy of \SI{4.2}{\kilo\electronvolt}, or by a pure $\beta$-decay (\SI{16(3)}{\%}) with an average $\beta$ energy of \SI{16.2}{\kilo\electronvolt}. When simulating electrons of this energy impinging on the detector with a \SI{200}{\micro\meter} glass window and \SI{5}{\micro\meter} of $\text{Si}$ thickness, there are zero hits on the active region out of $10^{6}$ simulated events. Since the energy of the other $\beta$ is even lower, the signature \pb{210} $\gamma$-ray of \SI{46.5}{\kilo\electronvolt} is the main focus of this simulation work whereas $\beta$-particles are not included.\\
The simulated particle source is located \SI{14.06}{\milli\meter} from the outermost layer of the surface, and fires mono-energetic $\gamma$- and x-rays directly at the detector.\\
To construct the energy variable used to compare with data, the simulated ionisation energy deposition in the \text{Si} layer is recorded for each incident particle, and then smeared according to the energy resolution function
\begin{equation}
\frac{\sigma_{\text{peak}}}{\varepsilon_{\text{peak}}} = p_0 + \frac{p_1}{\sqrt{\varepsilon_{\text{fit}}}}
\label{cmosPaper:sec:results:caliometry:energyResolution:eq:fit}
\end{equation} 
fitted to the data in \figref{cmosPaper:sec:results:caliometry:fig:allEnergyResults:energyResolution}.

\subsubsection{Analysis of the simulated spectra}
\label{sim:analysis}

\begin{figure}
\centering
\subfloat[]{\label{fig:sim_vs_data_100}\includegraphics[width=0.45\columnwidth]{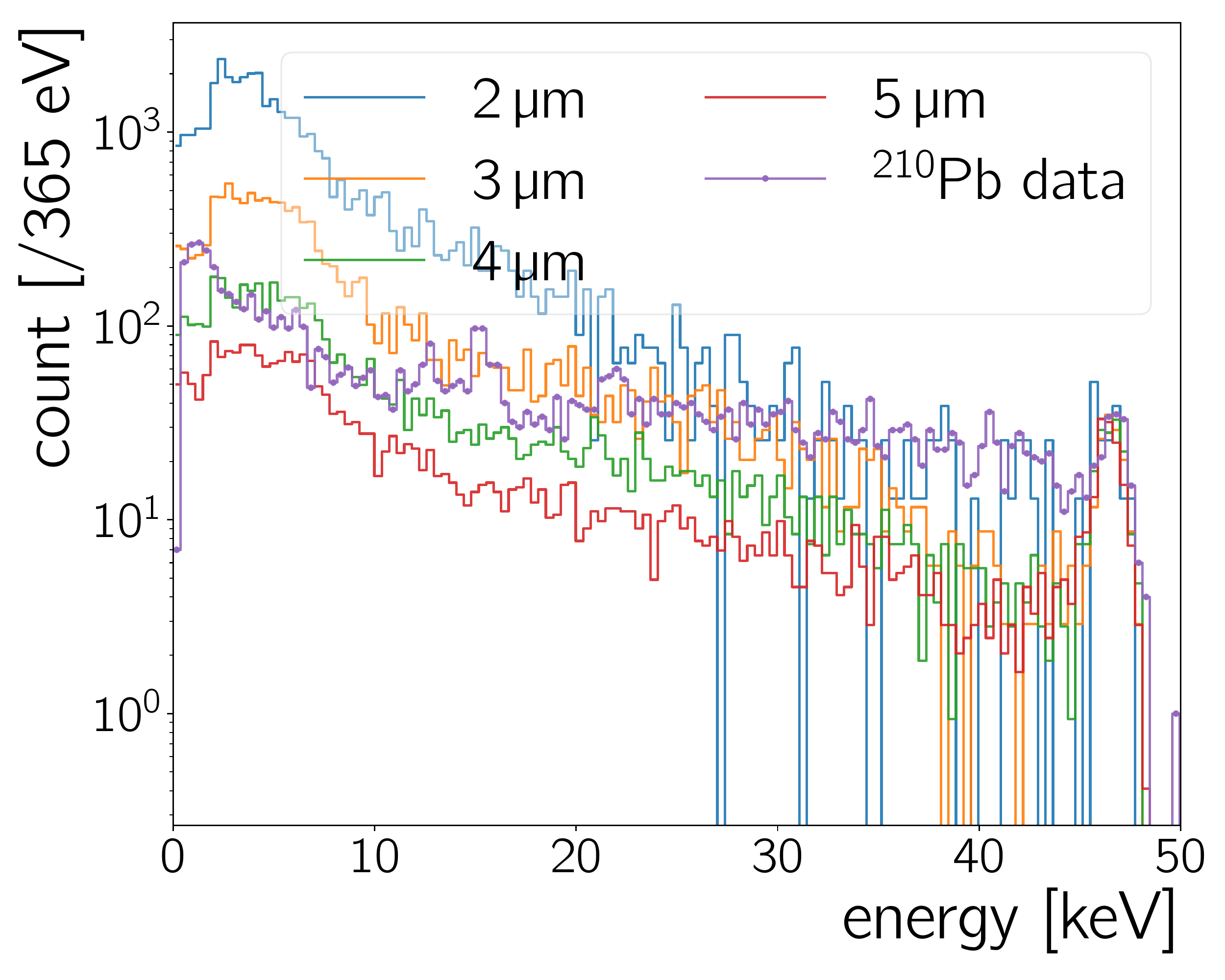}}
\subfloat[]{\label{fig:sim_vs_data_500}\includegraphics[width=0.45\columnwidth]{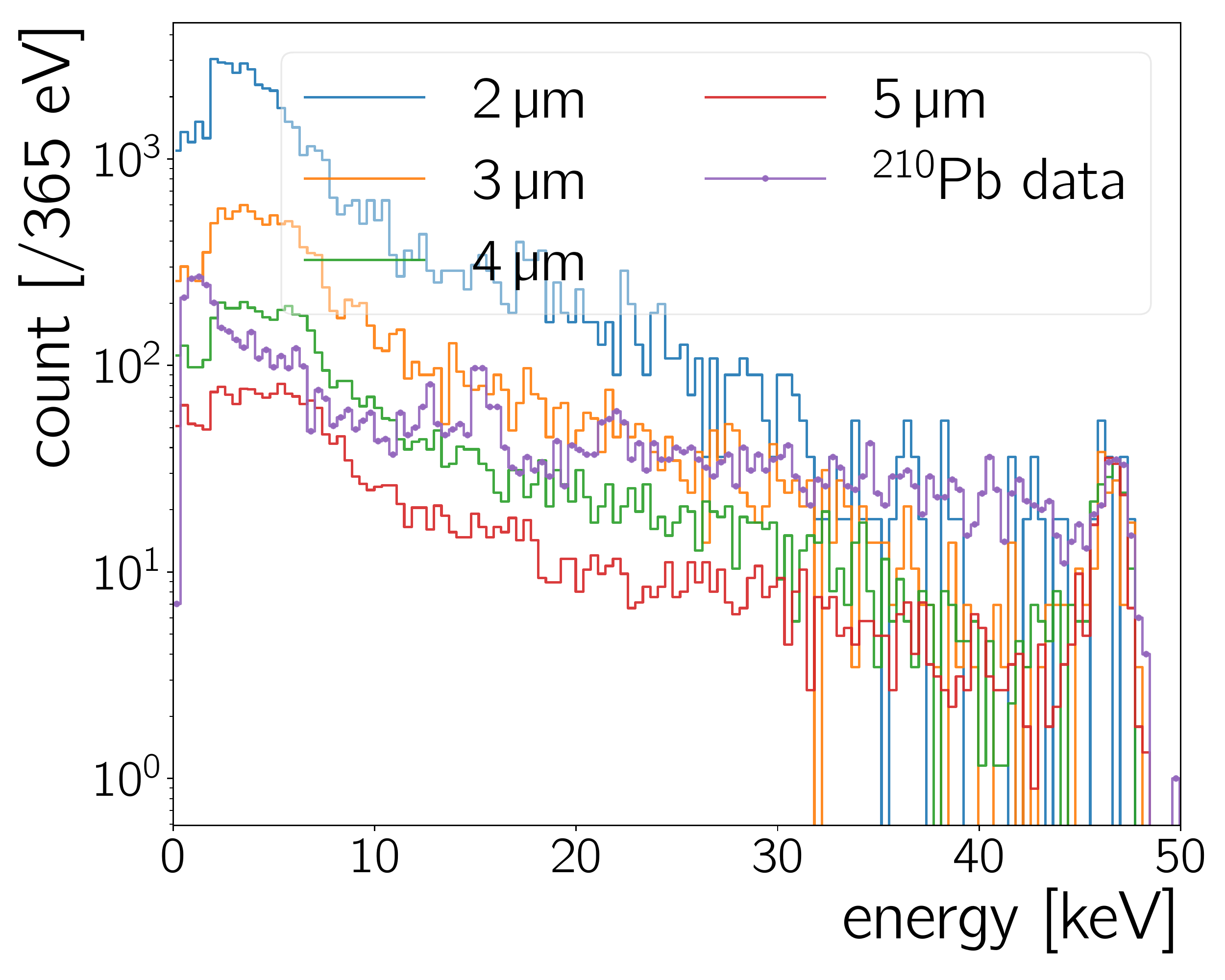}}\\
\subfloat[]{\label{fig:sim_vs_data_1000}\includegraphics[width=0.45\columnwidth]{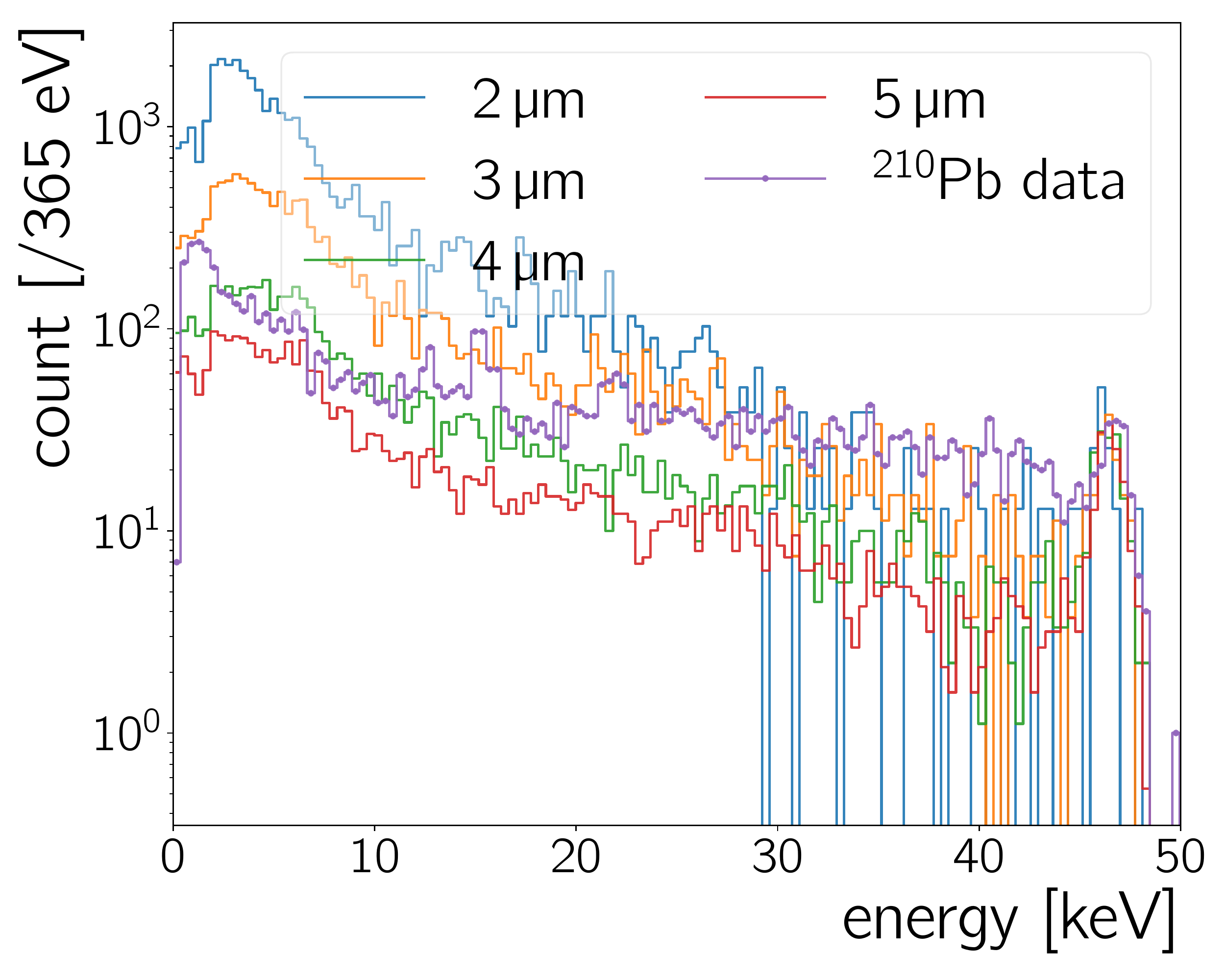}}
\caption{\label{fig:sim_vs_data}Simulated spectra of \SI{46.5}{\kilo\electronvolt} $\gamma$-rays shot towards the detector with varying active silicon layer thicknesses along with constant micro lens thickness of \SI{4}{\micro\meter} with following glass window thicknesses: \protect\subref{fig:sim_vs_data_100} \SI{200}{\micro\meter}, \protect\subref{fig:sim_vs_data_500} \SI{1000}{\micro\meter} and \protect\subref{fig:sim_vs_data_1000} \SI{2000}{\micro\meter}. Photopeak of each individual simulated spectrum is normalised to the photopeak of experimental data ($^{210}\text{Pb}$). The $\gamma$-rays' origin from a point source centred on the detector over the window, \textit{cf}. \secref{cmosPaper:sec:genatsimulations:subsec:gemoetry}.}
\end{figure}

A set of different thicknesses for the silicon layer and the glass window are simulated for comparison with the data. The list of thicknesses simulated ranges from $2-\SI{5}{\micro\meter}$ for the Si, and $200-\SI{2000}{\micro\meter}$ for the window, listed in \tabref{cmosPaper:sim:results:numbers}. \Figref{fig:sim_vs_data} shows example spectra of the energy deposition in the Si layer for \SI{200}{\micro\meter} (\figrefbra{fig:sim_vs_data_100}), \SI{1000}{\micro\meter} (\figrefbra{fig:sim_vs_data_500}) and \SI{2000}{\micro\meter} (\figrefbra{fig:sim_vs_data_1000}) glass thickness and a constant \SI{4}{\micro\meter} micro lens thickness. Colours represent different thicknesses of the silicon layer varying in between \SI{2}{\micro\meter} and \SI{5}{\micro\meter} while the thickness of other volume elements is kept constant. There are two visible differences between each simulated template:
\begin{itemize}
  \item The total number of events registered in the silicon layer increases with Si thickness which is caused by more particles being absorbed by a thicker Si layer.
  \item The number of events in the photo-peak and \textit{Compton continuum} increases and decreases, respectively, as the silicon thickness increases. This is because the fraction of events for which the incident photons' energy is fully contained in the \text{Si} rises as the thickness increases.
\end{itemize}
The information regarding the number of events in the photo-peak can be used to characterise the thickness of the silicon layer in conjunction with the amount of events in the Compton continuum. The ratio $\eta$ is used as a variable to compare the simulated templates with data:
\begin{equation}\label{eq:sim:photopeak}
  \eta = \frac{\int\limits_{\varepsilon_{\text{peak}} - \sigma_{\text{peak}}}^{\varepsilon_{\text{peak}} + \sigma_{\text{peak}}}N_{\text{events}}\ \text{d}N}{\int\limits_{0}^{\varepsilon_{\text{peak}} - \sigma_{\text{peak}}}N_{\text{events}}\ \text{d}N}
\end{equation}
where $\varepsilon_{\text{peak}}$ stands again for the energy of the photo-peak, $\sigma_{\text{peak}}$ for the peak's standard deviation and $N_{\text{events}}$ for the number of events. The exact value of the photo-peak and the energy resolution of the detector need to be known, to precisely define the integration limits in \eqref{eq:sim:photopeak}. We take $\varepsilon_{\text{peak}}$ to be \SI{46.539(1)}{\kilo\electronvolt} \cite{IAEA} and calculate $\sigma_{\text{peak}}$ from the energy resolution for this energy using \eqnref{cmosPaper:sec:results:caliometry:energyResolution:eq:fit} to be \SI{0.595}{\kilo\electronvolt}. $\eta$-values for different silicon layer and window thickness calculated from the simulation results can be found in \tabref{cmosPaper:sim:results:numbers}.

\subsubsection{Comparison of simulation with the experimental data}
\label{sim:comparison}

The thickness of the Neo sCMOS active layer, assuming Si as material, is estimated by comparing the spectra in \figref{fig:sim_vs_data} and the photopeak ratio $\eta$ with the data.
\begin{table*}
\centering
\begin{tabular}{c|c|c|c}
Silicon thickness $\ $ & $\ $ Glass window $\ $ & $\ $ Full absorption $\ $ & $\ $ $\eta$ $\left[10^{-2}\right]$ \\
  $\left[\si{\micro\meter}\right]$  & $\ $ thickness $\left[\si{\micro\meter}\right]$  & efficiency $\left[10^{-3}\:\%\right]$ &      \\ \hline
\multirow{3}{*}{2}&200& 0.13 $\pm$ 0.04 & 0.41 $\pm$ 0.11 \\
    &1000& 0.09 $\pm$ 0.03 & 0.29 $\pm$ 0.10 \\
    &2000& 0.10 $\pm$ 0.03 & 0.35 $\pm$ 0.01 \\
    \hline
\multirow{3}{*}{3}&200& 0.49 $\pm$ 0.07 & 1.1 $\pm$ 0.2 \\
    &1000& 0.40 $\pm$ 0.06 & 1.0 $\pm$ 0.1 \\
    &2000& 0.41 $\pm$ 0.06 & 1.1 $\pm$ 0.2 \\
    \hline
\multirow{3}{*}{4}&200& 1.7 $\pm$ 0.1 & 3.3 $\pm$ 0.3 \\
    &1000& 1.4 $\pm$ 0.1 & 2.8 $\pm$ 0.2 \\
    &2000& 1.4 $\pm$ 0.1 & 3.1 $\pm$ 0.3 \\
    \hline
\multirow{3}{*}{5}&200& 3.4 $\pm$ 0.2 & 5.7 $\pm$ 0.3 \\
    &1000& 3.3 $\pm$ 0.2 & 5.8 $\pm$ 0.3 \\
    &2000& 2.6 $\pm$ 0.2 & 5.0 $\pm$ 0.3 \\
\end{tabular}
\caption{\label{cmosPaper:sim:results:numbers}Corresponding full absorption efficiencies and $\eta$-values of 46.5 keV $\gamma$-rays for silicon layers of different thicknesses.}
\end{table*}
When this variable is calculated for the data (black line in \figrefbra{fig:sim_vs_data}), it is found to be $\eta_{\mathrm{data}} = (2.7 \pm 0.2) \times 10^{-2}$. Comparing $\eta_{\mathrm{data}}$ with the values calculated from simulated templates (\textit{cf.} \tabrefbra{cmosPaper:sim:results:numbers}), it can be seen clearly that a combination of \SI{4}{\micro\meter} silicon layer and \SI{1000}{\micro\meter} glass window comes closest to this value with $\eta = (2.8 \pm 0.2) \times 10^{-2}$.\\
As another check, one can investigate the spectral shape of simulation with respect to experimental data. Above \SI{30}{\kilo\electronvolt}, the spectral shape of the data are closest to the spectrum of the \SI{2}{\micro\meter} Si thickness simulation. Below that, the shape matches of the measured spectrum lies between the spectrum simulated for a \SI{3}{\micro\meter} and \SI{4}{\micro\meter} thick Si layer. All templates except for the \SI{4}{\micro\meter} and \SI{5}{\micro\meter} ones show a discontinuity before the photo-peak as the energies of the particles increase, which is an expected result due to thinner silicon layers being less efficient in stopping $\gamma$-rays and containing the particles energy.

\subsubsection{Closing remarks}

The close match of $\eta$-values and the spectral shape between the data and the Geant4 simulation suggest a silicon thickness between \SI{2}{\micro\meter} and \SI{4}{\micro\meter} for a window of \SI{1000}{\micro\meter}, where the comparison to $\eta_{\mathrm{data}}$ places the thickness at the high end of this range. \Tabref{cmosPaper:sim:results:numbers} shows that for this geometry only $(1.4\pm0.1)\times10^{-3}\,\%$ of the $\gamma$-rays of \pb{210} are fully absorbed in the silicon layer, \textit{i.e.} contribute to the photo-peak. In \secref{cmosPaper:sec:results:subsec:intrinsicAndAbsoluteEfficency} the intrinsic efficiencies for the \SI{26.3}{\kilo\electronvolt} and \SI{59.5}{\kilo\electronvolt} \am{241}-peak are determined to be $(80\pm20)\times10^{-3}\,\%$ and $(1.1\pm0.2)\times10^{-3}\,\%$, respectively. The efficiency for the \SI{59.5}{\kilo\electronvolt} $\gamma$-line is compatible with the value simulated here for \SI{46.5}{\kilo\electronvolt} $\gamma$-rays. Given the fact that the measured value is at a higher energy than the \SI{46.54}{\kilo\electronvolt} of the simulated $\gamma$-rays, the Geant4 simulation most likely underestimates the Neo sCMOS detection efficiency at \SI{46.5}{\kilo\electronvolt} slightly or the actual $\text{Si}$ thickness is larger than \SI{4}{\micro\meter}. This is, because the efficiency to detect \SI{46.54}{\kilo\electronvolt} photons is expected to be larger than the one for higher energy photons.\\
Both the Geant4 simulation of \pb{210} $\gamma$-rays and the toy MC simulation of \am{241} photons place the silicon layer thickness in the range between \SI{2}{\micro\meter} and \SI{4}{\micro\meter}, \textit{cf.} \secref{cmosPaper:sec:results:subsec:toyMC}, \figref{fig:eff_abs_peaks}. While the full Geant4 simulation gives more parameters to compare between the data and the simulation, the toy MC simulation is significantly faster as it runs in an instant. The agreement between the two gives confidence to use the less sophisticated toy MC simulation in instances where a full fledged Geant4 simulation is not easily accessible.

\section{Summary and discussion}
\label{cmosPaper:sec:summary}

An \textsc{Oxford Instruments} Neo 5.5 scientific CMOS \cite{neosmos} camera is examined as a detector for photons in the x-ray and low $\gamma$-ray energy regime. This camera is designed to image photons of optical wavelengths. The analysis (\secrefbra{cmosPaper:sec:anaProcedure}) of camera images identifies clusters (\secrefbra{cmosPaper:sec:anaProcedure:clusterFinding}) -- contiguous pixels with high charge values (unit: \si{ADU}) --  which are due to energy deposits of radiation impinging on the camera chip.\\
Requiring the cluster size to be larger than 2 pixel allows to sufficiently reject most of the clusters due to background radiation. We note a trend towards larger cluster sizes for increasing photon energy (\secrefbra{cmosPaper:sec:anaProcedure:clusterSize}).\\ 
The relation of the cluster charge in \si{ADU} to \si{\electronvolt} is measured to be \SI{2.467(7)}{\electronvolt\per ADU} or \SI{0.405(1)}{ADU\per\electronvolt} (\figrefbra{fig:conversion1}, \secrefbra{cmosPaper:sec:results:caliometry:energyMeasurement}). This relation is linear without an offset in the energy range from \SI{13.8}{\kilo\electronvolt} to \SI{59.5}{\kilo\electronvolt}.  It is in reasonable agreement with \SI{2.446}{\electronvolt\per ADU}, which is the expected value based on the supplier specified gain value of \SI{0.67}{electron\per ADU} and the energy of \SI{3.65}{\electronvolt} \cite{kolanoski2016teilchendetektoren} needed to create an electron-hole-pair in $\text{Si}$. The energy resolution is for the most part slightly better than \SI{2}{\%} (\secrefbra{cmosPaper:sec:results:caliometry:energyResolution}). We have not been able to measure any peaks below $\sim\!\!\SI{10}{\kilo\electronvolt}$, which is most likely due to the photon absorption in glass for energies $\leq\SI{10}{\kilo\electronvolt}$  (\textit{cf}. the comments on \fe{55} in \secrefbra{cmosPaper:sec:results:caliometry}).\\
The rate of background events detected without the presence of any source is measured to be \SI{20.4(8)}{\milli\hertz} (\secrefbra{cmosPaper:sec:results:caliometry:sensorEfficiency}). By increasing the distance between the camera and an \am{241} source we reduce the activity incident on the camera chip. The lowest detectable rate measured $5\sigma$ above background is \SI{40(3)}{\milli\hertz}, which corresponds to an incident activity of \SI{7(4)}{\becquerel}. Taking calorimetric information into account and integrating only around the \SI{26.3}{\kilo\electronvolt} and \SI{59.5}{\kilo\electronvolt} $\gamma$-peaks, the minimal detectable rate is \SI{4(1)}{\milli\hertz} and \SI{1.5(1)}{\milli\hertz}, respectively, which corresponds to incident activities of \SI{1.0(6)}{\becquerel} and \SI{57(33)}{\becquerel} (\secrefbra{cmosPaper:sec:results:subsec:minDetActivity}), respectively.\\
Comparing the measured rates and incident source activity allows to determine the intrinsic efficiency of the Neo sCMOS camera at the two \am{241} $\gamma$-lines. They are found to be \SI{0.08(2)}{\%} and \SI{0.0011(2)}{\%} for $\SI{26.3}{\kilo\electronvolt}$ and the $\SI{59.5}{\kilo\electronvolt}$ peaks, respectively (\secrefbra{cmosPaper:sec:results:subsec:intrinsicAndAbsoluteEfficency}). The efficiency drop from the lower to the higher energy follows roughly the drop of the photon-absorption efficiency in silicon of a few \si{\micro\meter} thickness as function of energy. The absolute value of the efficiency for \SI{59.5}{\kilo\electronvolt} is lower than expected for a silicon sensor with a few \si{\micro\meter} thickness, when only photo-absorption coefficients are considered (\secrefbra{cmosPaper:sec:results:subsec:toyMC}). Geant4 simulations (\secrefbra{cmosPaper:sec:genatsimulations}) show that for thin silicon layers the fraction of an absorbed photon's energy contained in the silicon decreases with increasing energy of the photon. Thus, lowering the fraction of counts in the photopeak and the detection efficiency measured at the energy of the photopeak. The increasing volume, in which a photon deposes its energy, for increasing photon energy fits the measurement of larger cluster sizes for high energy $\gamma$-lines (\secrefbra{cmosPaper:sec:anaProcedure:clusterSize}). The Geant4 simulations as well as the toy MC simulations indicate a thickness in the order of \SI{2}{\micro\meter} to \SI{4}{\micro\meter} for Neo sCMOS silicon layer.

\subsection{Lead detection capabilities assuming Lead-210 as a trace isotope}
\label{cmosPaper:sec:summary:subsec:pb210}

The ratio of the \textit{radioactive isotopes in \pb{}} to \textit{stable \pb{} isotopes} is required to estimate the capability of the Neo sCMOS to detect lead in drinking water. In \cite{HYang2016} the \pb{210} to stable lead ratio is used to monitor the lead intake of plants. They measure \SI{96(9)}{\becquerel\per\milli\gram} for \pb{210}$/$\pb{} in rainwater in London, corresponding to a ratio of \SI{34}{ppb} of \pb{210}$/$\pb{} (molar mass of lead: \SI{207.2}{\gram\per mol}, half-life of lead: \SI{22.3}{yr}). However, the origin of the stable lead and \pb{210} are not necessarily the same. In \cite{Orrell:2015zca} several \pb{} samples of different age -- from ancient lead to recently produced lead -- are analysed for their \pb{210}, $^{232}\text{Th}$ and $^{238}\text{U}$ contents. They find \pb{210}$/$\pb{} ratios from \SI{0.09}{\becquerel\per\kilo\gram} to \SI{68.7}{\becquerel\per\kilo\gram} (\SI{3.9E-8}{ppb} to \SI{2.4E-5}{ppb}) in their lead samples.\\
In the following we will estimate the incident activity on the Neo sCMOS, produced by \SI{10}{ppb} lead in water. Doing so the \pb{210}$/$\pb{} ratio of \cite{HYang2016} will be used, since it results from a measurement of \pb{} in water. However, the results in \cite{Orrell:2015zca} have to be kept in mind as caveat.\\
With the \pb{210}$/$\pb{} ratio of \SI{34}{ppb}, the WHO limit of \SI{10}{ppb} \cite{WHO2011} of lead in drinking water translates to a fraction of $3.4\times10^{-16}$ parts \pb{210} to one part of water. One gram of water contains at this ratio $9.9\times10^{5}$ \pb{210} atoms, which initially decay at a rate of \SI{0.96}{\milli\becquerel} ($R_{^{210}\text{Pb}}^{\text{decay}}$). A sample of water containing lead could be placed on the camera's window -- in \SI{1.75}{\centi\meter} distance from the silicon chip. Considering \SI{1}{\gram} of water as point source, the incident rate ($R_{^{210}\text{Pb}}^{\text{incident}}$) is only \SI{0.001}{\milli\hertz} after taking the geometric acceptance in \eqnref{geoaccet:eq:1} into account as well as the fact that the $\gamma$-yield for the \SI{45}{\kilo\electronvolt} $\gamma$ is only \SI{4}{\%}: 
\begin{align}
\begin{split}
R_{^{210}\text{Pb}}^{\text{incident}} &= R_{^{210}\text{Pb}}^{\text{decay}} \cdot \epsilon_{\text{G}}\left(\SI{1.8}{\centi\meter}\right) \cdot \left\{^{210}\text{Pb}\ \gamma\ \text{yield}\right\} \\
  &= \SI{0.96E-3}{\becquerel} \cdot \SI{3(1)}{\%} \cdot \SI{4}{\%}\\
  &= (1.1\pm0.4)\times 10^{-6}\;\si{\becquerel}
\end{split}
\label{cmosPaper:sec:summary:subsec:pb210eq:mrate}
\end{align}
$R_{^{210}\text{Pb}}^{\text{incident}}$ has to be compared to the measured value of the incident \am{241} source activity of \SI{1.0(6)}{\becquerel} and \SI{57(33)}{\becquerel} at the \SI{26.3}{\kilo\electronvolt} and \SI{59.5}{\kilo\electronvolt} lines, respectively (\secrefbra{cmosPaper:sec:results:subsec:intrinsicAndAbsoluteEfficency}). With $\sim\!\!\SI{46}{\kilo\electronvolt}$ the \pb{210} $\gamma$-line is located between these two energies. Thus, the sensitivity of the Neo sCMOS is a factor of $10^{6}$ to $10^{7}$ too low to detect the decay radiation of trace amounts of \pb{210} occurring with \SI{10}{ppb} lead in \SI{1}{\gram} of water. The fraction is even lower, given that the above calculations assume a point source and \SI{1}{\gram} of water measures a \SI{1}{\centi\meter\cubed}.\\
Independent from the actual \pb{210}$/$\pb{} ratio for commercial lead or lead in the water one can estimate the fraction of \pb{210} per mass (\textit{e.g.} \pb{210}$/$\pb{} or \pb{210} per gram of water) which corresponds to a $R_{^{210}\text{Pb}}^{\text{incident}}$ as the measured, minimal incident rate. Considering the same parameters as for Eqn. \eqref{cmosPaper:sec:summary:subsec:pb210eq:mrate} and using the two \am{241} $\gamma$-line energies we estimate a sensitivity between \SI{0.3(2)}{ppb} and \SI{166(11)}{ppb} \pb{210} per mass. This sensitivity is by itself not enough to reach the WHO limit, but makes the Neo sCMOS a competitive radiation detector.\\
Two more points need to be mentioned: the first estimate depends highly on the \pb{210}$/$\pb{} ratio; and, water can be concentrated by boiling it, potentially leaving heavy metals behind. In case of ashing of plants and acrylic \cite{GMM1994,HYang2016}, it has been found that heavy metals stay behind after the process. Furthermore, \pb{210} is not the only trace isotope potentially occurring together with stable lead. In case the radio-isotope to stable lead ratio is larger than the \pb{210}$/$\pb{} in \cite{HYang2016}, the $\gamma$ yield of these radio-isotope decays is larger than the one of \pb{210} -- the same is true when the water's volume has been reduced.

\section{Outlook and conclusion}

Based on the results above, we are currently examining how well \pb{210} is retained when water is concentrated by volume reducing it through boiling off. Furthermore, while our results do not allow to conclude that we can measure \pb{210} in low enough concentrations as needed to detect \SI{10}{ppb} of lead, our study has shown that a CMOS sensor optimised for optical wavelengths is well suited as $\gamma$- and x-ray detector for low energies in the range from $\sim\!\!\SI{10}{\kilo\electronvolt}$ to $\sim\!\!\SI{60}{\kilo\electronvolt}$. Follow up studies will establish the performance of commercial CMOS sensors and whether they are suitable for radio assay of materials.


\vspace{6pt} 



\authorcontributions{Conceptualization, A. Deisting and J. Monroe; 
methodology, A. Deisting and A. Dias and J. Monroe and J. Walding; 
software, A. Deisting and A. Dias and C. T\"urko\u{g}lu; 
validation, A. Deisting and A. Dias and C. T\"urko\u{g}lu; 
formal analysis, A. Deisting and A. Dias and C. T\"urko\u{g}lu; 
investigation, A. Deisting; 
resources, J. Monroe; 
data curation, A. Deisting and A. Dias; 
writing--original draft preparation, A. Deisting and A. Dias and C. T\"urko\u{g}lu;
writing--review and editing, A. Aguilar-Arevalo and X. Bertou and C. Canet and M. A. Cruz-P\'erez and A. Deisting and A. Dias and J. C. D'Olivo and F. Favela-P\'erez and E. A. Garc\'es and A. Gonz\'alez Mu\~noz and J. O. Guerra-Pulido and J. Mancera-Alejandrez and D. J. Mar\'{i}n-L\'ambarri and M. Martinez Montero and J. Monroe and S. Paling and S. J. M. Peeters and P. Scovell and C. T\"urko\u{g}lu and E. V\'azquez-J\'auregui and J. Walding; 
visualization,  A. Deisting and A. Dias and C. T\"urko\u{g}lu; 
supervision, J. Monroe and S. J. M. Peeters and J. Walding; 
project administration, J. Monroe and E. V\'azquez-J\'auregui; 
funding acquisition, X. Bertou and J. Monroe and E. V\'azquez-J\'auregui. 
All authors have read and agreed to the published version of the manuscript.}

\funding{This research was funded by STFC Global Challenges Research Fund (Foundation Awards, Grant ST/R002908/1), by STFC Grant ST/T506382/1, by DGAPA UNAM grants PAPIIT-IT100420 and PAPIIT-IN108020, and by CONACyT grants CB-240666 and A1-S-8960.}

\acknowledgments{The Authors wish to thank Ian Murray, Royal Holloway, University of London for his technical support.}

\conflictsofinterest{The authors declare no conflict of interest. The funders had no role in the design of the study; in the collection, analyses, or interpretation of data; in the writing of the manuscript, or in the decision to publish the results.} 



\appendixtitles{no} 

\reftitle{References}


\externalbibliography{yes}
\bibliography{cmos-bib.bib}




\end{document}